# State of
# Brain Emulation
# Report
# 2025

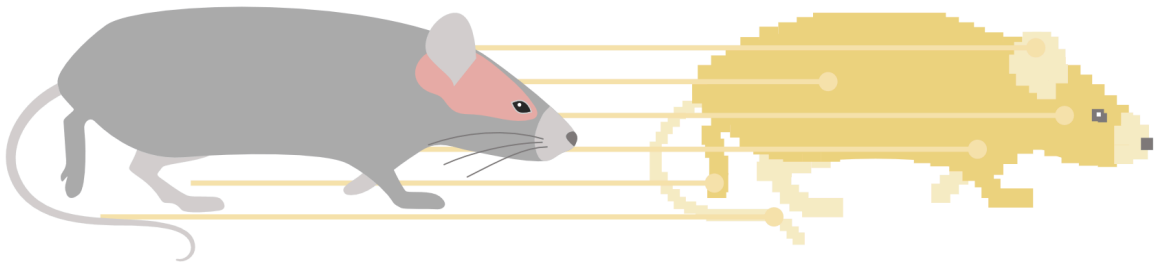

By Niccolò Zanichelli, Maximilian Schons, Isaak Freeman, Philip Shiu, Anton Arkhipov

Funded by

Fieldcrest Foundation    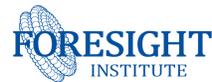

# Acknowledgements

We could not have written this report without the countless hours that experts worldwide gave us. We are deeply grateful for their pointers, repeated answers to our questions, and work in the field.

Adam Glaser, Adam Marblestone, Anders Sandberg, Andrew Payne, Andy McKenzie, Anshul Kashyap, Camille Mitchell, Christian Larsen, Claire Wang, Connor Flexman, Daniel Leible, Davi Bock, Davy Deng, Ed Boyden, Florian Engert, Glenn Clayton, James Lin, Jianfeng Feng, Jordan Matelsky, Ken Hayworth, Kevin Esvelt, Konrad Kording, Lei Ma, Logan Thrasher Collins, Michael Andregg, Michael Skuhersky, Michał Januszewski, Nicolas Patzlaff, Niko McCarty, Oliver Evans, Ons M'Saad, Patrick Mineault, Quilee Simeon, Richie Kohman, Srinivas Turaga, Tomaso Poggio, Viren Jain, Yangning Lu, Zeguan Wang

# Table of Contents

[Author Contributions](#)

[Competing Interests](#)

# Preface

It is hard to pinpoint numbers, but we estimate that less than 500 people globally are actively dedicated to the direct objectives of brain emulation. Even when considering those in overlapping areas of neuroscience, hardware, and software development, the total number likely still does not surpass 5,000 individuals worldwide. Such a small community means that every individual contributor's presence or absence can profoundly shape the field's trajectory. We hope this report will serve to attract new talent to this emerging and interdisciplinary endeavor.

This report offers a comprehensive overview of the field of brain emulation, detailing its current state, recent scientific and technological advances, and future challenges. It is intended for researchers, funders, and technical professionals seeking an understanding of this highly interdisciplinary domain.

It is important to note that this report does not claim exhaustive coverage. Our analysis focuses on five selected model organisms, offers only a glimpse into the vast topics of neuromodulation and neuroplasticity, and deliberately omits detailed discussions of transcriptomics or proteomics. As a primarily technical document, it also does not examine ethical and social considerations surrounding brain emulations, nor does it address questions pertaining to consciousness or personality preservation.

Future editions might aim to update the chapters below in light of scientific progress, providing a continuously evolving overview for interested audiences. All data referenced in this report is publicly available for further research and development (see [Data repositories](#)).

Niccolò Zanichelli, Maximilian Schons, Isaak Freeman, Philip Shiu, and Anton Arkhipov
September 2025

# Executive Summary

Whole brain emulation aims to create a computational model of a brain that matches its internal causal dynamics at a chosen level of biophysical detail. Achieving this requires three core capabilities: recording brain function, mapping brain structure, and emulation and embodiment. Since the publication of Sandberg and Bostrom's 2008 Whole Brain Emulation roadmap, all three capabilities have advanced substantially.

The scale and fidelity of neural activity recording have increased substantially. Electrophysiology has moved from a few hundred simultaneously recorded neurons to several thousand using high-density probes. Calcium imaging has expanded from a few thousand neurons to near-complete coverage of larval zebrafish brains and approximately one million cortical neurons in mice. Voltage imaging, capable of resolving single spikes, has progressed from hundreds of neurons to tens of thousands. Taking neuron count and sampling rate into account, these improvements represent about a two-order-of-magnitude increase in the effective data bandwidth of neural recordings since 2008. Beyond passive recording, causal perturbation methods, including optogenetics, now support proposals to systematically reverse-engineer neuron-level input-output relationships across entire small nervous systems such as *C. elegans*. Substantial obstacles remain across all modalities. Trade-offs persist among spatial coverage, temporal resolution, recording duration, invasiveness, and signal quality, and no method has yet delivered whole-brain, single-neuron, single-spike resolution in any organism when defined as capturing more than 90 percent of neurons simultaneously. Large-scale recordings from freely behaving animals also remain limited, constraining insight into neural function during naturalistic behavior. Tracking neuromodulators and neuropeptides at scale is even harder, with coverage still sparse.

Capabilities for mapping brain structure at the synaptic level have advanced in parallel. The field has moved beyond the original *C. elegans* connectome to produce multiple fully reconstructed adult fruit fly connectomes, and several larval zebrafish whole brains have been acquired at synaptic resolution, with segmentation complete and proofreading underway. This progress was enabled by higher-throughput electron microscopy and AI-driven reconstruction, which reduced human labor and pushed cost per reconstructed neuron from an estimated $16,500 in the original *C. elegans* connectome to roughly $100 in recent larval zebrafish projects. Expansion microscopy offers a complementary path that may scale more cost-effectively to mammalian tissue and enables dense molecular annotation, including protein barcoding for self-proofreading. However, substantial challenges persist. Proofreading remains an important bottleneck, particularly for mammalian neurons with large size and complex morphologies. Data volumes can reach exabyte scale even for small mammalian brains, creating storage and analysis burdens. Most current connectomes also lack rich molecular annotation, limiting their direct utility for functional interpretation and computational modeling.

The capacity to simulate and embody neural systems has also advanced. In *C. elegans*, connectome-constrained and embodied models now reproduce specific behaviors, while in the fruit fly, whole-brain models recapitulate known circuit dynamics. At the other end of the spectrum, feasibility studies on large GPU clusters have demonstrated simulations approaching human-brain scale, albeit with simplified biophysical assumptions. These achievements were driven by parallel advances in hardware and software. On the hardware side, the field has shifted from specialized CPU supercomputers toward more accessible GPU accelerators, which can simulate entire insect-scale nervous systems on a single device. On the software side, new frameworks have emerged that support not only more efficient simulation but also automatic differentiation, enabling more powerful, data-driven model fitting. The field's maturation is also reflected in the development of rigorous evaluation methods, including the first standardized benchmarks designed to predict future neural activity using both past functional recordings and the underlying structural connectome as inputs. Much like in neural recording and connectomics, however, significant obstacles remain. For mammalian-scale simulations, the primary hardware bottlenecks are now memory capacity and interconnect bandwidth, not raw processing power. A more fundamental limitation is that models remain severely data-constrained. Datasets that align structure and function from the same individual are exceptionally rare, and in the few instances where they exist, the structural connectome often lacks the molecular annotations needed to fully constrain models. Moreover, passive neural recordings alone are often insufficient to uniquely specify model parameters, and the ability to infer function directly from structure remains an unsolved challenge, highlighting the need for causal data from targeted perturbations. Finally, crucial biological mechanisms like neuromodulation are still largely omitted from models, and consistent integration with embodied, interactive environments is not yet standard practice.

Three cross-cutting challenges will shape the next phase of brain emulation research. First, the field lacks standardized evaluation criteria. To provide objective measures of progress, brain emulation requires stringent, multifaceted verification frameworks that move beyond simple activity prediction to include embodied behaviors and responses to controlled perturbations. Second, a central, unresolved question is which biological details are necessary for a faithful emulation. The functional importance of features ranging from gap junctions and specific proteins to non-neuronal elements like glial cells and neuromodulators remains poorly understood, making it difficult to prioritize data collection. Finally, a fundamental asymmetry exists between our ability to map structure and record function. While connectomes of ever-larger brains are becoming feasible, single-cell recording coverage in these same brains will remain limited for the foreseeable future. This necessitates the development of new methods to infer functional properties from structural and molecular data.

Addressing these challenges requires a two-pronged strategic approach. The first prong is a "detail-first" strategy in tractable organisms. By acquiring and integrating comprehensive functional, structural, and molecular datasets from the same individuals in *C. elegans*, larval zebrafish, and the fruit fly, researchers can empirically determine which biological factors are essential for faithful emulation, thereby de-risking efforts in more complex systems. Demonstrating validated emulations in these organisms will also be important for guiding and justifying the large-scale investments required for mammalian projects. The second prong is targeted technology development for the mammalian scale. This involves making connectome reconstruction cost-effective, developing robust methods to infer neural function from structure where comprehensive recording is impossible, and designing computational architectures that overcome memory and interconnect bottlenecks.

Successfully executing this strategy will require an evolution in how brain emulation research is organized and funded. The integrated, engineering-heavy nature of the challenges ahead calls for new organizational models to complement the vital contributions of academic labs. Focused research organizations (FROs) and specialized startups, supported by reliable long-term funding for talent and R&D, may be better suited to tackle the large-scale, coordinated efforts required to build the pipelines, tools, and datasets that will define the next era of brain emulation.

# Introduction

Seventeen years ago, Sandberg and Bostrom laid out an ambitious vision for whole-brain emulation, a concept then largely theoretical. The intervening years have seen substantial scientific and technological developments that now justify a careful reexamination of progress toward this objective. Here, we provide a detailed assessment organized around three fundamental capabilities: recording brain function, mapping brain structure, and emulation and embodiment. Our analysis concentrates on five commonly studied organisms, representing the primary systems that have driven most research relevant to whole-brain emulation, though we acknowledge that the organisms covered do not constitute an exhaustive list.

To emulate a brain, researchers must first understand its *neural dynamics*: the patterns of activity that emerge as organisms perceive, decide, and act. This functional mapping occurs *in vivo*, using recording methods that span a wide range of scales and invasiveness. Non-invasive techniques like fMRI and EEG can monitor whole-brain indicators of activity but lack the spatial and temporal precision needed to resolve individual neurons or single spikes. While valuable for human neuroscience and clinical applications, these approaches do not provide the detail necessary for emulation and receive limited coverage in this report. For that resolution, researchers turn to invasive techniques. Optical methods track activity in thousands of neurons simultaneously using fluorescent indicators and microscopy, while electrophysiological approaches insert electrode arrays to capture precise electrical signals from hundreds to thousands of cells. Beyond passive recording, perturbation methods like optogenetics allow researchers to selectively activate or silence specific neurons while observing network responses, enabling causal rather than merely correlational mapping of circuit function. These measurements fulfill two indispensable functions for emulation. First, they supply the constraints needed for parameter fitting: when integrated with structural data showing which neurons connect, functional recordings determine how those neurons behave, providing the firing rates, temporal patterns, and synaptic properties that connectivity alone cannot reveal. Second, they establish validation criteria: once model parameters have been fitted, the same recordings provide quantitative benchmarks that brain emulations must match to demonstrate biological fidelity.

Once functional data has been collected *in vivo*, the next step is mapping the brain's physical architecture. This is the domain of *connectomics*, which aims to reconstruct the organism's complete wiring diagram, and potentially its molecular composition, at synaptic resolution. This is an inherently destructive process conducted *post-mortem*. The brain is first chemically fixed to preserve its structure, then extracted and sliced into ultrathin sections, often just tens of nanometers thick. These sections are systematically imaged using techniques like electron microscopy (EM), which provides the resolution necessary to visualize individual synapses and subcellular structures. The resulting massive image datasets are computationally processed through AI-driven algorithms that align sections and trace neuron boundaries, followed by extensive human proofreading to correct segmentation errors. However, EM-based pipelines face two major limitations: proofreading dense mammalian tissue is labor-intensive and expensive, and EM captures minimal molecular information about neurotransmitter identities, synaptic receptor types, ion channel distributions, or the protein markers that define cell classes. These molecular features determine functional properties like synaptic strength and neural excitability, which are necessary for accurate model parameterization. These limitations have motivated the development of alternative approaches. X-ray microscopy can image much thicker tissue sections, which reduces the number of slices and simplifies reconstruction, though it remains an emerging technology. Expansion microscopy (ExM) has reached greater maturity and addresses both problems: it physically expands tissue to enable light-based imaging at synaptic resolution while preserving molecular labels that identify functional properties. Furthermore, molecular barcoding techniques assign unique identifiers to neurons, allowing computational algorithms to match disconnected fragments and substantially reduce manual proofreading. Together, these approaches produce structural maps that are essential for emulation: they define the network architecture by specifying which neurons connect to which others, and especially when enriched with molecular annotations, they constrain the parameter space so that functional models are built on a biologically plausible foundation.

With structural and functional data in hand, *computational neuroscience* seeks to instantiate the brain's causal dynamics *in silico*. This modelling work happens entirely in the computer, using experimental measurements as constraints. Neurons can be represented with varying complexity, from simple integrate-and-fire units to Hodgkin-Huxley formulations that explicitly track ion-channel kinetics. Synapses range from static weights to activity-dependent plasticity rules that strengthen or weaken according to spike history. When functional recordings are available for the target organism, parameters are tuned to reproduce observed firing rates, spike timing, and perturbation responses.

When such recordings are absent, parameters must be inferred from structural data alone, though this leaves many degrees of freedom unconstrained. The resulting network can then be evaluated in two ways: first, against held-out neural activity to test predictive accuracy; second, through embodiment in a simulated or robotic environment where sensory inputs drive the model and motor outputs are compared to the original behaviour. Discrepancies at either level reveal gaps in the model, whether missing structural details, insufficient functional constraints, or inappropriate modeling choices. This feedback is what makes emulation the integrative test of the entire pipeline: it identifies where data collection must be refined or expanded, closing the loop by directing future experimental work toward the measurements that are most likely to improve the next generation of emulations.

But why pursue brain emulation at all? The motivations for undertaking this complex, multi-stage process are diverse. For many scientists, the payoff is basic insight: a faithful model would provide unprecedented understanding of how perception, memory, and decision-making emerge from neural circuits under controlled conditions. Clinicians see a rapid-testing platform for neurological and psychiatric interventions, cutting the time and cost of *in vivo* trials. Others anticipate a path to AI systems whose native architecture mirrors our own, potentially easing the problem of aligning machine behaviour with human values.

Despite this transformative potential, fundamental questions remain unresolved, and experts hold disparate views on what will ultimately be required. Will faithful emulation demand neuron-by-neuron reconstruction, or must we descend to individual molecules and their conformational states? Should the scope encompass the entire body and nervous system, or can we isolate the brain, or even specific regions like the cortex or cerebellum? Numerous factors beyond neurons and synapses could prove necessary: glial cells, neuromodulatory peptides, hormones, ion concentrations, gap junctions, synaptic plasticity rules, and perhaps even the diffusion dynamics of signalling molecules. The relative importance of each remains uncertain, and this report does not attempt to cover every conceivable emulation possibility. Instead, we focus on the ingredients most commonly identified as necessary: electrical activity at single-neuron resolution, synaptic connectivity, molecular annotations that distinguish cell types and synaptic properties, and the computational frameworks needed to integrate them. Ultimately, determining which variables are essential is a strictly empirical question that must be resolved through the interplay between modeling and experimentation: the virtuous self-correcting loop described above. We present organism-level investigations first, followed by detailed chapters on neural dynamics, connectomics, and computational neuroscience. Readers new to the field may prefer to start with these methodological chapters before turning to the organism sections.

**Figure: Overview of the steps required for brain emulation** Using the complete toolkit of neurodynamics to measure variables influencing brain activity data while the organism is still alive, then destructive reconstruction by connectomic methods to digitally represent the brain's structure, and finally computational neuroscience approaches to use all of the data to create faithful emulation. The computational model predicts electrical neuronal activity constrained by reconstructed connections between neurons and potentially molecular substructure (individual transmitters and proteins, though other levels of detail are possible). During embodiment, brain activity needs to be decoded in behavior such as language or movement, and information from the environment, such as light and sound, as well as local or body-wide signal molecules like hormones or neuropeptides, needs to be encoded. This interplay must accurately capture the changes in brain structure and function over time, i.e., brain plasticity.

## 1 Neural Dynamics

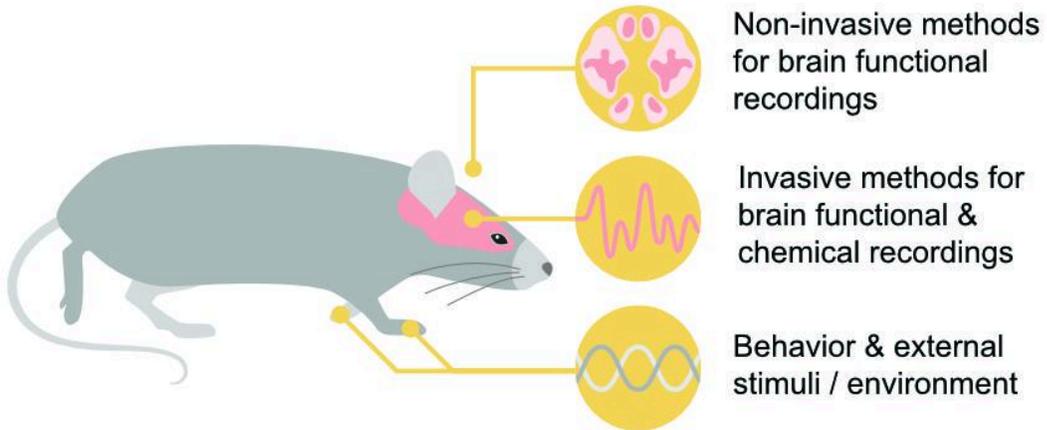

- Non-invasive methods for brain functional recordings
- Invasive methods for brain functional & chemical recordings
- Behavior & external stimuli / environment

## 2 Connectomics

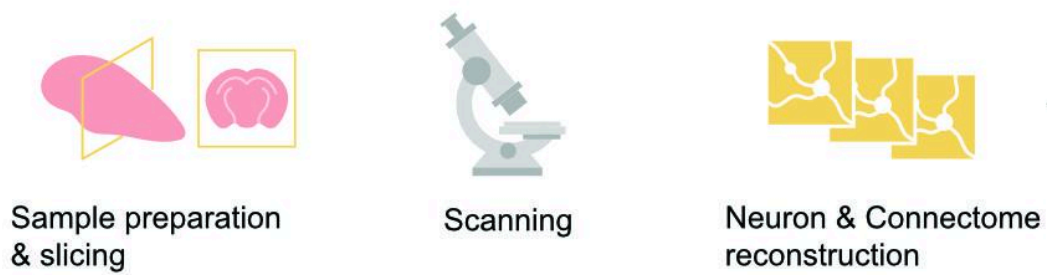

Sample preparation & slicing | Scanning | Neuron & Connectome reconstruction

## 3 Computational Neuroscience

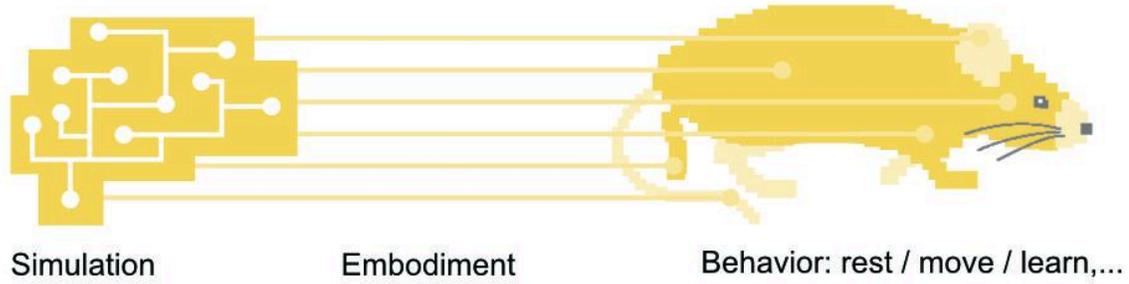

Simulation | Embodiment | Behavior: rest / move / learn,...

# Definitions

The terms 'simulation' and 'emulation' are often used interchangeably. In this report, however, we draw a sharp distinction, adopting the formal framework established in 2008 by Sandberg and Bostrom ([Sandberg & Bostrom, 2008](#)). In particular, we consider a model a *simulation* if, given the same inputs and initial conditions, it matches a target system's outputs without necessarily reproducing the internal causal dynamics that produced them at a chosen level of biophysical detail; and we consider a model an *emulation* if it matches the target system's outputs by *implementing* the same internal causal dynamics at a chosen level of biophysical detail, ensuring that it behaves like the original for the same underlying reasons.

The practical importance of this distinction is well-illustrated by considering a large language model trained on an individual's diary. Such a model functions as a simulation: its objective is to learn and reproduce the statistical patterns of the source text, enabling it to convincingly replicate the author's writing style and expressed views. It achieves high performance on this reference dataset (in-distribution behavior) but is likely to fail when generalizing to novel scenarios because it has not captured the causal structure of the brain that produced the text. This is due to a fundamental information bottleneck: the diary, as a low-bandwidth output, contains far less information than is required to uniquely specify the complex internal state of the brain that wrote it ([Sandberg, 2012](#)). As a consequence, when prompted about a childhood memory not mentioned in the diary, the simulation would most likely generate a plausible fabrication, whereas a true emulation would access the causal memory trace itself, or report its absence.

While the conceptual distinction is clear, operationalizing it requires specific criteria. To this end, we introduce the term *minimal brain emulation* to define the minimum properties we consider necessary for a model to be classified as an emulation rather than a simulation. It is crucial to note that meeting this baseline does not, by itself, say anything about whether an emulation is faithful or accurate. Faithfulness is a separate, quantitative measure of how well a model's predictions match a biological reference. Instead, this baseline establishes the necessary, but not sufficient, foundation for a model to even be considered a candidate for achieving high-fidelity emulation. The specific criteria that constitute a Baseline Emulation are detailed in the table below.

Finally, we wish to note that the term 'whole brain' lacks a standardized definition and is applied inconsistently across the literature. To ensure precision within this report, we adopt a quantitative threshold: a model is considered 'whole brain' if it incorporates at least 90% of the neurons present in the target biological brain.

**Table Scales for Defining Brain Emulations**: A) Observable Outputs of Simulations / Emulations. B) Dimensions of brain representations spanning from pure simulation to full emulation. The requirements in bold constitute a "minimal brain emulation" as defined by the authors.

| A) | Behavior Complexity | Personality-defining characteristics | In / Out-of-distribution | Learning horizon |
|---|---|---|---|---|
| No representation | None (no emergent behaviors modeled) | None (no personality traits represented) | No behavior | No Learning |
| | Simple outputs (basic stimulus-response patterns) | Basic traits (emerging from simple rules) | Limited in distribution behavior | Short term (seconds) |
| | Feedback loops (simple feedback between different neural systems) | Scattered traits (emerging from network interactions) | Full in-distribution behavior | Medium Term (days) |
| Complete representation | Complex behaviors (emerging from system interactions) | Complete profile (personality profile emerging from neural dynamics) | Out-of-distribution behavior | Long-term (years) |

| B) | Connectivity accuracy | Percentage of neurons | Cell types | Plasticity | Neuronal Accuracy | Neuromodulation | Temporal resolution |
|---|---|---|---|---|---|---|---|
| Pure simulation | Artificial (no real connectivity data, using random or artificial connections) | Small circuit – substantially less than a brain area. | No cell types (only one neuron type, e.g., like in typical ANNs). | Completely static network structure and synaptic parameters. | - | None (no neuromodulation signaling systems modeled) | No concept of time (ANNs) or very coarse representation of time (like "before" and "after"). |
| | Simple statistical rules (inspired by a few biological observations) | A whole brain area or multiple interacting brain areas. | Most basic cell types, e.g., excitatory vs. inhibitory neurons. | Some limited plasticity, such as short-term plasticity. | Neuronal networks (ANNs with units like ReLU, firing rate units, no spiking) | Single modulator (single neuromodulation signaling system, like dopamine only) | Slow relative to neuronal spiking activity (hundreds of milliseconds to seconds and more). |
| | Complex statistical rules (from systematic biological studies and/or partial connectomic information) | ~Whole brain | **Some cell type diversity, e.g., at the neuronal subclass level (such as Pvalb, Sst, Vip inhibitory subclasses, or differences between cortical layers).** | More extensive plasticity (e.g., STDP). | **Point Neurons (LIF)** | Multiple modulators (interacting neuromodulation signaling systems) | **Consistent with the scale of neuronal spiking activity.** |
| Minimal brain emulation | **Accurate connectome (complete, including verified synaptic connections)** | ~Whole body | Diverse cell types, possibly including non-neuronal cells. | Full dynamic plasticity, including growth and pruning of connections. | Neuronal Compartments | Complex system (interacting neuromodulation systems with feedback) | Fast (microseconds or faster, capturing details of fast subcellular processes). |

# Overview: State-of-the-art in brain emulation

## Neural dynamics

Despite impressive progress in recording capabilities, no organism has yet achieved whole-brain imaging at single-neuron resolution covering >90% of all neurons simultaneously. The closest achievements include larval zebrafish with approximately 80% brain coverage and C. elegans with roughly 50% of nervous system neurons recorded. These figures, however, come with substantial limitations: temporal resolution is typically well below neuronal firing rates (often 1-30 Hz for calcium imaging), recording durations remain short (minutes to hours), and the need for head-fixation severely constrains natural behavior. In larger organisms like mice, recordings focus on cortical regions or specific brain areas rather than whole-brain coverage, while human recordings are restricted to clinical settings and sample from extremely localized volumes.

Current methods fall into two broad categories, each with distinct trade-offs. Optical approaches, primarily calcium imaging, excel at capturing activity from large populations of neurons simultaneously (up to approximately one million in mouse cortex or tens of thousands in zebrafish and *Drosophila*) but suffer from slow temporal resolution that misses individual spikes in many neuron types. Electrophysiological methods like Neuropixels offer millisecond-precision spike detection but sample sparsely, typically recording from hundreds to a few thousand neurons along electrode trajectories. Voltage imaging with genetically encoded voltage indicators is emerging as a potential bridge between these extremes, with recent demonstrations approaching tens of thousands of neurons at spike-relevant speeds in larval zebrafish, though this technology remains in active development and recording durations are limited.

A fundamental challenge is that these methods primarily track electrical activity via calcium or voltage indicators. Monitoring the broader chemical context (neurotransmitters, neuropeptides, and other signaling molecules that critically shape circuit function) remains difficult. While genetically encoded neurotransmitter indicators have been developed for select molecules, they cover only a small fraction of the hundreds of neuromodulatory signals known to exist in these brains.

The path forward requires progress on several fronts. First, maturing voltage imaging to achieve spike-resolution whole-brain recordings in smaller organisms remains a primary goal, with extending recording durations being a particular challenge. Second, expanding behavioral freedom requires lighter microscopes, less invasive surgical preparations, and creative experimental setups that permit more natural movement patterns. Third, greater emphasis on causal rather than merely correlational data is needed. This means integrating large-scale recording with systematic perturbations: using anatomical connectivity as a scaffold to guide targeted optogenetic experiments that measure the effectome, a quantitative map of causal influence between neurons. Such experiments would ground computational models in measured functional interactions rather than inferred ones. Finally, developing molecular sensors for the broader range of neurotransmitters and neuropeptides present in these organisms will add the chemical dimension necessary to understand how circuit dynamics emerge.

However, obtaining comprehensive, single-neuron resolution recordings of whole mammalian brains faces severe physical constraints, and as a result will likely remain extremely challenging for the foreseeable future. The strategic priority should therefore be to create comprehensive, aligned datasets in accessible organisms where whole-brain recordings are feasible. These datasets must combine neural dynamics with causal perturbations and structural connectomes, enabling us to learn the mapping from structure to function. A robust model of this relationship, validated in a system where all variables can be measured, could then be applied to the more sparsely-sampled functional data produced from larger brains.

The following figure represents an aggregation of all studies available for this report. It highlights various trends: smaller organisms have more data available and are more likely to have significant data without fixation. Also, recording duration in all organisms is at least 2-3 orders of magnitude away from the entire life span. Finally, recording modalities with strong performance in one dimension will likely have poor performance in another.

**Figure: Heatmap plot of significant brain recording publications across different organisms.** The figure plots the relative distance from the respective organism's maximum value in a set of recording dimensions for a given publication. All papers referenced in the report and other noteworthy papers are listed. (data)

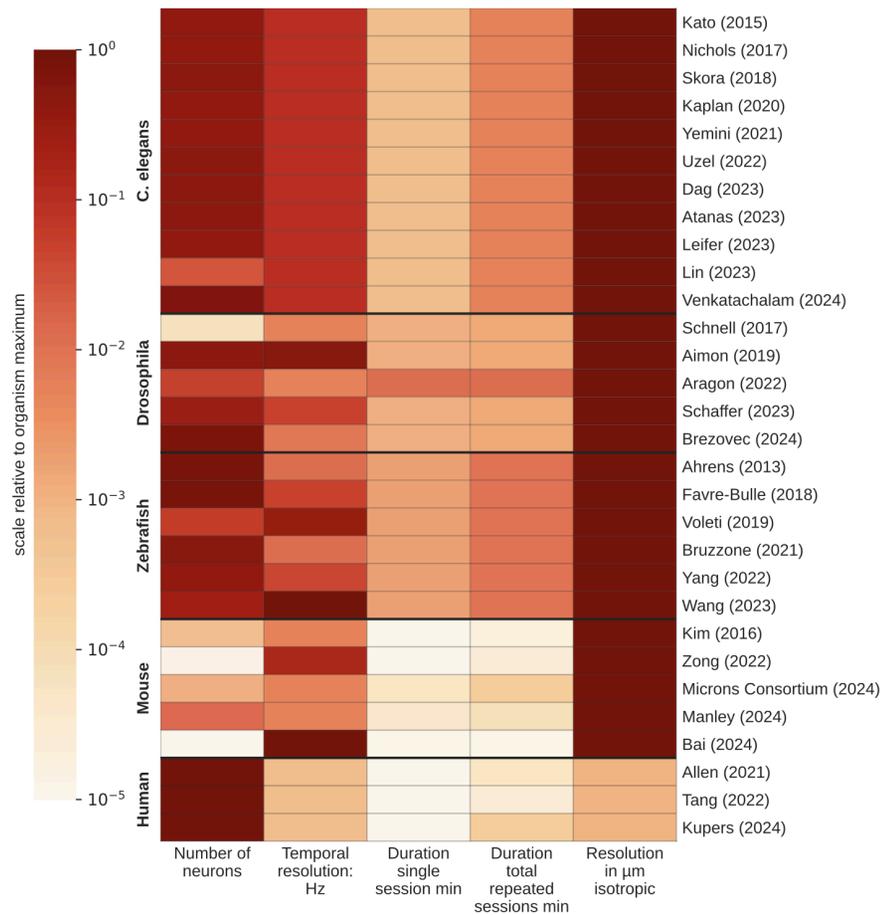

**Figure: Estimated instantaneous information rate of neural recordings over time.** Adaptation of Urai et al., 2022. This metric is defined as the number of simultaneously recorded neurons multiplied by their effective temporal resolution, capped at 200 Hz. This capped rate serves as a proxy for the instantaneous data bandwidth and provides a more consistent basis for comparison across different recording modalities. The 200 Hz cap balances the high-frequency capabilities of electrophysiology with the typical temporal dynamics of calcium imaging methods. Data points distinguish between Imaging (e.g., calcium imaging, light-sheet; blue circles) and Ephys (extracellular electrophysiology; red triangles), illustrating technological advancements. While this plot focuses on the simultaneous recording capacity, the total information acquired in an experiment also critically depends on the recording duration, a factor that varies widely and could be incorporated into future editions of this report. Horizontal dashed lines indicate theoretical maximum information rates for selected nervous systems (*C. elegans* body, fly brain, mouse cortex, whole mouse brain), calculated by multiplying their respective total neuron counts by the 200 Hz cap. These lines offer benchmarks for current experimental capabilities against the scale of these neural systems. (data)

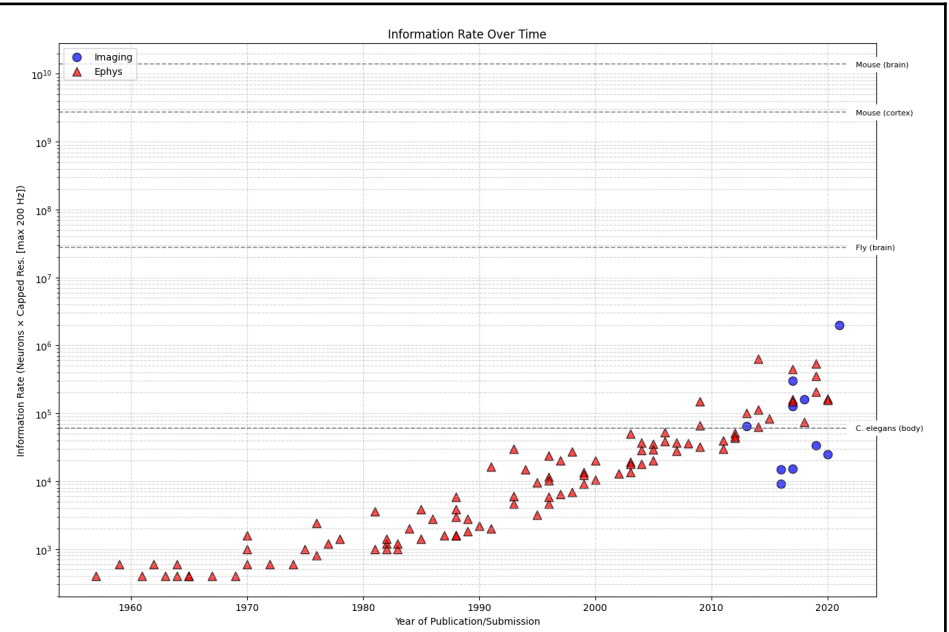

# Connectomics

Complete connectomes at synaptic resolution currently exist only for small organisms. C. elegans has multiple whole-nervous-system reconstructions from individual specimens, with approximately ten datasets available. Adult Drosophila has fully proofread connectomes for both the male central nervous system and the female brain, with another female CNS reconstruction in progress. Larval zebrafish has had its whole brain imaged at synaptic resolution multiple times, with reconstruction and proofreading efforts ongoing. For larger organisms, progress remains at the proof-of-concept stage. In mice, the largest densely reconstructed volume is a cubic millimeter of visual cortex, containing approximately 120,000 neurons and 523 million automatically detected synapses, with ongoing proofreading of a small fraction of neurons. Current efforts target 10 mm³ volumes, representing roughly 2-3% of the mouse brain. In humans, the largest synaptic-resolution volume is approximately 1 mm³ of temporal cortex (0.00007% of the whole brain), with only 104 neurons fully proofread from approximately 16,000 identified cells.

Electron microscopy provides the imaging foundation for nearly all existing synaptic-resolution connectomes. Recent technical advances include multi-beam systems that parallelize acquisition, achieving higher throughput for large volumes. However, EM workflows remain slow for mammalian-scale projects and provide minimal molecular information. Expansion microscopy is advancing rapidly on two fronts: high-expansion protocols have demonstrated effective lateral resolutions of approximately 20 nm, sufficient for dense reconstruction in mouse cortex while preserving protein-specific labels; separately, high-throughput systems have imaged entire mouse brains orders of magnitude faster than EM, albeit at lower resolution. X-ray microscopy offers another path to rapid, large-volume imaging. Synchrotron-based efforts have demonstrated cellular-resolution imaging of whole brains, and separate work has achieved the sub-40 nm resolution capable of resolving individual synapses under specialized laboratory conditions.

Irrespective of imaging modality, synaptic-resolution connectomics produces vast datasets that pose significant storage and analysis challenges. A mouse brain at 10 nm isotropic resolution would require approximately 1 exabyte of storage, while a human brain would require 2-2.8 zettabytes. These volumes necessitate specialized infrastructure and advanced, AI-based compression algorithms, with recent methods demonstrating storage reductions of up to 128x. Automated reconstruction has been a second major bottleneck. The latest AI-driven methods have improved key error rates by an order of magnitude or more compared to previous approaches, dramatically reducing the need for manual correction. This brings proofreading costs down to a level comparable with image acquisition, making exhaustive reconstruction of cubic-millimeter-scale mammalian brain volumes economically feasible. Expansion microscopy has also demonstrated proof-of-concept molecular barcoding techniques that enable automated matching of neuron fragments across spatial gaps, offering an alternative route to reducing manual proofreading requirements.

The path forward for connectomics requires advances on multiple fronts. Continued improvement in AI for automated segmentation, proofreading, intelligent imaging strategies, and data compression is essential for scaling to mammalian brains. Expansion microscopy holds significant promise for scalable, molecularly-annotated connectomics by integrating its demonstrated capabilities: high-throughput imaging, dense reconstruction with high expansion factors, and molecular barcoding for automated proofreading. These technical pursuits are particularly important because ex vivo structural mapping benefits from a key advantage over in vivo functional imaging: it is not constrained by the same physical limitations. Tissue can be chemically fixed, sectioned, expanded, and imaged over arbitrarily long timescales without the constraints imposed by maintaining a living organism. This makes mammalian-scale connectomics technically challenging but not fundamentally limited in the way that whole-brain functional imaging appears to be.

The strategic priority is twofold. First, learn and validate the structure-to-function mapping in small organisms where complete connectomes and whole-brain, causal recordings are feasible, yielding executable circuit models that reproduce dynamics and behavior. Second, scale connectomics to mammalian brains to produce the comprehensive structural blueprints on which those executable models can be instantiated, enabling faithful emulation of larger nervous systems.

> **Figure: Cost per quality-controlled reconstructed Neuron (inflation adjusted to 2025).** This plot uses best estimates on the end-to-end reconstruction costs (sample preparation & slicing, scanning, reconstruction & proofreading) for the three major connectomics initiatives of the past 40 years, *C. elegans*, Fruitfly, Zebrafish, and the estimates from experts for current costs. (plot, data)

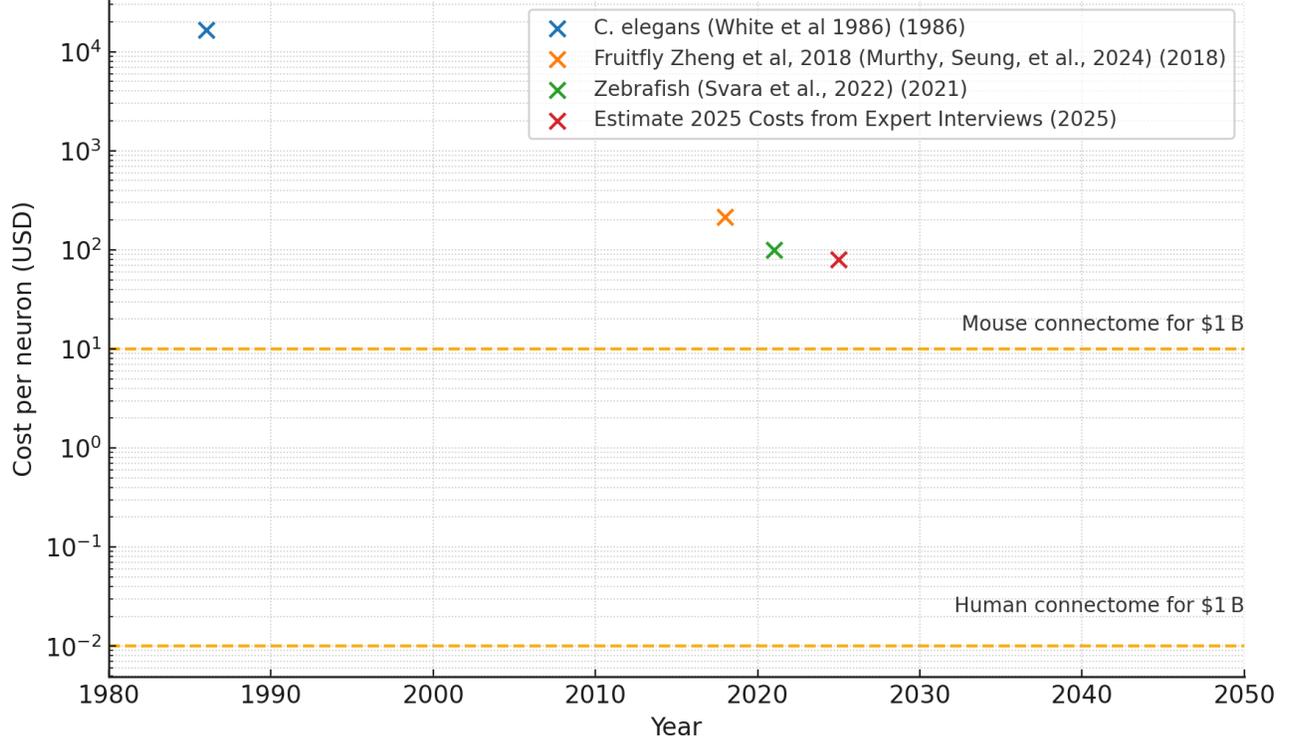

**Table Synaptic resolution Electron microscopy connectome reconstructions:** Complete overview of connectome reconstructions in the four model organisms. Additionally, multiple expansion and x-ray experiments are ongoing. Blue: scanned. Red: Scanned and traced.

| | | | |
|---|---|---|---|
| 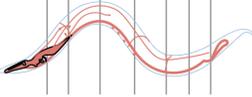 | "Original" composite *C. elegans* connectome. (EM). Synaptic resolution. ([White et al, 1986](#)) | 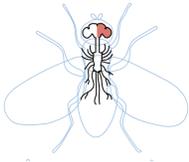 | *Drosophila*: One half-brain female individual. Synaptic resolution via EM ([Scheffer et al 2020](#)) |
| 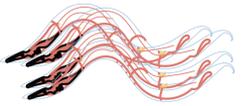 | *C. elegans*: 10 complete connectomes, one composite, including both sexes and five different developmental stages (EM). ([Varshney et al., 2011](#), [Cook et al., 2019](#), [Brittin et al., 2020](#), [Witvliet et al., 2021](#)) | 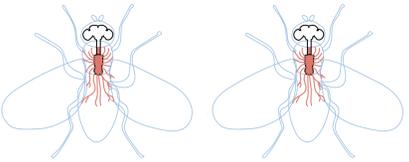 | *Drosophila*: Ventral nerve cord in female ([Azevedo et al., 2024](#)) and male individuals ([Takemura et al., 2024](#)) |
| 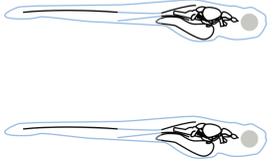 | Zebrafish: 10% of the spinal cord in an individual before sex differentiation. Brainstem ([Vishwanathan et al., 2024](#)) and spinal cord have also been reconstructed ([Svara et al., 2018](#)) | 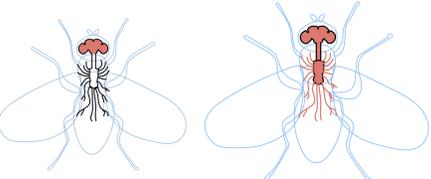 | *Drosophila*: One whole brain and one half of the central brain – in different female individuals ([Zheng et al, 2018](#), [Dorkenwald et al., 2024](#) and [Schlegel et al., 2024](#)). The entire male brain and nerve cord ([Berg et al., 2025](#)) Additionally, there is a complete connectome of the Drosophila larvae ([Winding et al, 2023](#)) |
| 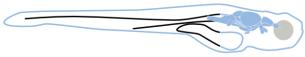 | Zebrafish: One whole brain in an individual before sex differentiation. ([Svara et al., 2022](#)) | 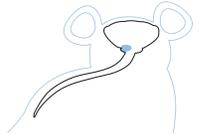 | Mouse: 1mm3 male mouse brain cortex (0.2% total brain volume). Synaptic resolution. ([Microns, 2025](#)) |
| 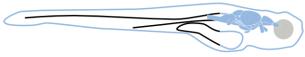 | Additional efforts are ongoing ([Lueckmann et al., 2025](#)). Note: during the editing process [two additional](#) projects were published: | 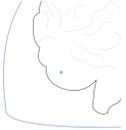 | 1 mm$^3$ female human brain cortex (not proofread, 0.000001% total brain volume). ([Shapson-Coe et al, 2024](#)) |

# Computational Neuroscience

Meaningful progress toward whole-brain emulation is currently confined to small organisms where comprehensive datasets are becoming available. In C. elegans, multi-scale, closed-loop simulations now reproduce basic behaviors by integrating neural dynamics, body mechanics, and environmental interaction. For Drosophila, the adult connectome has enabled models spanning the entire brain, successfully predicting neural responses and circuit functions for behaviors like feeding and grooming. Larval zebrafish modeling, while often circuit-specific, is driven by readily available whole-brain functional data, with proof-of-concept connectome-constrained simulations demonstrating accurate prediction of oculomotor integration dynamics, and embodied models replicating optomotor responses. With a full connectome for this organism expected soon, the field is poised for more integrated structural-functional models. In larger organisms like mice and humans, however, comprehensive emulation remains at the proof-of-concept stage. These efforts demonstrate building biophysically detailed cortical circuits by algorithmically inferring connectivity, or running human-scale simulations on supercomputers as feasibility tests. Across all scales, a central challenge is the scarcity of sufficiently dense functional and structural data to constrain model parameters. Furthermore, even with simplified neuron and synapse models, real-time mammalian-scale simulation strains computational resources, memory capacity, and interconnect bandwidth.

The path forward for computational neuroscience involves three coordinated strategic objectives. First, achieve high-fidelity emulations in small, tractable organisms by fully integrating their complete connectomes with rich, whole-brain functional and causal perturbation datasets. Second, within these same systems, develop and validate generative models that map molecularly-annotated structure to function. Such a mapping allows functional parameters to be inferred for larger mammalian nervous systems where comprehensive functional recordings are largely unavailable. Third, in parallel, optimize simulation software (leveraging modern accelerators and event-driven paradigms) and specialized hardware to significantly reduce computational and memory requirements for mammalian-scale emulations. These efforts collectively aim to bridge the gap from experimental data to biologically grounded, large-scale brain emulations.

As part of this report, simulation attempts for different organisms were rated on the following 0-3 point scale across 10 dimensions. No simulation attempt scores highly across all dimensions, and some cannot be found in any simulation attempt at all.

**Figure: Heatmap plot of computational brain models across different organisms.** The figure plots the score of a brain model across various dimensions. All papers referenced in the report and other noteworthy papers are listed. ([data](#))

**Figure Computational demands across organisms:** This figure illustrates the computational demands across compute and storage for various organisms and compares current state-of-the-art hardware against it. It uses point neurons and 5-compartment neurons estimates. For mice, a neuron is between 0.3 to 4 million FLOP/s and 15-30KB. This totals around 0.1 PetaFLOP/s and 1-2 TB of memory. Single GPUs like the Blackwell Ultra can calculate this fast, but as of today, they max out at 288 GB memory. Interconnect speeds depend on many setup variables and accurate connectomes, which is why the figure does not include estimates.([data](#))

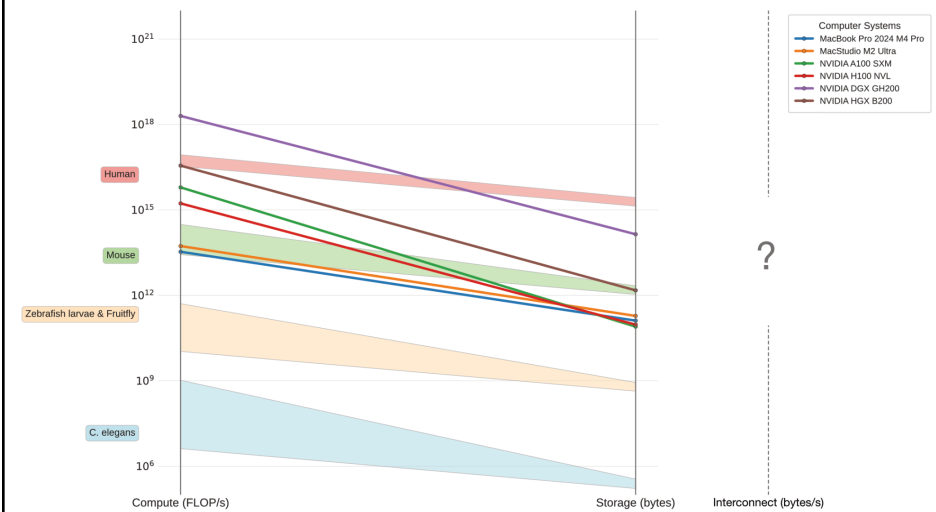

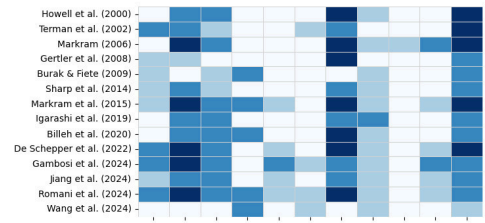

# State of Brain Emulation across Organisms

In the following chapters, we juxtapose key anatomical and behavioural features of *C. elegans*, larval zebrafish, *Drosophila*, the mouse and the human brain. The former constitute four commonly used animal models in neuroscience. However, many other relevant species exist that this report was not able to cover: from other small fish like Danionella ([Hoffman et al., 2023](#)), bees, ants, cockroaches, rats ([Herculano-Houzel and Lent, 2005](#); [Welniak-Kaminska et al., 2019](#), see also these [two](#) [videos](#)), to marmosets ([Herculano-Houzel et al., 2007](#); [Seki et al., 2017](#)) and rhesus macaques ([Dash et al., 2023](#); [Herculano-Houzel et al., 2007](#)) – among others.

The number of neurons varies dramatically across the animal kingdom. Some multicellular animals, like sponges, possess no neurons at all. Among those with nervous systems, counts range from organisms like Intoshia variabili with as few as 4-6 neurons ([Slyusarev et al., 2023](#)) to elephants with an estimated 257 billion neurons ([Herculano-Houzel et al., 2014](#)). Current single-neuron recording technologies allow for monitoring approximately 1,000,000 neurons in mice (~1% of the brain), and a key challenge for the upcoming decade in neuroscience is to scale this to 100 million neurons. However, further increases may eventually encounter physical limitations, potentially capping the maximum number of neurons that can be simultaneously recorded at single-cell resolution ([Marblestone, 2013](#)).

The information presented in the chapters is limited to data and methods that, in theory, allow for single-neuron resolution. Accordingly, barely any non-invasive recording modalities experiments or meso-scale microscopy with resolutions below synaptic resolution are discussed. Comprehensive discussions of general methodological details are discussed in the respective chapters after the organism section. Readers new to the field might want to first read those chapters, as organism chapters assume fluency across Neural Dynamics, Connectomics, and Computational Neuroscience.

Based on outlook sections of papers referenced and conversations with experts, we conclude each organism chapter with a model organism overview and gap analysis:

- Pros and Cons of Anticipated Scientific Insights and Experimental Tractability

- A non-exhaustive list of gaps and illustrative project opportunities

# C. elegans

## Anatomy & Behavior

*Caenorhabditis elegans (C. elegans)* is a worm the size of a grain of salt (~1mm long, 50 μm in diameter), or 0.002 mm³. Hermaphrodite worms have a total of 300 neurons across their bodies ([White et al, 1986](), [Shuhersky et al., 2022]()), with neurons distributed between a dense anterior "brain" region containing roughly half, and the remainder organized in ganglia and as motor neurons throughout the body ([Arnatkevičiūtė et al., 2018]()). Only a few of these neurons are thought to spike. For example, motor neurons are believed to rely on graded signals rather than firing action potentials, due to their reported lack of voltage-gated sodium channels ([Goodman et al., 1998]()). Specific *C. elegans* neurons, including sensory types and even certain motor neurons involved in behaviors like defecation, do fire all-or-none, calcium-based action potentials. However, the typical firing rates of these spiking neurons are not yet well characterized across the nervous system, especially under natural conditions. The *C. elegans* nervous system is known for its overall stereotypy. However, even in this system, studies indicate individual variability in synaptic wiring, with some chemical synapses differing between genetically identical animals ([Witvliet et al., 2021]()). This variability may contribute to individual behavioral differences. The typical worm lives for roughly two weeks at 20 °C ([Mack et al., 2018]()).

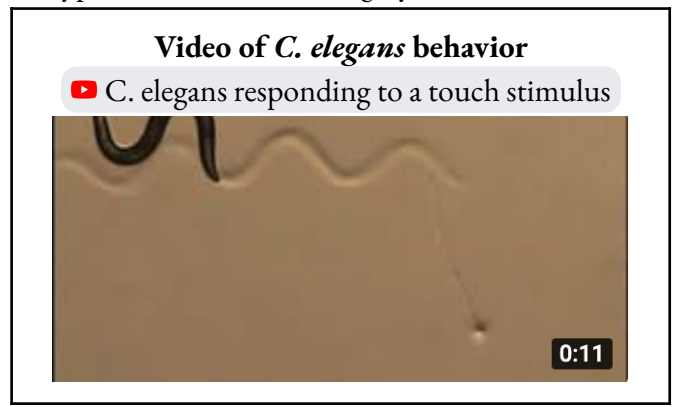

**Video of *C. elegans* behavior**
▶ C. elegans responding to a touch stimulus

*C. elegans* exhibits a diverse and adaptive behavioral repertoire, enabling it to thrive in dynamic environments. Its locomotion includes forward crawling, reversals, head movements, and pauses, which combine into higher-order states like "roaming" (rapid movement with few turns) and "dwelling" (slower movement with frequent turns). When food is removed, the worm executes a precise search strategy, starting with local exploration through sharp turns before transitioning to wider-ranging movement. It navigates chemical gradients to locate attractive substances like NaCl and temperature gradients to find its preferred thermal conditions. Remarkably, *C. elegans* can learn from experience, associating specific cues with food or harm. For example, it increases attraction to odors linked to food and avoids chemicals or temperatures associated with negative experiences. Males display additional behaviors, such as leaving food to search for mates, highlighting the worm's behavioral flexibility ([Sterling and Laughlin, 2015](); [Bainbridge et al., 2023]()).

Beyond wakeful behaviors, *C. elegans* exhibits sleep-like states, including developmentally timed sleep during larval molting and quiescence in response to stress or starvation. These states involve periods of reduced movement and responsiveness, akin to sleep in other animals. The worm also displays brief spontaneous pauses during feeding and prolonged quiescence after satiation, mirroring behavioral sequences seen in higher organisms. Together, these behaviors – ranging from basic locomotion and navigation to learning, memory, and sleep – underscore the remarkable adaptability of *C. elegans* and its ability to respond effectively to environmental challenges ([Flavell et al., 2020]()).

## Neural Dynamics

### Neural activity recording

Typical GCaMP-based approach at moderate laser power and short epochs allows repeated short sessions (5–15 minutes each) over several days, totalling a few hours of imaging across the life of a worm without severely compromising the animal. Proof-of-concept whole-brain or whole-body calcium imaging of *C. elegans* has been achieved, e.g., using light field microscopy ([Prevedel et al., 2014]()), a fully automated tracking platform to enable freely-moving imaging ([Li et al., 2021]()), improved detection, tracking, and segmentation algorithms ([Wu et al., 2022](); [Lanza et al., 2024]()). Recent advances in live animal labeling, such as NeuroPAL ([Yemini et al., 2021]()), create unique fluorescent labels of each neuron in the worm, allowing for almost unambiguous identification of each of the 300 neurons in freely behaving worms, though activity recording itself requires separate, co-expressed functional indicators (e.g., GCaMP). No large-scale whole-body neuron (i.e., recording from 90% of all neurons) activity datasets of *C. elegans* are available today ([Sprague et al., 2024](); [Simeon et al., 2024]()). Atanas et al. performed imaging of approximately 60 freely moving animals for a total of ~1,000 minutes of imaging, measuring the activity of approximately 150 head neurons ([Atanas et al., 2023]()). Individual sessions per animal were 16 minutes, often split into two 8-minute segments, and temporal resolution was 1.7 Hz. They used a nuclear calcium indicator (NLS-GCaMP7f) and a spinning disk confocal for volumetric fluorescence imaging, coupled with brightfield imaging of the worm's behavior. After processing, this generated about 30 GB.

**Figure: Overview of the optical neural recording landscape in *C. elegans*:** Radar plots based on the optical recording literature cited in the report. We plot the following dimensions of brain recordings: spatial resolution, brain volume, temporal resolution, and (estimated) individual and cumulative recording duration. The plots split recordings from fixated (A) and freely moving experiments (B). The outer ring is normalized to the maximum known values. Each ring represents one order of magnitude. The data for this figure is available in the linked data repository.

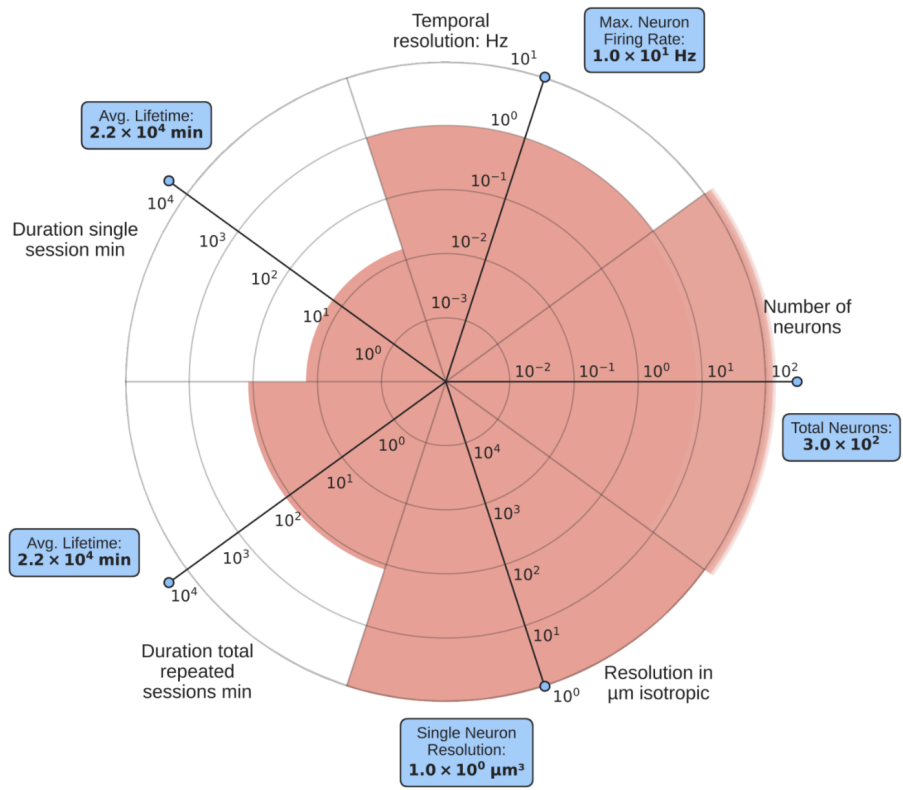

A) fixated

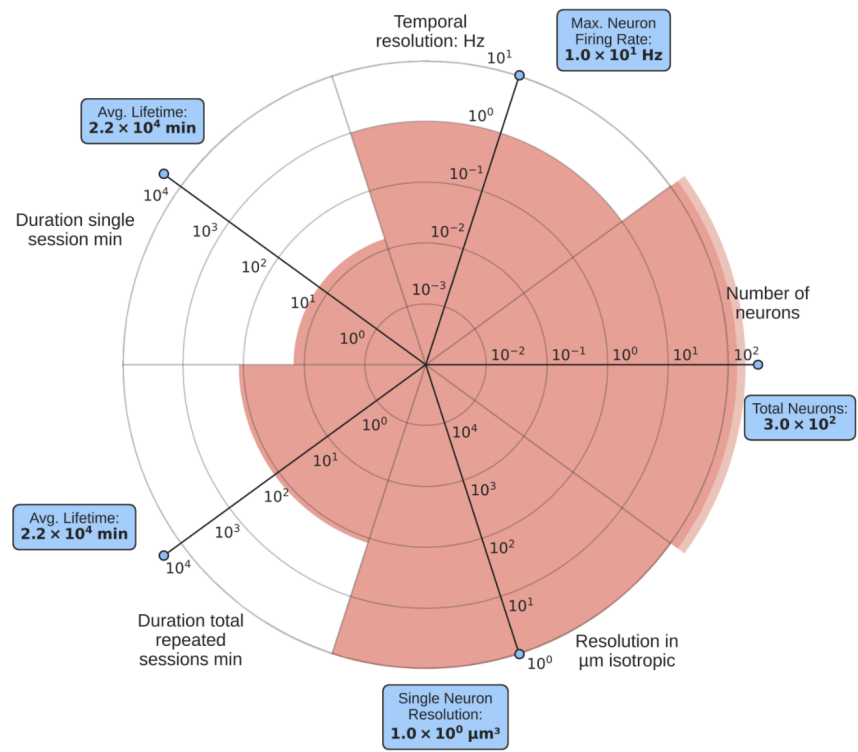

B) moving

While patch-clamp recordings in *C. elegans* offer millisecond precision for directly measuring electrical activity in neurons and muscles, this technique is practically limited to recording from only one or a minimal number of cells at a time, making it best suited for detailed studies of specific ion channels and synaptic connections. The main drawbacks are that it requires careful dissection of the worm's protective outer layer and can only record from immobilized animals, making it impossible to study neural activity during behavior ([Goodman et al., 2012](#)). Voltage imaging with GEVIs offers a complementary approach, enabling researchers to simultaneously measure electrical activity across multiple cells in living, behaving worms. The main limitation is that GEVIs produce relatively weak fluorescent signals compared to calcium indicators. However, recent advances using rhodopsin-derived indicators, such as Arch-based GEVIs, have improved their brightness and sensitivity ([Hashemi et al., 2019](#)). Tokunaga et al. recorded voltage in individual neurons from several hundred individuals for 30-60s at 20-250 Hz ([Tokunaga et al., 2024](#)).

## Neurotransmitters and Neuromodulators

The *C. elegans* genome encodes over 300 neuropeptides processed from approximately 160 precursor genes, along with around 150 peptide G protein-coupled receptors (GPCRs) ([Taylor et al., 2021](#)). Recent large-scale deorphanization efforts have identified 461 peptide-GPCR signaling pairs, providing unprecedented insight into the organization of peptidergic signaling ([Beets et al., 2023](#)). This work revealed that while some peptide-GPCR interactions are highly specific, others display complex combinatorial patterns - individual neuropeptides can activate multiple receptors, and single receptors can respond to various peptides ([Beets et al., 2023](#); [Ripoll-Sánchez et al., 2023](#)). Functional studies indicate that neuropeptides operate across various timescales in *C. elegans*, from seconds to hours. While typically considered slow modulators, some neuropeptides like NLP-40 can trigger responses within seconds. Recent whole-brain imaging combined with optogenetics has revealed that many rapid functional connections between neurons depend on dense-core vesicle release, suggesting neuropeptides extensively shape fast neural dynamics ([Randi et al., 2023](#)). The peptidergic network appears organized into distinct functional modules, including specialized "hub" neurons rich in dense-core vesicles and highly connected through peptidergic signaling ([Ripoll-Sánchez et al., 2023](#)).

## Perturbation

Due to its anatomy, *C. elegans* represents an exemplary model organism for perturbation studies. Since the first expression of optogenetic proteins in *C. elegans* in 2005 ([Boyden et al., 2005](#)), optogenetic studies have revealed numerous insights into neural circuit function ([Piatkevich and Boyden, 2023](#)): neurons controlling locomotor rhythms ([Fouad et al., 2018](#)), interneurons that integrate multiple olfactory inputs to represent valence (Dobosiewicz et al., 2019), single neurons that can both encode chemotactic memory (Luo et al., 2014) and regulate multiple behavioral outputs (Li et al., 2014), circuits mediating behavioral state switching in response to oxygen levels (Laurent et al., 2015), and specific interneurons controlling chemotaxis programs (Kocabas et al., 2012). Additionally, optogenetics has revealed how synaptic energy demand regulates glycolytic protein clustering and identified neurons that contribute oscillatory activity controlling backward locomotion (Piatkevich and Boyden, 2023). Recent technical advances have dramatically expanded the scale and precision of perturbation studies in the worm. Sharma et al. developed TWISP (Transgenic Worm for Interrogating Signal Propagation), a strain expressing both optogenetic actuators and calcium indicators in all neurons while avoiding optical crosstalk –unwanted interference between light signals used for neural activation and activity measurement (Sharma et al., 2023). Moreover, Randi et al. systematically measured signal propagation between 23,433 pairs of neurons in the worm's head through direct optogenetic activation combined with whole-brain calcium imaging, revealing that actual signal flow often differs from predictions based on anatomical connectivity due to extrasynaptic signaling (Randi et al., 2023).

## Connectomics

The relative simplicity and high degree of stereotypy in *C. elegans* neuroanatomy made it the target for the first complete connectome mapping effort (Brenner, 1974, Brenner, 2002). In 1986, White and colleagues published this landmark reconstruction, detailing approximately 5000 chemical synapses, 2000 neuromuscular junctions, and 600 gap junctions (White et al, 1986). However, as discussed in later analyses, this connectome was a composite, painstakingly assembled from electron microscopy sections of multiple, different individuals: specifically, three adult hermaphrodites, one L4 larva, and one adult male. This mosaic approach, necessary due to the technical challenges of the time, combined with potential preparation-induced distortions and significant manual curation, resulted in a generalized model. While groundbreaking, this method inherently introduced variability and did not fully capture the precise neuron positions or individual idiosyncrasies now being addressed by modern atlases (Skuhersky et al., 2022). The underlying assumption of largely consistent neuronal positions and connectivity across individuals (Varshney et al., 2011) was nonetheless crucial for this effort.

This initial mapping was later complemented by the reconstruction of the male-specific nervous system, which contains approximately 80 additional neurons primarily involved in mating behaviors (Jarrell et al, 2011). A comprehensive re-mapping effort by Cook et al. (2019) produced updated connectomes for both sexes, correcting earlier errors and introducing synaptic strength weights based on measuring how many consecutive tissue sections each synapse spanned in the electron microscope images ([Cook et al, 2019](#)). More recent work has expanded into developmental neurobiology, with reconstructions across five different developmental stages ([Witvliet et al, 2021](#)), while continuing refinements driven by recent advances in neuron identification, alignment, processing, and scale have improved both accuracy and completeness ([Skuhersky et al, 2022](#); [Brittin et al, 2021](#)). To date, *C. elegans* remains the organism with the most individual connectomes reconstructed. This includes approximately ten datasets detailing its complete brain ([Witvliet et al., 2021](#)) or main neuropil ([Brittin et al., 2020](#)), complemented by two whole-animal connectomes ([Cook et al., 2019](#)) and a near-complete somatic nervous system wiring diagram ([Varshney et al., 2011](#)), all from distinct individual specimens. However, due to the highly stereotyped nature of *C. elegans* neural connectivity, the marginal scientific value of additional individuals may be limited compared to other organisms with more variable nervous systems.

While electron microscopy provided the foundational *C. elegans* connectome maps, these structural maps alone do not capture the rich molecular complexity that supports the worm's behaviors. Despite having only 300 neurons, *C. elegans* relies heavily on sophisticated molecular machinery at each synapse, with the worm's postsynaptic proteins amounting to about half the number for mammals ([Emes et al., 2008](#)). Optical approaches like expansion microscopy have emerged as crucial complementary tools to understand this molecular complexity. The first *C. elegans*-specific expansion microscopy protocol (ExCEL) enabled immunostaining for molecular identification and tissue expansion through innovative cuticle permeabilization techniques ([Yu et al., 2020](#)). With standard ExCEL achieving 3.5x expansion and iterative ExCEL reaching up to 20x expansion (25 nm resolution), these methods may resolve individual synaptic connections while preserving the critical molecular information that EM cannot capture. However, achieving effective and reliable expansion microscopy in *C. elegans* remains challenging, with the worm's cuticle often acting as an impermeable barrier, severely limiting chemical labeling and detection to organs directly exposed to the external environment ([Kuo et al., 2024](#)).

Given a total body volume of 2-6 million μm³ for an adult hermaphrodite worm, imaging at 10nm isotropic resolution would theoretically generate a total of 2-6 × $10^{12}$ voxels. At 2 bytes per voxel for a single channel, this would theoretically require 4-12 terabytes of storage, scaling linearly with the number of colors. Historically, early attempts at computer-assisted reconstruction in the 1970s by White et al. highlight the immense computational hurdles of the era. The 'Modular I' computer they employed, a machine with only 64 KB of memory and 22 MB of storage that required custom-written operating systems in assembly language, proved insufficient for the task ([Ermons, 2015](#)). Consequently, they resorted to the painstaking manual annotation of printed electron micrographs using colored Rapidograph pens to trace neurons through serial sections. The tracing and reconstruction of the *C. elegans* required years of painstaking work, with the project taking 15 years to complete. More recent efforts, such as Cook et al., have leveraged specialized software tools like Elegance (Xu et al., 2013), resulting in speedups of several orders of magnitude, and recent advances in machine learning and computational processing have sped up the process further.

## Computational Modelling

The history of *C. elegans* emulation is marked by ambitious efforts that, while pioneering, were ultimately constrained by the nascent state of neurotechnology and a critical lack of vertical integration. Early conceptual work, such as biophysical locomotion modeling (Neibur and Erdös, 1991) and planned, though unrealized, comprehensive models in the late 1990s, signaled the field's aspirations. The Virtual *C. elegans* project (Suzuki et al., 2005) made headway by simulating motor control but had to approximate missing biophysical parameters using machine learning, highlighting the data gaps. Later, more comprehensive initiatives like the community-driven OpenWorm (Szigeti et al., 2014), initiated around 2011, and academically-led projects such as Nemaload (2011-2013), which aimed to leverage the then-new optogenetic tools, also faced these fundamental limitations. At the time, even with the static connectome (White et al., 1986) available, the technologies for detailed, dynamic data acquisition were immature. Optogenetic perturbation for causal inference only emerged in *C. elegans* in 2005. As discussed above, large-scale neural recordings in the worm were, and to a large degree still are, in their infancy (Stiefel and Brooks, 2019) – partly due to the dynamic nature of the worm's nervous system, which, with its constant motion and deformation, made tracking of individual neurons a challenge (Nguyen et al., 2017). This confluence of technological limitations in neural recording, perturbation, tracking, and imaging meant that the kind of comprehensive, correlated dynamic data needed to truly understand and emulate the worm's nervous system simply was not available during those times, directly contributing to the development of Focused Research Organizations (FROs, Marblestone et al., 2022), designed to be capable of tackling projects requiring such a high degree of systems integration.

Only now, with the advent of high-throughput connectomics, advanced imaging techniques, and increasingly sophisticated computational tools, the field is beginning to amass the data necessary to make meaningful progress towards *C. elegans* emulation. While no FRO dedicated to C. elegans exists, recent integrative models like BAAIWorm demonstrate important initial successes, such as replicating basic sensory integration and realistic zigzag locomotion behavior (Zhao et al., 2024). In the following, we discuss the most prominent computational modelling attempts.

## OpenWorm

OpenWorm, launched in 2011, represents one of the most comprehensive attempts to create an integrative biological simulation of *C. elegans*, implementing a multi-scale biophysical approach (see [Virtual Worm Project](#) "worm body") that spans from ion channel dynamics to whole-organism behavior ([Sarma et al., 2018](#)).

The simulation architecture integrates multiple types of experimental data, though significant gaps remain. Connectome data from electron microscopy provides the basic network structure ([White et al., 1986](#)), while calcium imaging data informs neural dynamics. However, a major challenge is the limited availability of electrophysiological data - patch-clamp recordings exist for only a small subset of ion channels, necessitating homology-based inference from other organisms for parameter estimation ([Sarma et al., 2018](#)).

The project's core simulation stack consists of several interconnected frameworks (see figure below), enabling simulation at multiple levels of abstraction - from basic integrate-and-fire neurons to detailed multi-compartment models incorporating Hodgkin-Huxley dynamics ([Gleeson et al., 2018](#)). A critical limitation in the current implementation is the unidirectional flow of information from neural simulation to body mechanics, though work is ongoing to incorporate sensory feedback ([Sarma et al., 2018](#)).

> **Replication from Figure 1 from ([Sarma et al, 2018](#)). Overview of OpenWorm Simulation stack** a) A component diagram describing the relationships between inputs and outputs of sub-projects within OpenWorm b) A highly simplified schematic view of the system of equations executed in the combined c302/Sibernetic system.

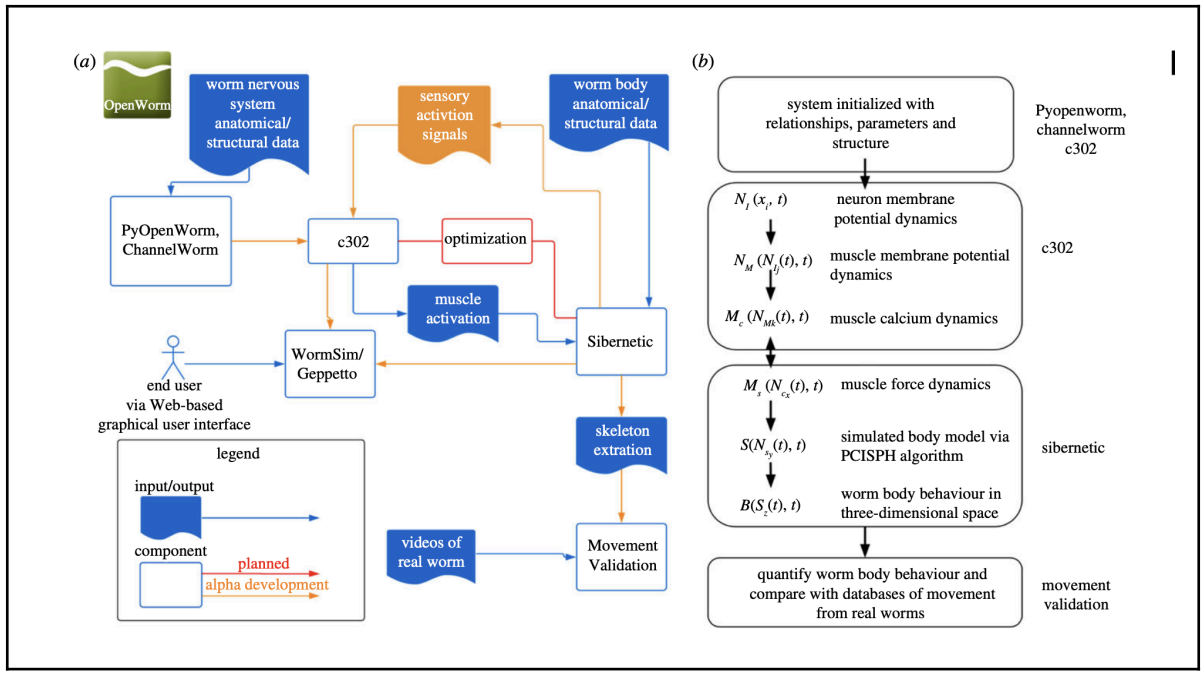

The project employs sophisticated validation frameworks, particularly SciUnit, to systematically compare model outputs against experimental data across multiple scales (Omar et al., 2014). Behavioral validation compares simulated movement patterns against real worm tracking data, while neural validation involves comparing activity patterns with calcium imaging recordings (Javer et al., 2018). The current implementation can demonstrate basic locomotion, though matching the full repertoire of *C. elegans* behaviors (including chemotaxis, thermotaxis, and learning) remains a future goal. These limitations reflect computational challenges and gaps in biological understanding of how sensory information is processed to generate behavior.

> **Figure of OpenWorm simulation:** Demonstration of a computational model of the *C. elegans*. (Github page, 2024)

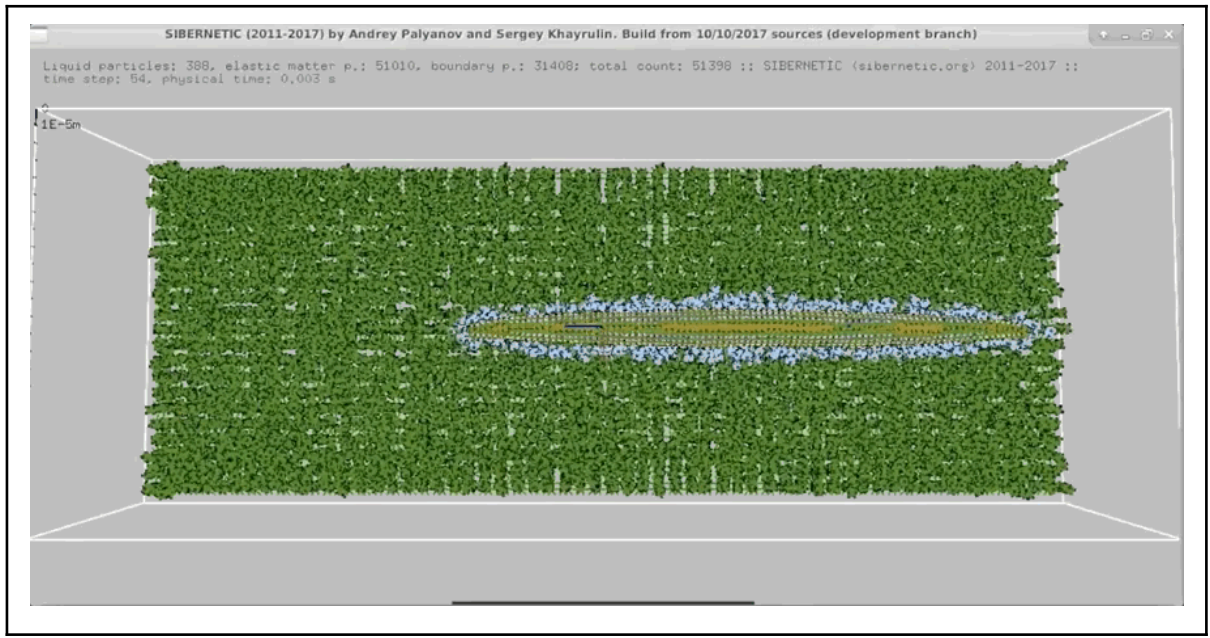

The OpenWorm project makes its tools and simulations accessible through the Geppetto visualization platform (Cantarelli et al., 2018), enabling web-based exploration of models and data. This infrastructure, combined with the project's open-science approach, provides a foundation for community-driven refinement of whole-organism simulations (Larson, 2021).

Simeon et al, 2024

Recent approaches leveraging artificial neural networks (ANNs) offer a complementary data-driven strategy for modeling *C. elegans* neural dynamics, with particular emphasis on capturing the inherent predictability of neural activity patterns without requiring explicit biophysical modeling (Simeon et al. 2024). This represents a shift from traditional bottom-up approaches toward letting the underlying structure emerge from experimental data.

Simeon et al.'s approach makes extensive use of calcium imaging datasets, integrating recordings from multiple experimental sources spanning different behavioral contexts - from freely moving to immobilized worms, and from sleep to optogenetically stimulated states. The combined dataset represents neural activity from 284 worms. A key innovation is the standardization of data processing across sources, including z-scoring of calcium signals and temporal alignment, enabling the pooling of data despite varying experimental conditions.

The approach currently focuses on neural activity prediction without direct connection to behavioral outputs. The simulation framework compared multiple neural network architectures, including Long-Short Term Memory (LSTM) networks, Transformer networks, and Feedforward networks. All architectures shared a common structural framework consisting of an embedding layer for neural state representation, a core processing module, and a readout layer for prediction.

Model validation focuses on next-time-step prediction accuracy, using a teacher-student framework where the biological nervous system serves as the teacher and ANN models as students. The approach employs a 50:50 temporal split for validation, where models are trained on the first half of neural recordings and tested on the second half.

Current implementations demonstrate success in short-term prediction of neural activity patterns for up to approximately 20 seconds (accurately predicting roughly 20-30 timesteps of ~0.7s each), with recurrent models (e.g., LSTM) showing superior performance compared to other architectures with the current dataset. However, long-horizon predictive capabilities remain limited. It is worth noting that this represents a prediction of autonomous neural activity without direct modeling of sensory input or behavioral context. The models show consistent scaling properties across different experimental conditions, suggesting they capture fundamental aspects of neural dynamics. CTRNN models, in particular, exhibited the best scaling properties; for instance, their prediction error consistently decreased with more training data, such that doubling the amount of training data reduced the prediction error by a factor of about 1.57. This scaling behavior allows for an estimation of data requirements for longer predictions. To extend predictive success from a baseline of approximately 20 timesteps to the full 180 timesteps the dataset would need to be approximately 30 times larger than the current one. This translates to requiring recordings from roughly 8,400 worms (up from the current 284), highlighting the substantial data requirements for achieving robust long-horizon predictions of autonomous neural dynamics using this approach.

### Zhao et al, 2024

BAAIWorm, introduced in 2024, represents a significant advance in integrative biological simulation of *C. elegans* by implementing a closed-loop system bridging brain dynamics, body mechanics, and environmental interactions ([Zhao et al, 2024](#)). The project builds upon OpenWorm's foundational contributions, particularly its cell model morphologies, synaptic dynamics, and 3D body representations, while introducing crucial new capabilities for real-time simulation and behavioral feedback.

The model integrates diverse experimental data, including ion channel dynamics, neural morphologies, electrophysiology, and whole-brain calcium imaging. A notable strength is validating single-neuron models against patch-clamp recordings for five representative neurons (sensory, inter-, command, and motor neurons), with parameters for other neurons derived through functional grouping. The body-environment component leverages detailed anatomical data to construct a biomechanical model comprising 3,341 tetrahedra and 96 muscles.

The neural network model implements 136 neurons as multicompartmental models with sub-2μm compartments, incorporating 14 types of ion channels. Rather than enforcing strict neurotransmitter constraints, the model uses an optimization approach to determine synaptic properties that reproduce observed dynamics. The body-environment simulation employs simplified but efficient hydrodynamics, enabling real-time simulation at 30 frames per second while maintaining behaviorally realistic movement patterns.

The model demonstrates multi-scale validation, from single-neuron current-voltage characteristics to network-level correlation matrices (achieving 0.076 mean squared error against calcium imaging data) and behavioral reproduction of chemotaxis. While the current implementation focuses specifically on zigzag locomotion, synthetic perturbation experiments reveal important insights about neural circuit function, notably, the absence of neurites or synaptic/gap junctions disrupts global neural dynamics and impairs forward motion. A novel Target Body Reference Coordinate System enables precise quantification of movement patterns, providing a stable framework for comparing simulated and biological behavior.

While BAAIWorm represents a significant advance in integrating different scales of biological simulation, several challenges remain, including expanding to the complete 300-neuron network and incorporating additional behaviors beyond chemotaxis.

**Replication from Figure BAAIWorm moving towards simulated chemicals**. Video showing the simulated worm (ref)

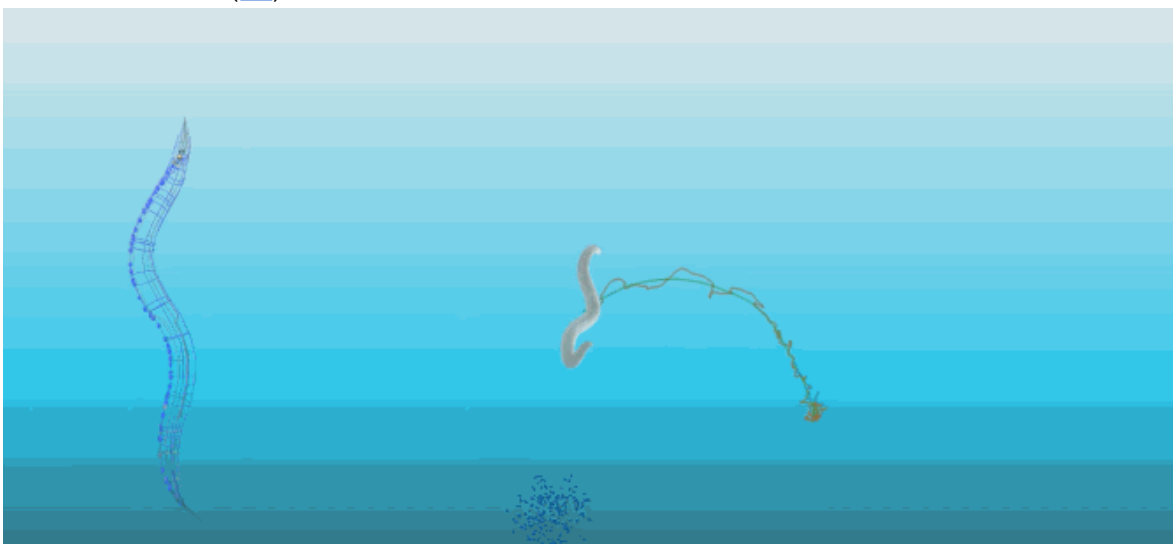

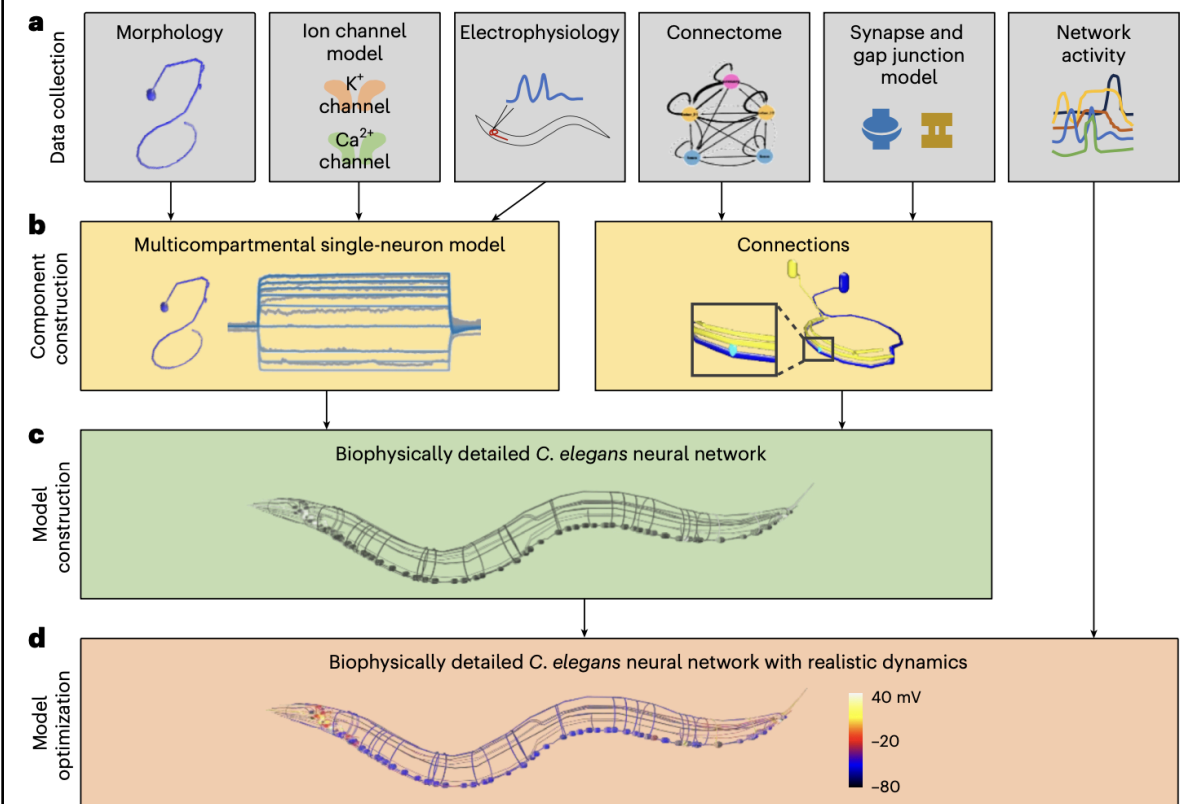

**Replication from Figure 1 from Zhao et al, 2024** A) Experimental data collection to constrain models. These data include neural morphologies, ion channel models, electrophysiological recordings of single neurons, connectome, connection models and neural network activities. b, The construction of multicompartmental neuron models and connection models (synapses and gap junctions). c, The biophysically detailed *C. elegans* neural network model without functional neural activities. d, Optimization of the biophysically detailed *C. elegans* neural network model to achieve realistic network dynamics. Neurons are color-coded to represent membrane potential.

# Gap Analysis

In a 2021 Interview, the long-time OpenWorm lead Stephen Larson advocated that the *C. elegans* emulation ecosystem would benefit massively from an integrative approach, the "Allen Institute for *C. elegans* research", that brings the broad range of interdisciplinary experts together ([Larson, 2021](#)). In 2023, ([Haspel et al., 2023](#)) proposed systematically determining the input-output functions of every neuron through comprehensive perturbation experiments in order to fully reverse engineer the entire *C. elegans* nervous system and to simulate its full range of behaviors. The project assumes it will need thousands of hours of experiments, but has not disclosed specific numbers or funding proposals publicly.

A concentrated, interdisciplinary and sustainably funded effort could likely provide conclusive answers to many of the above mentioned points and could mark a transition away from grants aimed at narrow research questions (e.g., genetics, single-neuron knockouts) towards a large-scale integrative project across subject matter experts.

| Model Organism Overview: *C. elegans* | Pros | Cons |
|---|---|---|
| **Anticipated Scientific Insights** | <ul><li>**Completeness Threshold:** What anatomical, molecular, and physiological level of detail is sufficient to recapitulate worm behavior in silico? Likely the only current model organism where ultra-high resolution models are possible.</li><li>**Allows to determine the relative importance of different variables to overall emulation success:** Specifically, non-synaptic signaling in whole-brain models, gap-junctions, deriving function from structure, extrapolation via cell types, etc</li><li>*C. elegans* **promises to be a Rosetta Stone** for converting imaging data into functional models.</li></ul> | <ul><li>**Substantial differences to mammalian nervous systems:** Some of the worm's neurons primarily use graded potentials rather than action potentials, making direct translation to other organisms challenging (though synapses seem more similar to other organisms). Additionally, it makes particularly heavy use of non-neuronal signaling, which may limit the generalizability. Despite being a model organism in neuroscience and medicine (e.g., longevity research), it is hard to extrapolate from the worm to other organisms.</li><li>**Limited behavioral repertoire:** While sophisticated for its size, the behavioral repertoire remains relatively constrained compared to vertebrates. Additionally, the worm lacks clear analogues for many higher cognitive functions of interest (though it has many neurotransmitters implicated in human cognitive function).</li><li>**Little individualism:** The opposite side of a highly conserved organism is that it is hard to study how individual defining characteristics arise.</li></ul> |
| **Experimental Tractability** | <ul><li>**Substantial existing infrastructure**: A decade of OpenWorm and substantial datasets make it the most mature emulation ecosystem of any organism.</li><li>**Data acquisition is relatively cheap:** The organism is cheaply maintained and can be produced in large numbers. Genetically modifying the organism is possible. No highly specialized equipment is necessary, and *C. elegans* are relatively cheap to maintain. The scope of all variables is orders of magnitude smaller than in other organisms.</li><li>**Simulation is computationally cheap:** Memory and compute requirements are easily covered by modern consumer hardware. Everyone can participate.</li><li>**Organisms are highly stereotyped:** Data can be collected from multiple individuals without getting too many interindividual differences.</li></ul> | <ul><li>**Challenges of Calcium Imaging:** The deformable body and brain make tracking for calcium imaging more difficult</li></ul> |

| Gaps and opportunities: *C. elegans* | Gaps (non-exhaustive selection) | Illustrative Project Opportunities |
|---|---|---|
| Neural dynamics | <ul><li>**Voltage Imaging Technology:** Current rhodopsin-based voltage sensors are limited to small fields of view, limiting whole-brain voltage imaging capabilities. Additionally, voltage imaging can detect rapid events like action potentials, but the signal-to-noise ratio remains a challenge, particularly in *C. elegans* neurons, which often exhibit only small membrane potential changes ([Hasehmi et al., 2019](#))</li><li>**Calcium Imaging:** Nuclear-localized GCaMP, which enables brain-wide recordings, provides more restricted spatial and temporal resolution than other approaches. This affects our ability to capture fast neural dynamics and the precise timing of neural events. ([Atanas et al. 2023](#))</li><li>**Integration of Multiple Modalities:** While some progress has been made in simultaneous calcium and voltage imaging ([Tokunaga et al., 2024](#)), current approaches require complex optical setups and careful correction of photobleaching effects. The ability to correlate different measures of neural activity (calcium, voltage, and neurotransmitter release) remains limited.</li><li>**Long-term Recording Capabilities:** The development of nuclear calcium indicator strains has improved long-term imaging capabilities, but there remains a need for better tools to study neural dynamics across different timescales and behavioral states ([Atanas et al., 2023](#)).</li><li>**Circuit Manipulation Precision:** There is a need for improved tools that allow simultaneous manipulation and recording from multiple identified neurons ([Piatkevich, 2023](#)).</li><li>**Neuropeptide Signaling Resolution:** While recent advances have provided insights into peptide-receptor interactions ([Beets et al., 2023](#)), the technology is still in its infancy, and there is a need for tools to visualize neuropeptide signaling in vivo and understand its temporal dynamics in neural circuit function ([Watteyne et al., 2024](#)).</li></ul> | <ul><li>**Perturbation-free experiments:** By automating the handling of individual *C. elegans* using microfluidics and using a spinning-disk confocal microscopy, labs would perform *C. elegans* whole-brain imaging of large counts of freely behaving animals.</li><li>**Optogenetic perturbation experiments:** Activate single neurons, and measure the behavior and changes in all other neurons at using a lightsheet microscope.</li></ul> |
| Connectomics | <ul><li>**Relevance of synaptic variability:** Researchers have observed that even for identically aged hermaphrodites, as many as forty to fifty percent of the synaptic connections can differ from worm to worm ([Brittin et al., 2021](#); [Cook et al., 2019](#)). This means that two worms will share most of the same underlying "scaffolding" (i.e., which neurons physically touch), but their synaptic connectivity at those touches may differ slightly. The relevance of simulation approaches is unclear at this point.</li><li>**Gap Junctions**: Gap junctions are often incompletely captured. Improved methods are needed for consistent visualization and annotation of electrical synapses across samples. ([Witvliet et al., 2021](#))</li><li>**Integration of Multiple Modalities:** Current approaches typically analyze either connectivity or activity patterns separately. Integrated datasets that combine connectomic reconstruction with activity imaging and behavioral correlates are needed to bridge structure-function relationships.</li></ul> | <ul><li>**The "definitive" gap junction resolution connectome:** A connectome scanned at a resolution sufficient to determine gap junctions reliably.</li><li>**The 100 worm connectome project:** to study and determine actual variability amongst what are suspected to be highly stereotyped individuals.</li><li>**Establish expansion microscopy and protein labelling:** ensure this technology reliably works on worms.</li></ul> |
| | - **Dynamic connectivity**: the molecular underpinnings that guide the selective strengthening and formation of specific synaptic contacts, alongside the relative stability of others, remain only partially understood (Brittin et al., 2021; Cook et al., 2019). Finally, there is growing recognition that extrasynaptic signaling (including neuropeptides) and gap junctions can be just as critical for circuit function (Randi et al., 2023). However, legacy datasets often have incomplete elements (Witvliet et al., 2021). | |
| **Computational Neuroscience** | - **Limited Experimental Electrophysiological Data:** Large portions of the worm's nervous system have only partial biophysical parameterization, i.e., missing parameters like time constants and amplitudes of Hodgkin-Huxley-type conductances of different types, which determine the intrinsic electrophysiological properties of neurons. Likewise, its parameters of synaptic conductances - latencies, rise and decay times, short-term plasticity, EPSP and IPSP (or EPSC and IPSC) amplitudes- are used in simulation approaches. (Zhao et al., 2024; Sarma et al., 2018)<br>- **Limited behavior repertoire and sensory feedback loops:** Collect richer behavioral and sensory feedback loop datasets that enable respective modeling and verification of computational *C. elegans* models.<br>- **Multi-lab Synergy instead of Heterogeneous and uncoordinated datasets:** Data sourced for emulation attempts is scattered rather than systematically acquired in standardized formats. This slows the overall process substantially. Although many groups research *C. elegans*, no lab has all the expertise and technology (e.g., advanced calcium or voltage imaging, computational physics, etc.) required to build a fully integrated simulation. (Larson, 2021)<br>- **Integrate neuropeptide data:** Neuropeptides are not accounted for in any simulation approaches.<br>- **Long-Horizon Predictive Capabilities:** Current computational brain models are limited to seconds and could be expanded substantially. | - **Official *C. elegans* computational model data Backlog:** Concrete list of necessary variables, listed by computational neuroscientists that can be created by students / PhDs / or research groups. This includes electrophysiological parameters as well as behaviors and sensory feedback loops.<br>- **Filling the *C. elegans* computational model data backlog:** Follow-up project to fill in the gaps in electrophysiology and behaviors.<br>- **Proposal for "*C. elegans* Emulation Institute":** Integrative proposal / FRO for advancing *C. elegans* emulation efforts. |

# Larval Zebrafish

## Anatomy & Behavior

The larval zebrafish (*Danio rerio*), defined as the developmental stage preceding 30 days post-fertilization (dpf), possesses a transparent body ranging from 3.5 to 7.8 mm in length at day 5 (Kimmel et al., 1995; Svara et al., 2022), roughly 500x the size of *C. elegans*. The larval brain, excluding its spine, measures roughly 0.08mm$^3$ (400 × 800 × 250 μm, W × L × H) at this stage, contains approximately 100,000 neurons, a number that increases by two orders of magnitude to 10 million in adults (that is, 90-day old or older individuals), whose brains measure 0.4–2 mm in thickness and 4.5 mm in length (Wullimann et al., 1996; Bruzzone et al., 2021; Hill et al., 2003; Hinsch and Zupanc, 2007). These neurons utilize action potential-based signaling, similar to mammals, enabling the study of vertebrate-like neural dynamics. However, comprehensive data on natural *in vivo* firing rates across the entire larval brain remain limited. Characterized examples reveal significant diversity: *embryonic* primary motoneurons exhibit rhythmic bursts containing high-frequency (40-50 Hz) action potentials (Saint-Amant and Drapeau, 2000), while sensory afferents show low spontaneous rates of ~9 Hz (Levi et al., 2015). Central neurons also vary, with cerebellar Purkinje cells displaying distinct tonic simple (~9 Hz) and phasic complex (~0.3 Hz) spiking (Hsieh et al., 2014), cerebellar output neurons showing a baseline spontaneous rate around 4 Hz (Najac et al., 2023), and other central neurons like vestibulospinal cells being largely silent at rest (Hamling et al., 2023).

The transparency of the larval brain, particularly in pigmentless mutants, facilitates high-resolution optical imaging of neural activity across the entire brain (White et al., 2008; Antinucci and Hindges, 2016). However, increasing pigmentation in wild-type larvae typically limits the practical window for such optical approaches to the first 1-2 weeks post-fertilization (Volkov et al., 2022). Additionally, larval zebrafish respire through their skin until approximately day 15, allowing for unparalyzed and unanesthetized imaging in agarose-embedded preparations. This feature, combined with the ability to survive for extended periods without external food sources, enables long-duration in vivo neural recordings with minimal maintenance (Hasani et al., 2023). These anatomical and developmental traits make the larval zebrafish a powerful model for studying neural dynamics, structural connectomics, and structure-function relationships in vertebrate neural circuits.

Larval zebrafish display a diverse and adaptive behavioral repertoire, enabling them to navigate and thrive in dynamic environments (Zocchi et al., 2025). After day 3-4, they exhibit a range of stereotyped movements, including slow swims (scoots), rapid escape responses (C-starts), and specialized maneuvers like J-turns for predation (Privat et al, 2020).

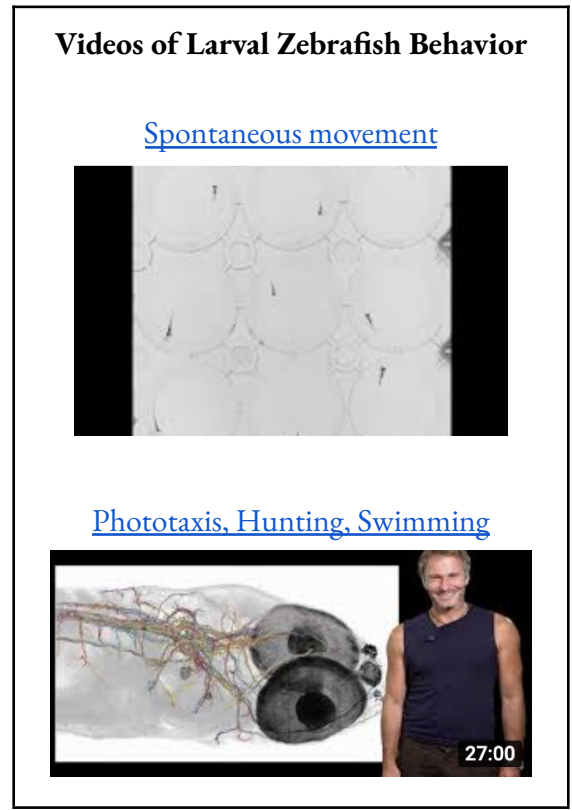

These behaviors are finely tuned to sensory inputs: larvae respond to visual stimuli with optomotor and optokinetic reflexes, acoustic/vibrational cues with escape movements, and tactile stimuli with highly directional turns. They also display adaptive behaviors such as phototaxis, visual background adaptation, and alarm responses to chemical cues, which help them navigate, camouflage, and avoid predators. Social behaviors, such as shoaling and aggregation, emerge by day 9–10 and are influenced by learned preferences for conspecifics. Additionally, larvae exhibit circadian rhythms, with periods of activity during the day and immobility at night, resembling sleep-like states (Fero et al., 2012).

Zebrafish larvae are also capable of multiple forms of learning and memory. Starting at five days post fertilization, they begin to habituate to repeated stimuli, showing distinct forms of rapid, short-term, and long-term habituation that depend on specific neural mechanisms (Roberts et al., 2013). For example, long-term habituation of the C-start escape response requires protein synthesis and shares mechanistic similarities with learning in other species. Zebrafish larvae can also undergo associative learning - for instance, they learn to move their tails in response to light signals and to avoid areas associated with electric shock. They also display social learning, as shown by their specific preferences for shoaling partners based on their early-life exposure to the appearance of other fish ([Roberts et al., 2013](#)).

## Neural Dynamics

Neural activity recording

Neural activity in larval zebrafish can be recorded using head-fixed or freely-moving preparations ([Hasani et al., 2023](#)). In head-fixed preparations, the fish's head is immobilized in agarose, which limits natural behavior but enables stable imaging. Several approaches have been developed to increase behavioral output: the tail can be freed to monitor movement intentions, and fish can perform fictive navigation in virtual environments by using their tail movements to control the change in the virtual environment ([Ahrens et al., 2012](#); [Trivedi and Bollmann, 2013](#); [Vladimirov et al., 2014](#); [Torigoe et al., 2021](#)).

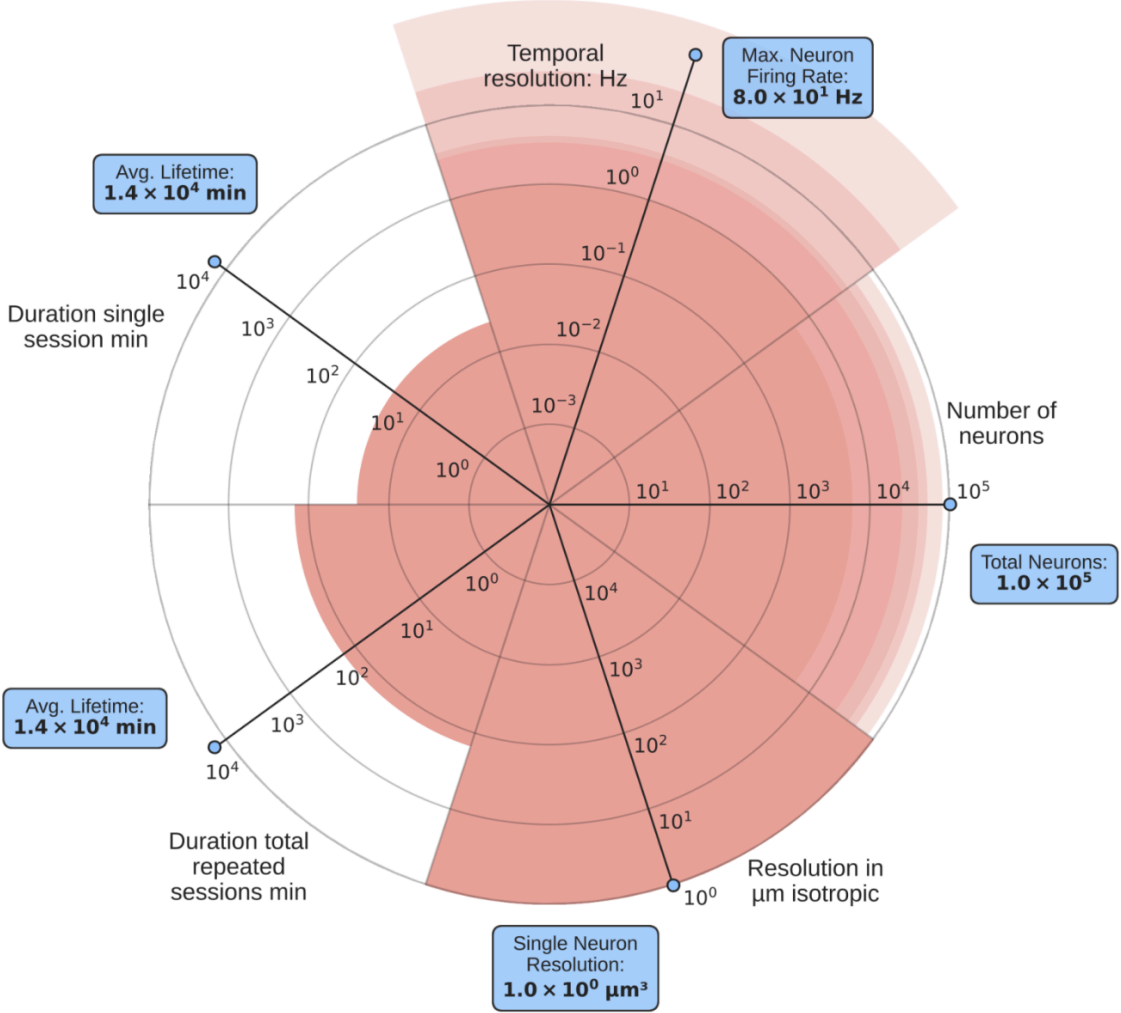

**Figure Overview of the optical neural recording landscape in Larval Zebrafish:** Radar plots based on the optical recording literature cited in the report. We plot the following dimensions of fixated brain recordings: spatial resolution, brain volume, temporal resolution, and (estimated) individual and cumulative recording duration. Only recordings from fixated (A) and no freely moving experiments are available. The outer ring is normalized to the maximum known values. Each ring represents one order of magnitude. The data for this figure is available in the [linked data repository](linked data repository).

Initial progress in whole-brain imaging came from one-photon techniques like light-sheet microscopy, which achieved simultaneous recording of ~80,000 neurons (about 80-90% of all neurons) at 1 Hz (Ahrens et al., 2013), later improved to 4 Hz using electrically tunable lenses (Favre-Bulle et al., 2018). Recent advances include DaXi microscopy, achieving 3.3 Hz over 0.03 mm$^3$ (37.7% of total brain volume, approx. 40,000 neurons), and SCAPE microscopy reaching 25.75 Hz over smaller volumes of 0.00481 mm$^3$ (6% total brain volume, approx. 6,000 neurons) (Yang et al., 2022; Voleti et al., 2019). More recently, the ZAPBench dataset (Lueckmann et al., 2025) showcased light-sheet microscopy recording of over 70,000 neurons throughout nearly the entire larval zebrafish brain at approximately 1 Hz. However, scattered light from the excitation laser is visible to the fish in all one-photon approaches and can interfere with presented visual stimuli. Two-photon microscopy uses infrared light that is invisible to the fish, though care must be taken as fish eyes absorb infrared light, and exposure can be lethal. While two-photon imaging eliminates visual interference, its main limitation is speed: current two-photon technology can record 52,000 neurons distributed throughout 80% of the brain's volume at 1 Hz (Bruzzone et al., 2021). However, even with these advances, head-fixed preparations fundamentally constrain the fish's natural behavior, particularly during complex tasks like hunting.

Recording neural activity in freely moving zebrafish presents unique technical challenges, as larvae move in rapid bursts reaching speeds of 10 cm/s and angular velocities of 600°/s, i.e., close to 2 revolutions around themselves per second (Johnson et al., 2020; Mearns et al., 2020). These movements require sophisticated real-time tracking to keep the fish within the microscope's field of view. Higher imaging speeds typically come at the cost of either reduced spatial resolution or a smaller field of view. Current tracking approaches either move the stage to cancel the fish's motion (Kim et al., 2017) or use mirror systems to keep a stationary microscope focused on the fish (Symvoulidis et al., 2017). Several imaging techniques have been developed, including differential illumination focal filtering microscopy (DIFF), light field microscopy (LFM), and hybrid approaches combining LFM with confocal or light-sheet methods (Cong et al., 2017; Zhang et al., 2021). Additional challenges include calcium indicators being too slow to track rapid neural changes during behaviors like hunting, and current setups restricting the fish's vertical movements, which are crucial for natural hunting where fish prefer to strike at prey from below (Bolton et al., 2019; Mearns et al., 2020). Additionally, counteracting the fish's movements by moving the imaging chamber can significantly alter their behavior, reducing how far, fast, and often they move (Kim et al., 2017).

Calcium imaging speeds still fall roughly two orders of magnitude behind the fastest neurons in zebrafish larvae (Lueckmann et al., 2025). Voltage imaging in larval zebrafish has also made remarkable strides in recent years, bringing the field closer to the goal of whole-brain voltage imaging at cellular resolution. Innovations in genetically encoded voltage indicators (GEVIs), such as ASAP3-Kv,

Voltron2-Kv, and Positron2-Kv, have significantly improved brightness and signal-to-noise ratio, enabling the detection of single action potentials across large populations of neurons. Concurrently, advances in microscopy, particularly light-sheet techniques like remote refocusing and oblique plane microscopy, have pushed volumetric imaging rates to over 200 Hz, allowing researchers to capture cellular-resolution neural activity across approximately 25,000-33,000 neurons (25-33% of total brain) in head-fixed zebrafish brain even when at maximum firing rates ([Wang et al., 2023](#)), although these techniques are still in active development, with data from only a single fish presented.

## Neurotransmitters and Neuromodulators

Optical approaches for monitoring neurotransmitter dynamics are particularly well-suited for larval zebrafish given their transparency and accessibility for whole-brain imaging. Several GENIs have been successfully validated in zebrafish for glutamate ([Marvin et al., 2013](#)), GABA ([Marvin et al., 2019](#)), acetylcholine ([Borden et al., 2020](#)), dopamine ([Sun et al., 2018](#)), noradrenaline ([Feng et al., 2019](#)), and ATP ([Wu et al., 2021](#)), but developing sensors for the more than 100 neuropeptides identified in the zebrafish brain ([Van Camp et al., 2016](#)) remains a challenge.

## Perturbation

The larval zebrafish represents an attractive model organism for optogenetic perturbation due to its transparency and amenability to genetic modifications. One-photon optogenetic approaches face limitations from light scattering, lack of z-axis resolution, and unwanted visual stimulation of the fish ([Chai et al., 2024](#)). Two-photon optogenetics offers improved spatial precision and reduced scattering ([Turrini et al., 2024](#)), though reliable single-neuron manipulation remains a technical challenge. Current approaches typically target anatomically defined regions rather than individual neurons, particularly when the fish is relatively stationary ([Chai et al., 2024](#); [Turrini et al., 2024](#)). Despite these technical limitations, optogenetics studies in the fish have led to several discoveries ([Piatkevich and Boyden, 2023](#)), including neural populations and activity patterns responsible for saccadic eye movements ([Schoonheim et al., 2010](#)), for increasing sleep ([Oikonomou et al., 2019](#)), for controlling swim turn direction ([Dunn et al., 2016](#)), for providing sensory feedback to spinal circuits during fast locomotion ([Knafo et al., 2017](#)), for producing coordinated swimming patterns ([Ljunggren et al., 2014](#)), for stopping ongoing swimming ([Kimura et al., 2013](#)), and for contributing to movement in response to noxious stimuli ([Wee et al., 2019](#)). Most recently, Chai et al. developed a system capable of enabling whole-brain calcium imaging in freely swimming larvae while simultaneously allowing targeted optogenetic stimulation of specific brain regions during stationary periods ([Chai et al., 2024](#)).

Further development of tools that permit genetic access to limited populations of neurons would allow for more specific perturbation of individual populations of neurons.

## Connectomics

Early electron microscopy efforts in larval zebrafish demonstrated the feasibility of whole-brain imaging through multi-scale imaging approaches. Hildebrand et al. achieved imaging of an entire larval zebrafish brain and portions of the spinal cord, though at resolutions insufficient for dense reconstruction ([Hildebrand et al., 2017](#)). A significant advance came with Svara et al., who achieved whole-brain electron microscopy imaging at synaptic resolution in a 5-day post-fertilization larval zebrafish, enabling tracing of neural connections across nearly the entire brain except for the retinae ([Svara et al., 2022](#)). Parallel efforts are ongoing, including as part of integrated functional & structural studies that imaged another whole larval brain ([Lueckmann et al., 2025](#)) and as part of work at the Brain Science and Intelligent Technology Innovation Center of the Chinese Academy of Sciences, where a team led by Du Jiulin is combining both optical and electron microscopy approaches for cellular and synaptic level mapping respectively ([Du Jiulin, 2022](#)). Small pieces of the brainstem ([Vishwanathan et al., 2024](#)), hindbrain ([Boulanger-Weill et al., 2025](#)), and spinal cord have also been reconstructed ([Svara et al., 2018](#)).

Beyond electron microscopy, expansion microscopy (ExM) has emerged as a powerful complementary approach, first demonstrated in zebrafish by Freifeld et al., who resolved putative synaptic connections between fluorescently labeled cell populations ([Freifeld et al., 2017](#)). Subsequent work has revealed detailed protein organization at specific synapses, such as those on Mauthner cells ([Cárdenas-García et al., 2024](#)). Recent ExM protocols enable the expansion of intact, several-millimeter-long animals up to 5 dpf while maintaining compatibility with genetically-encoded protein labeling, providing an essential link between structure and function by revealing both subsynaptic structures and intricate signaling pathways throughout the nervous system ([Steib et al., 2023](#); [Behzadi et al., 2024](#)). While x-ray-based approaches have been demonstrated using synchrotron radiation to achieve cellular resolution ([Osterwalder et al., 2021](#)), they have not yet been widely adopted for neural circuit mapping in larval zebrafish.

The Hildebrand et al. dataset requires approximately 2.7 terabytes of storage across all resolution levels ([Hildebrand et al., 2017](#)). In comparison, the Svara et al. dataset likely requires approximately 15 terabytes of storage (assuming 2 bytes per voxel at 7.33 teravoxels, the size of the current effort is ~11TB at 8 bits) ([mapZebrain, 2025](#)). Neuron reconstruction efforts for the larval zebrafish are ongoing: while automated segmentation and synaptic detection of the Svara et al. dataset have been completed, comprehensive proofreading efforts continue to this day and are scheduled to be completed most likely in a few years.

## Computational Modeling

The landscape of larval zebrafish simulation efforts is distinctly shaped by three key factors: the current absence of a fully reconstructed connectome; its relative novelty as a brain simulation model organism in computational neuroscience compared to *C. elegans*; and the optical transparency of its brain, making whole-brain calcium imaging data readily accessible. These characteristics have led to simulation approaches that are generally not whole-brain, not connectome-constrained, and revolve around functional data as the key to model fitting and validation.

### Vishwanathan et al., 2024

Vishwanathan et al. implemented a highly simplified linear rate model as a proof of concept for connectome-constrained circuit simulation in larval zebrafish. It was based on a reconstructed 0.0024 mm³ brainstem volume (approximately 2.4% of the total brain volume) containing circuits involved in oculomotor and velocity-to-position integration ([Vishwanathan et al., 2024](#)). Based on the identified ~3,000 neurons and ~75,000 synapses, they derived a directed graph weighted by synaptic counts between neurons, with weights normalized by the total synaptic input to each postsynaptic neuron and with prior physiological studies of neurotransmitter identities and anatomical mapping of circuit components used to constrain the model further.

Model validation leveraged previously collected two-photon calcium imaging data from 20 different larval zebrafish, combined with simultaneous eye position recordings. The validation focused on two key aspects of oculomotor integration: how neuron firing rates correlated with eye position during fixations, and how neural activity changed following rapid eye movements. For each neuron population, the authors compared how strongly neurons respond to eye position changes in the model versus experimental measurements from calcium imaging across 20 different zebrafish - the model, despite being based on a single specimen's connectivity, correctly predicted the characteristic distribution of position sensitivities for each population.

## Liu et al., 2024

simZFish, developed by Liu et al., combines an embodied model that bridges sensory processing, neural control, and biomechanics to reproduce behavior. It is based on their previous behavioral and calcium imaging recordings of the optomotor response - a behavior where fish adjust their swimming to maintain position when their entire visual environment appears to move around them (Liu et al., 2024; Naumann et al., 2016). The model was fit using data from their 2016 work, where they identified distinct neural types in the pretectal circuit by analyzing calcium imaging recordings from 3,070 neurons across 12 fish, characterizing their firing patterns during visual stimulation, and establishing their functional connectivity through analysis of concurrent behavioral recordings from 38 fish.

This characterization was used to initialize a simplified, biologically inspired rate-coding network. In such networks, a common approach in computational neuroscience, the detailed spiking activity of individual neurons is not simulated. Instead, a large population of similar neurons is often abstracted as a single computational node (or unit). The activity of this node is then characterized by a single, continuous time-varying signal representing the average firing rate of that entire population. These rate-based units, whose outputs are these average firing rates, then interact with each other to form the network. This contrasts with the more granular, single-neuron spiking models that are a primary focus of other sections in this report. While this rate-coding neural simulation formed the control system, the model's key contribution lies in its detailed biomechanical implementation, comprising seven body segments with realistic hydrodynamics and a virtual visual system with two cameras mimicking the fish's eyes.

The model underwent comprehensive validation through multiple approaches: behavioral analysis comparing simulated and real fish responses to visual stimuli, comparison of artificial neural activity patterns to calcium imaging recordings, and finally, implementation in a physical robot tested in both controlled and natural environments. Both the simulation and the robot – see some of the videos from Liu et al. (Liu et al., 2024) – demonstrated the ability to maintain position in moving water through visual input alone.

## Immer et al., 2025

Lueckmann et al. developed a black-box machine learning model trained on approximately two hours of 1 Hz calcium recordings from the ZAPBench dataset (Lueckmann et al., 2025), which captures the activity of approximately 70,000 neurons across the whole brain (2048 x 1152 x 72 voxels at 406 nm x 406 nm x 4 μm resolution, likely 70-80% of all neurons of the organism) from a single head-fixed but tail-free larval zebrafish during nine different behavioral tasks. The recordings were obtained using light-sheet microscopy while the fish was exposed to various visual stimuli, including forward-moving gratings to test gain adaptation, random dot patterns for decision-making, alternating light/dark flashes for startle responses, asymmetric illumination for phototaxis, and various other motion patterns to probe turning behavior and positional homeostasis. The fish's tail movements were monitored throughout the experiments via electrical recordings from motor nerves. This allowed real-time coupling between the fish's attempted swimming behavior and the visual stimuli presented (Immer et al., 2025).

Unlike traditional approaches that first segment individual neurons and extract their activity traces, this method worked directly with the raw volumetric video data to predict future frames of brain-wide activity. The authors implemented a 4D UNet architecture operating on three spatial dimensions plus time, trained end-to-end to minimize the mean absolute error between predicted and actual (voxel-based) calcium traces. The model was trained on approximately 1.4 hours of data (70% of the full 2-hour recording) spanning eight different behavioral conditions, with performance evaluated on held-out test segments from these conditions and a completely held-out ninth condition. Attempts to improve performance through pre-training on data from two other specimens proved unsuccessful. While these initial attempts at pre-training on data from two other specimens were unsuccessful, this single outcome on a new benchmark with few baselines offers limited insight into the ultimate viability of transfer learning across individuals, which remains a promising avenue for future exploration.

When validated against held-out recordings, the model achieved higher prediction accuracy for short temporal contexts, though it showed comparable performance to trace-based approaches when using longer temporal contexts. In completely held-out experimental conditions, the model demonstrated better generalization for one-step-ahead predictions than for longer forecast horizons. Successful prediction, at least when defined in terms of mean absolute error, thus remains a challenge for trace-based and video-based models, especially with longer temporal contexts. Importantly, this particular modeling approach focuses solely on predicting voxel-level neural activity from past activity and does not incorporate embodiment; it has no simulated body or environment to interact with, nor does it directly model how sensory inputs (like the visual stimuli presented to the fish) translate into neural activity or how neural activity translates into motor outputs (like tail movements).

While this work demonstrates the feasibility of purely ML-based approaches to neural activity prediction in Zebrafish larvae, the main challenge remains sample efficiency. These models lack biological priors constraining their predictions, and they require substantially more data to learn effectively. Importantly, the authors have made this valuable dataset publicly available in a convenient format. Future work will likely focus on improving sample efficiency through better architectures and finding ways to leverage data across individuals through cross-specimen pre-training. The connectome for this exact fish is expected to be released in a year or so, thus enabling a variety of approaches that incorporate serious biological priors.

## Gap Analysis

The larval zebrafish represents a unique convergence of experimental tractability and biological sophistication, making it a unique target for integrated brain emulation efforts. While larval zebrafish have minimal clinical / industrial utility, success in fully modeling a vertebrate brain could broadly validate whole-brain emulation efforts.

| Model Organism Overview: Larval zebrafish | Pros | Cons |
|---|---|---|
| **Anticipated Scientific Insights** | <ul><li>**Similarity to Mammalian Brain Structure and Physiology:** Spiking neurons, layered structures, more pronounced plasticity – offering richer testbeds than *C. elegans*.</li><li>**High-Resolution Structure – Function Mapping:** Near-whole-brain calcium/voltage imaging can be paired with EM or ExM for synaptic-level detail, enabling correlational and causal studies.</li><li>**Diverse behaviors even in larvae:** Prey capture, startle, and associative learning make the modeling of somewhat sophisticated behaviors theoretically possible.</li></ul> | <ul><li>**Developmental & Inter-individual Variability:** The rapid development of larval zebrafish brains (typically studied at 5-7 days post-fertilization) creates significant registration challenges, as small age differences of days or even hours can lead to substantial physiological and structural changes This complicates efforts to create a "standard" reference connectome and hinders long-term studies. It also raises questions about circuit stereotypy across individuals.</li><li>**Ecological Validity Gaps:** 3D naturalistic behaviors (e.g., full hunting, social interaction) are hard to record at high spatiotemporal resolution, raising questions about real-world relevance.</li><li>**Somewhat limited behavioral repertoire for advanced cognitive functions**: At the age where typical whole brain recordings are performed (~5-6 days post fertilization), memory formation, individuality/"personality" and behaviors are limited ([Roberts et al., 2013](#)).</li></ul> |
| **Experimental Tractability** | <ul><li>**Strong & Growing Community:** Active consortia, open resources (atlases, transgenic lines), imminent first vertebrate connectome release.</li><li>**Feasibility–Complexity Sweet Spot:** ~100k neurons with vertebrate circuitry and small enough for near-complete connectomics and whole-brain imaging.</li><li>**Manageable Scale & Cost:** Smaller facility/outlay than rodents, feasible HPC demands for ~100k neurons, and simpler housing.</li><li>**Whole-Brain Perturbation Feasibility:** Structure-informed whole-brain optogenetics with full-brain imaging is realistic, potentially enabling systematic "perturbation atlases."</li></ul> | <ul><li>**No Standardized Pipeline:** No universal protocol for connecting imaging, connectomics, and behavioral data from the same specimen.</li><li>**Time-Limited Transparency:** Beyond ~7-14 days, reduced optical clarity and morphological changes hinder long-term/adult study (though special lines can maintain transparency).</li><li>**Molecular Tool Gaps:** Rapid larval development complicates viral barcoding or extended protein-expression protocols; the genetic toolkit for barcoding is relatively underdeveloped compared to other model organisms. Cell-type specific genetic access tools still need substantial development.</li><li>**Current Proofreading Bottleneck:** While the first EM connectome completion is expected within 1-2 years, manual validation limits availability.</li><li>**Perturbation Coverage:** Systematic perturbation of all ~100,000 neurons remains intractable within the brief larval stage.</li></ul> |

| Gaps and opportunities: Larval Zebrafish | Gaps (non-exhaustive selection) | Illustrative Project Opportunities |
|---|---|---|
| Neural dynamics | - **Limited Functional Data:** Current imaging techniques face fundamental trade-offs between spatial coverage and temporal resolution. One-photon light-sheet microscopy can image nearly all ~100,000 neurons but only at 1-4 Hz, while faster techniques like SCAPE microscopy (up to ~25 Hz) or voltage imaging (millisecond-scale) achieve higher temporal resolution at the cost of observing less than 10% of the brain. No current method achieves both whole-brain coverage and spike-timing resolution simultaneously.<br>- **Behavioral Recording Constraints:** Most high-resolution neural recordings rely on head-fixed preparations that restrict natural behavior. While recent advances enable tail movement in virtual environments or use mirror-based tracking systems for freely moving fish, these methods compromise either spatial or temporal resolution. Current setups particularly struggle with complex three-dimensional behaviors like hunting, limiting the ecological validity of collected neural data.<br>- **Scarce data on neuromodulation:** Very limited data exist on the effect of different neuromodulators in larval zebrafish. | - **Extended ZAPBench 2.0:** Expand current whole-brain calcium imaging efforts to include recordings from more individuals and simultaneous high-resolution behavioral tracking (tail kinematics, eye movements) in head-fixed preparations.<br>- **Whole-Brain Calcium / Voltage Imaging Scale-up:** Increase the imaging rates and coverage of calcium and voltage imaging, ideally in freely behaving individuals.<br>- **3D Unrestricted Swimming Microscopy:** Develop imaging systems specifically addressing the vertical movement limitation in current setups, enabling natural hunting behaviors where fish strike from below.<br>- **Comprehensive Perturbation Atlas:** Systematic optogenetic perturbations with whole-brain activity recording. Map circuit-wide effects of activating/inhibiting defined neuron populations during specific behaviors. |
| Connectomics | - **Reconstruction Bottlenecks:** Although automated segmentation and synaptic detection of the Svara et al. dataset are complete, comprehensive proofreading remains ongoing. The scale of manual intervention required for accurate proofreading continues to delay the availability of a fully reconstructed connectome.<br>- **Molecular Information Limitations:** EM datasets provide primarily morphological and connectivity information, in addition to limited data to distinguish excitatory vs. inhibitory neurons, lacking crucial details about neurotransmitter identities and receptor distributions. While expansion microscopy (ExM) could enable structural mapping and molecular annotation, no comprehensive whole-brain ExM effort has been completed in zebrafish.<br>- **Lack of neuroplasticity datasets:** No substantive neuroplasticity datasets exist. | - **Transgenic Barcoding Feasibility Study:** Develop rapid expression strategies for molecular barcoding compatible with larval stage timing.<br>- **Whole-Brain ExM with Molecular Profiling:** Create a comprehensive expansion microscopy dataset with molecular annotations at synaptic resolution.<br>- **10x the connectome data, including** brain and spine, to better understand different stages and interindividual differences. |
| **Computational Neuroscience** | <ul><li>**Limited Experimental Electrophysiological Data:** Similar to *C. elegans*.</li><li>**Missing Structural Foundation:** The absence of a fully proofread connectome forces whole-brain models to rely primarily on functional data or partial circuit reconstructions. Even when structural data exists, it lacks crucial molecular information about neurotransmitter types and receptor distributions.</li><li>**Immature Model Development:** Existing whole-brain models are few and currently lack detailed biophysical implementation. The lack of systematic perturbation data (e.g., comprehensive optogenetic studies) further complicates model validation.</li></ul> | <ul><li>**Port OpenWorm to the Larval Zebrafish:** Collaborate with OpenWorm on shared infrastructure (3D environments, databases with receptors, etc.)</li><li>**Connectome-Constrained Biophysical Simulator:** Build pilot-level compartmental models using the current version of the larval zebrafish whole-brain connectome and available ephys patch-clamp recordings.</li><li>**Embodied ML Models of Zebrafish Behavior:** Develop physics-based differentiable simulation environments and train biophysical models supporting auto differentiation to replicate specific behaviors (hunting, escape responses) using zebrafish behavioral data.</li></ul> |

# Drosophila

## Anatomy & Behavior

The development of *Drosophila melanogaster* proceeds through four stages: egg, larva, pupa, and adult. The larval stage represents the primary growth period, with a ~200-fold increase in body mass ([Church and Robertson, 1966](#)) and brain expansion from approximately 10,000 neurons ([Jiao et al., 2022](#)) to roughly ~130-140,000 neurons in adults. Adult flies at day 7-15 measure 2.5-3 mm in body length and approximately 2 mm in width, with males slightly smaller than females. Under laboratory conditions at 25°C, their mean lifespan typically ranges from 25 to 75 days for females and 25 to 85 days for males. However, this can vary significantly depending on factors like genetics and environment ([Lints et al., 1983](#)). The adult nervous system consists of two main parts: the brain and the ventral nerve cord (VNC). The male CNS occupies about 0.054 mm³ ([Berg et al., 2025](#)), of which roughly 75% (~0.04 mm³) corresponds to the brain – including the central brain and optic lobes – and about 25% corresponds to the VNC ([Janelia, 2023](#)). The adult brain contains ~130-000 ([Dorkenwald et al., 2024](#)) to ~140,000 neurons ([Berg et al., 2025](#)), while the male VNC contributes ~33,000 neurons within a ~166,000-neuron CNS ([Berg et al., 2025](#)). All neural structures are enclosed by the cuticle – a multilayered exoskeleton partially transparent to visible light and transparent to infrared wavelengths ([Hsu et al., 2018](#)) – and are oxygenated by the tracheal system, a network of air-filled tubes that branch progressively finer until directly contacting brain tissue.

Neuronal firing rates in *Drosophila* are diverse. For example, while some sensory neurons fire spontaneously at only 1-2 Hz, others can exceed 200 Hz in response to stimuli ([de Bruyne et al., 2001](#), [Dweck et al., 2023](#)). However, the brain's overall energy budget limits the average network activity. While comprehensive empirical data on typical, whole-brain average in-vivo firing rates in *Drosophila* are currently scarce, we can perform a highly approximate, illustrative calculation to explore this metabolic constraint. We come up with a rough estimate of this limit using the measured resting brain oxygen consumption rate of ~160 pmol O2/min ([Neville et al., 2018](#)), which translates to a total resting ATP budget of ~5.1 x 10$^{12}$ ATP/sec (assuming standard energy conversion factors: ~450 kJ/mol O2 oxycaloric equivalent, ~35% metabolic efficiency, ~50 kJ/mol ATP hydrolysis energy). Distributing this budget across ~140,000 neurons using an estimated cost per spike of ~9 × 10$^6$ ATP, calculated for a hypothetical action potential in a related blowfly neuron's axon ([Laughlin et al., 1998](#)), yields a maximum average firing rate of approximately 4 Hz. This theoretical upper bound neglects significant synaptic and resting potential energy costs, meaning the actual sustainable average rate is likely lower.

*Drosophila melanogaster* displays many behaviors ([Kohsaka, 2023](); [Caldwell et al., 2003]()). Larvae show movements like crawling forward and backward, sweeping their head, or rolling over ([Kohsaka, 2023](); [Clark et al., 2018](); [Caldwell et al., 2003]()). Sensory feedback modulates these movements, allowing larvae to respond to light, physical obstacles, and food availability ([Clark et al., 2018](); [Busto et al., 1999]()). Adult *Drosophila* engage in various behaviors, including walking, running, grooming, aggression, mating, and flying. Automated postural analysis has defined over 100 distinct behavioral states ([Berman, 2014]()), with undoubtedly more to be defined in social encounters or the natural environment. Adult flies exhibit a substantially more complex range of movements than *C. elegans* or larval zebrafish, such as rapid banked turns (body saccades) during flight, which help minimize motion blur and avoid collisions ([Mujires et al., 2015]()).

Videos of Fruitfly Behavior

Fighting

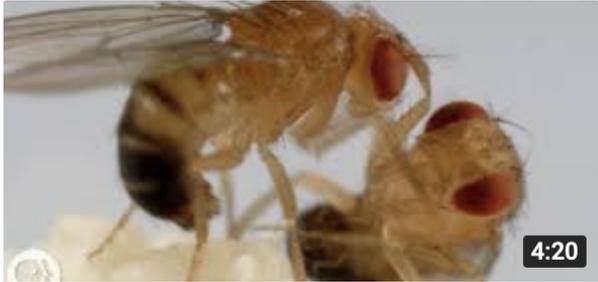

A male fruit fly is courting a female fruit fly.

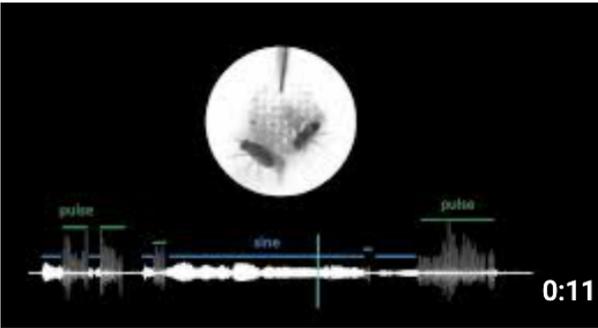

Social mating behavior

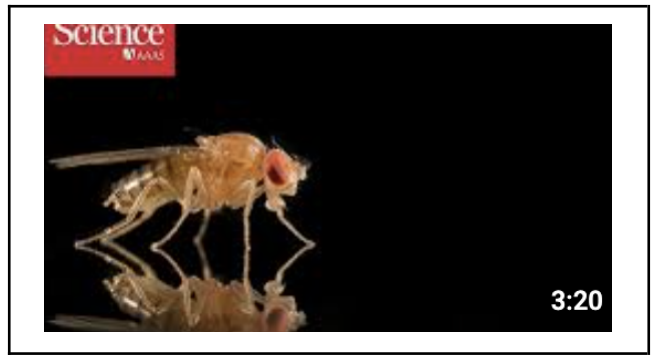

*Drosophila* demonstrates notable capabilities for learning, memory, and communication. It can habituate to repeatedly presented stimuli (a form of non-associative learning), form aversive courtship memories after rejection, and recall outcomes of past aggressive encounters (Durrieu et al., 2020; Zhao et al., 2018; Yurkovic et al., 2006, YouTube Video). Social learning also appears in mate choice, foraging decisions, and predator avoidance, with individuals copying behaviors observed in their peers (Kacsoh et al., 2018; Battesti et al., 2012). Additionally, even genetically identical flies display stable idiosyncratic "personalities" – for example, consistent left- or right-turning biases and phototactic preferences that persist for days (de Bivort et al., 2022).

Social behavior is another major facet of *Drosophila*'s ethology and includes aggression, territory defense, and group formation (Monyak et al., 2021; Schretter et al., 2020). Both males and females exhibit aggression, although intensity and tactics differ based on sex and experience - males typically engage in higher-intensity boxing and fencing. In contrast, females display more subtle aggressive behaviors (Zwarts et al., 2012, YouTube video). In some species, territoriality is more pronounced, with males defending leaves or fruit patches, whereas in *Drosophila melanogaster*, territorial claims tend to be conditional and triggered by factors like female presence or food resources (Zwarts et al., 2012; White et al., 2015). Flies also gather on food sources in aggregations that may improve foraging efficiency and promote social information sharing (Philippe et al., 2016), while maintaining individual spacing that adjusts with density and social context (McNeil et al., 2015). Courtship in *Drosophila* involves an intricate behavior sequence that showcases their social communication capabilities. Males perform elaborate courtship rituals, including following, wing vibration to produce species-specific songs, tapping, and attempted copulation (Pavlou et al., 2013). The courtship song, produced by precisely controlled wing vibrations, contains specific patterns of pulses and sine waves that are crucial for species recognition and female choice. These acoustic signals work with chemical cues - males detect female pheromones through specialized receptors, while females evaluate male quality through acoustic and chemical signals (Fernandez et al., 2013; Lillvis, 2024). The successful integration of these multimodal cues - acoustic, chemical, and behavioral - determines mating success and helps maintain species barriers (Dweck et al., 2015).

Finally, internal states and social context further influence these behaviors. Isolation can heighten aggression, disrupt sleep and change locomotor activity, suggesting that *Drosophila* depends on social cues for regulating stress and energy balance (Eddison., 2021; Li et al., 2021). After mating, females temporarily become non-receptive and increase egg production, a shift mediated by seminal fluid proteins (Mackay et al., 2005). Temperature preferences also depend on feeding status – hungry flies often choose cooler temperatures, while sated flies gravitate toward warmth (Umezaki et al., 2024). Crowding triggers further behavioral changes, with flies drawing on a sense of "group size" to modify spacing, movement, and interactions (Rooke et al., 2020). These findings highlight how *Drosophila*'s behavioral repertoire is highly plastic, shaped by physiological, environmental, and social factors.

## Neural dynamics

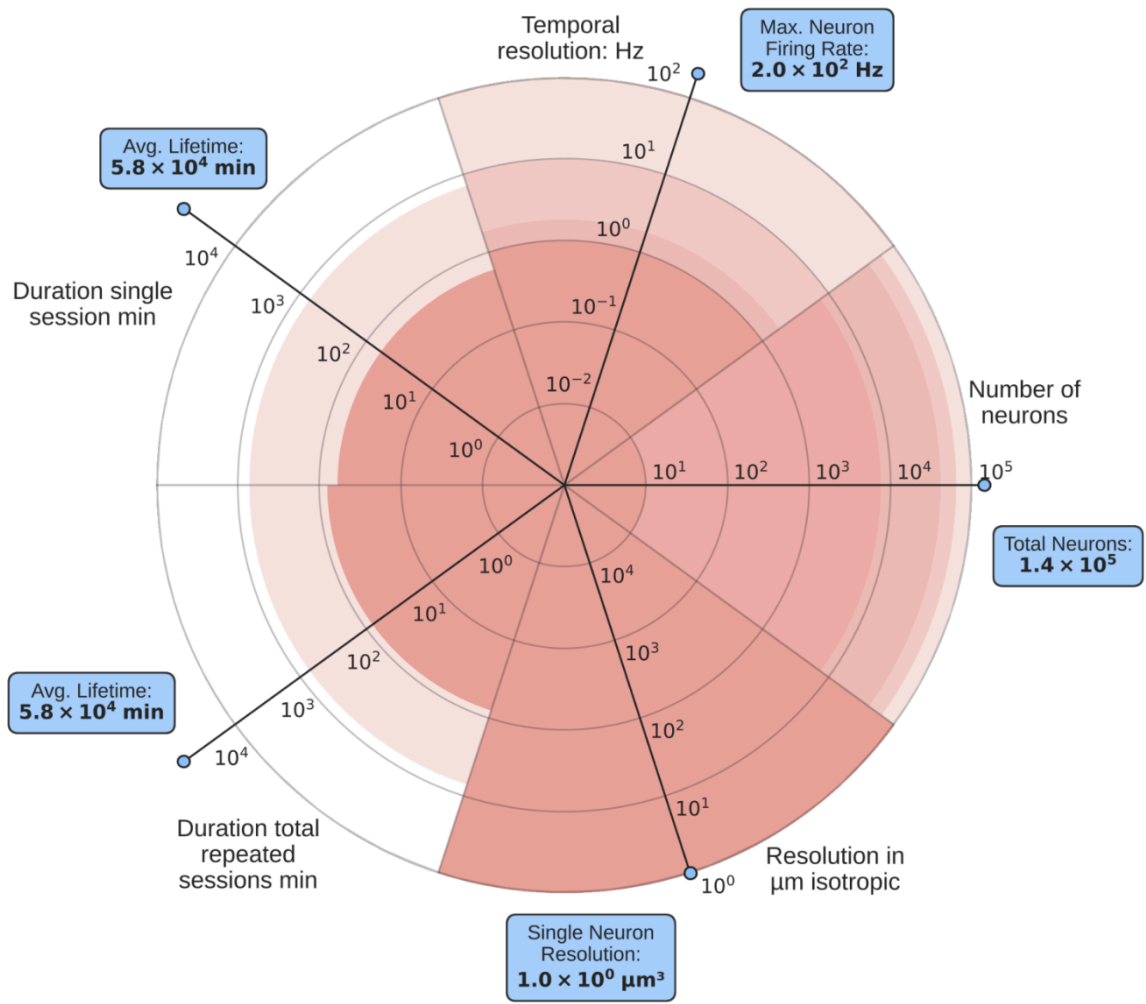

**Figure Overview of the optical neural recording landscape in *Drosophila*:** Radar plots based on the optical recording literature cited in the report. We plot the following dimensions of brain recordings: spatial resolution, brain volume, temporal resolution, and (estimated) individual and cumulative recording duration. Only recordings from fixated (A) and no freely moving experiments. The outer ring is normalized to the maximum known values. Each ring represents one order of magnitude. The data for this figure is available in the [linked data repository](linked data repository).

A) fixated

## Neural activity recording

Calcium imaging of small populations of neurons, or all the neurons with a brain region is routine and facilitated by the wealth of genetic tools available in *Drosophila*. The *Drosophila* brain's small size and partial transparency suggest the feasibility of whole-brain imaging (i.e., >90% of all neurons) at single-neuron resolution. In practice, however, the presence of air-filled tracheae creates significant challenges through different mechanisms: for single-photon imaging, light scattering is the primary limitation, while for multiphoton approaches, optical aberrations from the tracheae are the dominant obstacle (Hsu et al., 2018). Indeed, while two-photon microscopy can image at depths of 600-800 μm in mouse brain tissue (Xu et al., 2024), it often struggles with imaging depths beyond about 0.1 mm in *Drosophila* when attempting to image through the intact cuticle or with minimally invasive preparations, largely due to air-filled tracheae. However, with more invasive preparations involving cuticle and superficial trachea removal, functional 2P imaging depths of 120-245 μm (Aragon et al., 2022; Brezovec et al., 2024) have been demonstrated. Single-neuron resolution in adult Drosophila is also intrinsically challenging because somata are somewhat smaller than in other model organisms. Indeed, typical central brain neurons have soma diameters of 2–6 μm, and mushroom body Kenyon cells are only ~2–3 μm across and densely packed in clusters (Tuthill, 2009; Gu et al., 2006).

In head-fixed preparations, current techniques navigate these challenges through different trade-offs: light field microscopy achieves fast (~100Hz) imaging of a large portion (600 x 300 x 200 μm3, ~90% of brain tissue by box volume assuming a total brain volume of 0.04mm$^3$) of the brain, though at resolution insufficient to identify individual cells (Aimon et al., 2019). Schaffer et al. used swept confocally-aligned planar excitation (SCAPE) microscopy to image the dorsal third of the *Drosophila* brain (450 × 340 × 150 μm$^3$, ~57%) at single-cell resolution at 8-12 Hz, though scattering makes resolving individual neurons in deeper brain structures difficult (Schaffer et al., 2023). Two-photon microscopy offers superior penetration power, but its point-scanning nature creates an inherent trade-off between imaging speed and volume coverage – although advances like light beads microscopy may permit significantly faster imaging (Demas et al., 2021). Indeed, Gauthey et al. recently demonstrated this potential by achieving whole-brain imaging (e.g., 295 x 675 x 235 μm$^3$, whole brain by box volume) at 28 Hz in preparations where tracheae were surgically removed (Gauthey et al., 2025). This approach, with a spatial resolution of ~1 x 1 x 10 μm, was able to capture fast-timescale auditory responses missed by standard volumetric imaging and could even be pushed to 60 Hz for the central brain alone. Recent work has pushed these limits by imaging the brain (~665 x 333 x 245 μm$^3$, whole brain by box volume) at 1.8 Hz in preparations where the cuticle and superficial trachea were removed (Brezovec et al., 2024).

Overcoming the significant optical challenges posed by air-filled tracheae is an important step towards achieving comprehensive whole-brain imaging in *Drosophila* at single-neuron resolution. Several strategies show promise for mitigating these tracheal effects. Three-photon microscopy is one such avenue due to its inherent advantages in reducing scattering and aberrations (Hsu et al., 2019), though it often faces trade-offs with imaging speed. Other approaches, including advancements in light-sheet techniques (like SCAPE or light beads microscopy), the development of improved dissection preparations or optical window techniques that minimize tracheal interference (Aragon et al., 2022; Brezovec et al., 2024) also offer potential solutions. The goal of these ongoing efforts is to enable consistent, whole-brain functional imaging at single-cell resolution and behaviorally relevant speeds. However, simultaneously achieving optimal depth, speed, resolution, and minimal invasiveness across the entire brain remains a significant technical hurdle. It is worth noting, though, that such advanced imaging may not be strictly required. Indeed, research in *Drosophila* has typically emphasized specific neural circuits rather than whole-brain imaging, in which defined neural populations express calcium indicators. For imaging of large populations of neurons, researchers generally prefer imaging neuropil despite sacrificing single-cell resolution because it enables consistent registration between specimens. However, structural imaging and reconstruction of the same specimen after functional recording could enable a different approach: imaging the neural cell bodies that form an outer "rind" around the *Drosophila* brain. This would provide single-cell resolution data while avoiding the need to penetrate deeply through the problematic tracheal system, potentially achieving close to whole-brain functional imaging with 1- or 2-photon approaches (Harris et al., 2015).

Neural recording capabilities are more limited in freely moving or flying *Drosophila*. Two-photon calcium imaging systems compatible with tethered preparations have been developed (Seelig et al., 2011) and successfully used to record from small numbers of neurons, such as 4-5 pairs of descending neurons (Schnell et al., 2017), during wing flapping and steering behaviors that closely mimic natural flight. However, these techniques remain restricted to highly constrained experimental conditions.

Another key challenge is achieving sustained recordings over behaviorally-relevant timescales. Several innovative approaches have recently emerged to tackle this challenge. Huang et al. developed a cranial window preparation that allows repeated imaging sessions over up to 50 days by using laser microsurgery to create and then reseal an optical window in the cuticle (Huang et al., 2018), while Aragon et al. demonstrated continuous two-photon imaging through the intact cuticle for up to 12 consecutive hours, though this was limited to bright, superficial neurons (Aragon et al., 2022). Most recently, Flores-Valle et al. achieved long-term calcium imaging by collecting 30-second recordings every 5 minutes over 7 days. This intermittent imaging strategy allowed them to accumulate over 15 hours of calcium imaging throughout the week-long experiment (Fores-Valle et al., 2022). These long-term imaging approaches are, however, typically focused on recording from small numbers of neurons, prioritizing stability and duration over coverage.

Progress is also being made in voltage imaging, where several genetically encoded voltage indicators have been applied to *Drosophila* brain imaging, including ArcLight, ASAP, Ace2N-mNeon, Varnam, and, more recently, Voltron (Jin et al., 2012; Yang et al., 2016; Gong et al., 2015; Kannan et al., 2018; Abdelfattah et al., 2019). Using light field microscopy with the ArcLight indicator, Aimon et al. achieved fast (200 Hz) voltage imaging across a considerable portion of the brain at subcellular resolution, though with fewer extractable activity components compared to calcium imaging due to lower signal-to-noise ratios (Aimon et al., 2019). Indeed, voltage imaging faces even greater technical challenges than calcium imaging due to poorer signal-to-noise ratios, and the same limitations regarding imaging depth and resolution apply.

## Neurotransmitters and Neuromodulators

Several GENIs have been successfully validated in *Drosophila*, including sensors for acetylcholine (Jing et al., 2018; Borden et al., 2020), dopamine (Sun et al., 2018; Sun et al., 2020), and serotonin (Wan et al., 2021). A 2009 review identified 119 predicted neuropeptide precursor genes in *Drosophila*, with 46 neuropeptides biochemically confirmed from just 19 of these precursors (Clynen et al., 2009); more recent work suggests approximately 50 neuropeptide precursor genes and a similar number of peptide GPCRs (Nässel & Zandawala, 2019). In any case, significant progress remains to be made in developing tools to monitor these neuropeptides' dynamics in vivo.

## Perturbation

The adult fruit fly represents an attractive model organism for perturbation studies, having a well-characterized connectome, a relatively small number of neurons (~140,000), and an extensive genetic toolkit for targeted manipulation. Single-photon optogenetic stimulation using red-shifted opsins like Chrimson is most commonly used in *Drosophila*, as these wavelengths can penetrate the cuticle without requiring dissection (Klapoetke et al., 2014). Likewise, individual neurons can easily be optogenetically silenced in behaving flies (Mohammed et al., 2017). The powerful genetic toolkit available represents perhaps the strongest aspect of *Drosophila* neuroscience. The split-Gal4 system enables precise targeting of specific cell types, often achieving single-cell-type resolution that surpasses capabilities in other model organisms (Meissner et al., 2025). This genetic specificity, combined with high-throughput behavioral analysis, allows systematic investigation of neural function across the brain. Upon identification of a neuron of interest, the availability of the connectome and of existing tools permits identification and testing of the role of specific neurons upstream and downstream of that neuron (Meissner et al., 2025). Indeed, these genetic tools have revealed neural populations and activity patterns responsible for a wide range of *Drosophila* behaviors (Piatkevich and Boyden, 2023): acquired feeding preferences (Musso et al., 2019), chemotactic navigational decision-making (Hernandez-Nunez et al., 2015), courtship control (Seeholzer et al., 2018), sleep promotion and locomotor activity suppression (Guo et al., 2016), long-lasting internal states in the female brain that regulate multiple behaviors (Deutsch et al., 2020), touch signal processing (Tuthill and Wilson, 2016), context-appropriate walking programs (Bidaye et al., 2020), complex behavioral sequences (Vogelstein et al., 2014), and heading direction representation through ring attractor dynamics (Kim et al., 2017).

Notably, while the recently reconstructed FlyWire connectome provides a complete map of anatomical connections, it does not directly reveal how strongly neurons affect each other's activity in vivo. Pospisil et al. propose that perturbation studies could enable recovery of the "effectome" - a quantitative model of the causal interactions between neurons during brain function (Pospisil et al., 2024). Their key insight is that while studying all possible pairwise interactions between 140,000 neurons would be intractable, the connectome's extreme sparsity (only 0.01% of neuron pairs form synaptic contacts) provides a powerful prior: neurons without direct anatomical connections are unlikely to have substantial direct causal effects on each other. This dramatically reduces the number of potential interactions that need to be measured through perturbation experiments to recover the fly's "effectome," by about four orders of magnitude (~$10^4$-fold) if the approach's sparsity assumption holds.

## Connectomics

Synaptic resolution electron microscopy efforts in *Drosophila* achieved an early milestone in 2015 with the complete imaging of a female first instar larva nervous system (Ohyama et al., 2015), followed by another significant advance in 2018 with the complete imaging of an adult female fly brain using a custom high-throughput serial-section transmission electron microscopy (ssTEM) platform (Zheng et al, 2018). This was followed by focused-ion-beam scanning electron microscopy (FIB-SEM) imaging of the female hemibrain (Scheffer et al, 2020). Subsequent efforts also mapped the female (Azevedo et al., 2024) and male (Takemura et al., 2024) ventral nerve cords from different individuals. Signifying a major step towards holistic maps from single specimens, imaging acquisition has now also been completed for an entire adult male CNS (Berg et al., 2025)– a Janelia FlyEM project in collaboration with the Cambridge Drosophila Connectomics Group – and for an entire adult female CNS as part of the BANC project ([FlyWire Blog, 2024](#)). These initiatives provide the raw data for comprehensive structural maps of individual nervous systems.

Beyond electron microscopy, expansion microscopy (ExM) has emerged as a complementary approach for *Drosophila* connectomics. When coupled with lattice light-sheet microscopy, ExM has enabled the comparison of neural circuit connectivity across multiple specimens while preserving molecular contrast information ([Gao et al., 2019](#)). This protein-specific molecular labeling capability has been further enhanced through transgenic approaches like Bitbow, which enables combinatorial protein barcoding to label neurons uniquely. By targeting five spectrally distinct fluorescent proteins to three subcellular compartments (membrane, nucleus, and Golgi apparatus), Bitbow has demonstrated the ability to generate up to 32,767 distinct molecular barcodes for studying neural circuits in the fly brain ([Li et al., 2021](#)). Most recently, advances in expansion microscopy using potassium acrylate-based hydrogels have achieved expansion ratios exceeding 40x, enabling light microscopy visualization of features like mitochondria within presynaptic compartments at resolutions approaching those of electron microscopy while maintaining whole-brain coverage ([Tian et al., 2024](#)). X-ray-based techniques also offer another avenue for large-volume imaging, with X-ray holographic nano-tomography (XNH) achieving 140-170 nm resolution across millimeter-scale volumes of *Drosophila* nervous tissue ([Kuan et al., 2020](#)).

Assuming a brain volume of 0.04 mm³, imaging at 10 nm isotropic resolution would theoretically generate approximately $4 \times 10^{13}$ voxels. At 2 bytes per voxel for a single channel, this would require approximately 80 terabytes of storage per channel.

The progression of *Drosophila* neuron reconstruction reflects advancing capabilities in automatic neuron tracing and machine-assisted reconstruction proofreading. The initial complete brain imaging by Zheng et al. demonstrated feasibility with a proof-of-concept reconstruction of 120 neurons, with reconstruction requiring, on average, 11.2 person-hours per neuron ([Zheng et al., 2018](#)). This was followed by a significant improvement in the hemibrain project ([Scheffer et al., 2020](#)), which reconstructed over 22,000 neurons and 20 million synapses using automated segmentation supplemented by an estimated 50-100 person-years of human proofreading. The recent FlyWire achievement ([Dorkenwald et al., 2021](#), [Dorkenwald et al., 2024](#)) represents another leap forward, reconstructing the central brain containing 140,000 neurons and over 50 million synapses with 33 person-years of human proofreading, bringing down proofreading time to 19 minutes per neuron ([Dorkenwald et al., 2021](#)). Beyond just the central brain, Azevedo et al. reconstructed approximately 15,000 neurons and 45 million synapses in the ventral nerve cord ([Azevedo et al., 2024](#)), and a fully proofread connectome of an entire adult male CNS has been released ([Berg et al., 2025](#)). Further datasets covering the entirety of the fly's central nervous system are imminent, with the BANC project, also aiming to reconstruct the entire CNS of a single specimen, expected to require several years to complete neuron reconstruction and proofreading ([FlyWire Blog, 2024](#)).

## Computational Modeling

The relative abundance of connectomics data has profoundly shaped *Drosophila* brain emulations, from earlier partial reconstructions in FlyCircuit to the recent release of a full adult connectome. This detailed structural knowledge has enabled increasingly comprehensive models, from circuit-specific implementations to whole-brain simulations attempting to capture system-wide dynamics.

### Huang et al., 2019 and Higuchi et al., 2022

Before a full connectome became available, a few efforts sought to model the whole *Drosophila* brain by inferring connectivity from the partial FlyCircuit database ([Chiang et al., 2011](#)). For instance, Huang et al. and Higuchi et al. developed simulations based on ~14-15% of the fly's neurons, employing methods like spatial proximity to estimate connections from morphological data ([Huang et al., 2019](#); [Higuchi et al., 2022](#)). These groups adopted various modeling approaches: Huang et al. used simplified leaky integrate-and-fire neurons, while Higuchi et al. implemented biophysically detailed Hodgkin-Huxley models, comparing simulated activity to experimental data and known circuit behaviors. Lacking a complete connectome, these early whole-brain models had limited predictive power for the detailed activity of specific neurons or circuits, a challenge subsequent connectome-based models aimed to address.

### Lappalainen et al., 2024

Another influential example of *Drosophila* modeling leveraging partial connectomics data came from Lappalainen et al., who developed a "differentiable mechanistic network" (DMN) constrained by anatomical connectivity ([Lappalainen et al., 2024](#)). Their approach combined two electron microscopy datasets spanning different regions of the optic lobe. This connectivity data was complemented by transcriptomics to infer synaptic signs based on neurotransmitter expression, resulting in predictions for both excitatory and inhibitory connections. Since the fly's visual system is organized in repeating columns of similar neural circuits (akin to a convolutional neural network), they used their detailed reconstructions of a small region to build a larger model spanning approximately 45,000 neurons across 721 columns of the central visual field. The team trained their model to detect visual motion using synthetic visual inputs, converting frames from the Sintel film database ([Butler et al., 2012](#)) into hexagonal arrays of photoreceptor activations that matched the fly's eye structure. Type-specific parameters like membrane time constants, resting potentials, and unitary synapse strengths (a scaling factor for each type-to-type connection, applied to the anatomically measured synapse counts) were optimized for motion detection, specifically, to predict the direction and speed of movement for each point in a visual scene.

Each neuron was treated as a leaky, non-spiking node, with graded synapses encoding excitatory or inhibitory interactions based on known transmitter identities. To translate the network's neural activity into motion predictions, they used a convolutional neural network that took the activity of a subset of neurons as input and produced estimates of movement direction and speed. Model validation involved comparing these motion predictions against ground truth from the Sintel database, alongside testing whether the model reproduced known properties of the fly visual system (e.g., ON/OFF selectivity, T4/T5 direction tuning) established by prior experimental studies. Although training successfully reproduces core visual computations, ablation analyses revealed strong dependence on correct synaptic signs and connectome structure, underscoring the importance of biologically grounded constraints.

## Shiu et al., 2024

The release of the Flywire reconstruction of the adult *Drosophila* brain connectome ([Dorkenwald et al., 2024](#)) enabled a new generation of simulations characterized by both whole-brain coverage and connectome-derived structure constraints. Shiu et al. developed one of the first comprehensive models incorporating this data, working with over 127,400 proofread neurons and their 50-plus million synaptic connections ([Shiu et al., 2024](#)). To assign neurotransmitter identities, they leveraged prior large-scale predictions ([Eckstein et al., 2024](#)), broadly classifying neurons as excitatory (primarily cholinergic) or inhibitory (GABAergic or glutamatergic). Dopaminergic, octopaminergic, and serotonergic neurons were also incorporated and treated as excitatory. Connection weights were derived directly from the Flywire connectome, with sign determined by the assigned neurotransmitter identity. The model required fitting only a single free parameter - the global synaptic weight scale (Wsyn). This global scaling factor parameter determined how strongly each synapse influenced the postsynaptic membrane potential. This single Wsyn value was applied globally, meaning its magnitude was independent of the particular identities of the neurons forming any given connection. It was calibrated using known feeding circuit dynamics, specifically tuned so that 100 Hz activation of sugar-sensing gustatory receptor neurons produced approximately 80% of maximal firing in motor neuron 9 (MN9), a key neuron controlling extension of the fly's proboscis (feeding appendage). This calibration point was chosen based on extensive prior experimental characterization of the sugar sensing in the feeding initiation pathway.

The model employed a leaky integrate-and-fire (LIF) framework with α-synapse dynamics, incorporating synaptic conductance, membrane resistance, and time constants derived from previous *Drosophila* modeling and electrophysiological studies. Overall, the model made several simplifying assumptions: neurons had zero basal firing rates, gap junctions were not available and thus excluded, and neuromodulatory effects beyond basic excitation/inhibition were not incorporated.

Despite those simplifications, validation efforts yielded encouraging results. Shiu et al. primarily focused on simulating two well-characterized circuits: the feeding circuit and the antennal grooming circuit, to validate their model ([Shiu et al., 2022](), [Hampel et al., 2015]()). They computationally activated subsets of gustatory receptor neurons (sugar, water, bitter, and Ir94e-expressing) for the feeding circuit and analyzed the resulting network activity patterns. Similarly, they simulated the antennal grooming circuit by activating mechanosensory neurons in the Johnston's organ, a sensory structure in the antenna that detects antennal movements. The authors employed three primary approaches to validate these predictions: direct comparison of computationally predicted neural activity with experimental calcium imaging data, in silico silencing experiments to assess the necessity of specific neurons (validated against genetic silencing experiments in real flies), and optogenetic activation studies testing whether specific neurons were sufficient to elicit the behaviors predicted by the model (without embodiment). Through this multifaceted validation strategy, they demonstrated their model could accurately recapitulate known feeding and grooming behaviors, achieving 91% accuracy across 164 experimental predictions and generating novel insights like the inhibitory role of Ir94e neurons in feeding. A companion study further validated the model's predictive power by using it to successfully identify distinct neural circuit mechanisms underlying context-specific halting behaviors in the *Drosophila* locomotion system ([Sapkal et al., 2024]()).

### Cowley et al., 2024

Unlike the connectome-driven approaches described above, Cowley et al. developed their model primarily through behavioral and functional constraints ([Cowley et al., 2024]()). Their approach centered on understanding how specific neuron types contribute to behavior by systematically silencing neurons and incorporating these perturbations into model training. They collected behavioral data from 459 male-female fly pairs during courtship interactions for model fitting. They recorded six behavioral variables from the male: three movement parameters (forward velocity, lateral velocity, and angular velocity) and three measures of song production (sine song, fast pulse song, and slow pulse song). In each experimental condition, they genetically silenced one of 23 different visual projection neurons (LC) types that form a bottleneck between the optic lobe and the central brain. Their model consisted of three components: a convolutional network processing the male's reconstructed visual experience, a bottleneck layer of 23 units (each representing one LC type), and a decision network producing behavioral outputs. The model processed sequences of 10 frames (~300ms) of visual input to predict behavior, and its training involved "knockout training": when training on data from flies with a silenced LC type, Cowley et al. set the corresponding model unit's activity to zero, forcing the network to learn how each LC type contributes to behavior. To validate their model, they performed two-photon calcium imaging in head-fixed males from five LC types, testing responses to both artificial and naturalistic visual stimuli. Despite being trained only on behavioral data, their model achieved a 35% correlation with neural responses. The model's prediction that LC types work in combination rather than as independent channels was further supported by analysis of the FlyWire connectome, which revealed shared inputs and outputs among LC types.

### NeuroMechFly and Vaxenburg et al., 2025

Parallel to these advances in brain simulation, there has also been significant progress in developing detailed embodied simulations of *Drosophila*. While these models currently employ relatively simple neural controllers compared to the simulations discussed above, they represent an important complementary approach that could eventually enable studying a broader range of fly behaviors when combined with more sophisticated neural models. Two notable efforts in this direction are the NeuroMechFly project and the recent work by Vaxenburg et al. The NeuroMechFly team developed a morphologically accurate model derived from X-ray microtomography data with 65 body segments and 122 degrees of freedom, where each degree of freedom represents an independent type of movement or rotation possible at a joint. The model initially focused on walking and grooming behaviors ([Lobato-Rios et al., 2022](#)). Their first version learned to produce stable walking patterns matching those seen in real flies – see the [video](#) – including the typical three-legged walking pattern where legs move in alternating groups of three. Their 2024 update expanded the model's capabilities to include vision and smell, demonstrating more complex behaviors like following scent trails while avoiding obstacles in their path ([Wang-Chen et al., 2024](#)). Notably, this updated version integrated the connectome-constrained model developed by Lappalainen et al. to simulate visual processing during a fly-following task. Vaxenburg et al. took a different approach, building their model from high-resolution confocal microscopy data with 67 body segments and 102 degrees of freedom, along with sophisticated physics modeling including fluid dynamics for flight and special features that allowed the simulated fly to stick to surfaces like real flies do ([Vaxenburg et al., 2025](#)). Through reinforcement and imitation learning, their ANN-driven model successfully replicated both walking and flight behaviors from real fly trajectories, including complex maneuvers like rapid turns and sudden changes in direction. Their model also demonstrated behaviors such as maintaining altitude over uneven terrain and navigating through winding trenches without collision.

## Gap Analysis

Like larval zebrafish, *Drosophila* represents a unique convergence of experimental tractability and biological sophistication, making it an attractive target for integrated brain simulation efforts. Scheffer and Meinertzhagen recognized the need for integrated approaches in a comprehensive 2021 analysis, where the authors outlined 15 key areas requiring coordinated investigation beyond connectomics, spanning biochemistry, cell physiology, and whole-animal concerns ([Scheffer and Meinertzhagen, 2021](#)).

A core technical challenge facing *Drosophila*-based approaches stems from challenges of imaging all neurons simultaneously. The presence of air-filled tracheae and fat deposits in *Drosophila* creates significant challenges for whole-brain imaging at cellular resolution. However, the ability to efficiently image defined populations of neurons, as well as easily perturb those defined neurons while measuring both neural activity and behavior, makes *Drosophila* a powerful model for neuron modeling.

Similar to larval zebrafish, adult *Drosophila* has only minor clinical and industrial applications. However, fruit flies *do* represent stable adult-stage organisms rather than rapidly developing larvae. As a result, they could potentially support neural recording in the same individuals over longer time horizons, without the complications caused by rapid neural development. This eliminates many of the constraints that characterize larval zebrafish work, where the brief developmental window limits experiment duration and complicates data integration across modalities. Furthermore, adult *Drosophila* exhibit a substantially richer behavioral repertoire, including sophisticated social behaviors. This behavioral sophistication could provide more stringent validation criteria for whole-brain *Drosophila* simulations, including better tests of out-of-domain generalization.

| Model Organism Overview: *Drosophila* | Pros | Cons |
|---|---|---|
| Anticipated Scientific Insights | <ul><li>**Rich Adult Behavioral Repertoire:** Complex social behaviors, learning, individuality and motor sequences provide sophisticated validation targets.</li><li>**Structure-Function Bridge:** Could illuminate approaches to inferring causal models in systems where exhaustive recording/perturbation is not feasible.</li></ul> | <ul><li>**Evolutionary Distance:** Further from mammalian brain architecture than other tractable models like larval zebrafish.</li></ul> |
| Experimental Tractability | <ul><li>**Mature Research Ecosystem:** Drosophila has a broad research community, with sophisticated genetic tools including frontier technologies such as facile single cell-type specific genetic access, combinatorial barcoding libraries already being close to achieving whole-brain coverage (Bitbow's 32,000+ barcodes) and expansion microscopy protocols achieving electron microscopy-like spatial resolution (>40× expansion with Re-PKA-ExM), and was able to leverage this for larger consortia.</li><li>**Multiple Technical Paths Forward:** Complementary advances in EM, ExM, and X-ray microscopy provide diverse routes to molecular and structural mapping</li><li>**Stable Adult Platform:** Unlike larval models, it allows extended experiments and complex behavioral studies</li><li>**Practical Advantages:** Low-cost maintenance and manageable computational scale (~140k neurons)</li><li>**Feasibility–Complexity Sweet Spot:** ~140k neurons, still small enough for whole-brain connectomics and near-whole-brain imaging.</li></ul> | <ul><li>**Challenges performing whole-brain imaging.** Compared to the transparent larval zebrafish brain, high speed, whole brain imaging at cellular resolution is more challenging.</li><li>**Behavioral Recording Constraints:** Current imaging setups (head-fixation, etc.) somewhat limit the observable behavioral repertoire, particularly for organisms capable of complex behaviors like courtship or flying.</li></ul> |

| Gaps and opportunities: *Drosophila* | Gaps (non-exhaustive selection) | Illustrative Project Opportunities |
|---|---|---|
| | | |
| **Neural dynamics** | <ul><li>**Imaging Trade-offs:** Current approaches navigate these challenges through different compromises. Light field microscopy achieves fast (~200Hz) whole-brain imaging but at sub-neuropil resolution. Two-photon microscopy provides cellular resolution but usually requires head fixation and has a limited field of view.</li><li>**Advance the frontier of maximum recording time in *Drosophila*:** Currently, up to ~15h are possible. Given that the flies can get up to 75 days old, potentially up to an order of magnitude more is theoretically possible.</li><li>**Neuromodulation in Drosophila**: bigger datasets on neuromodulation are needed</li></ul> | <ul><li>**Deep-Brain Three-Photon Performance:** Systematically characterize resolution, signal-to-noise ratio, and photodamage limits of three-photon microscopy in deepest regions of the *Drosophila* brain. Establish performance benchmarks for imaging through trachea-filled tissue at cellular resolution.</li><li>**Whole-Brain Voltage Dynamics**: Optimize next-generation voltage indicators and imaging preps for simultaneous recording from >10,000 neurons across multiple *Drosophila* brain regions at millisecond resolution.</li><li>**Develop some way of imaging flies in complex behaviors**, including extremely lightweight threads for continuous brain recording.</li><li>**Expand single brain recording horizon:** Develop sophisticated repeated imaging abilities for fruit flies that allow for 25 h+ of recording per individual</li></ul> |
| **Connectomics** | <ul><li>**Molecular Limitations:** The FlyWire connectome, derived from electron microscopy, provides primarily morphological and connectivity information, in addition to limited data to distinguish excitatory vs. inhibitory neurons,, lacking crucial details about neurotransmitter identities, receptor distributions, and other molecular properties that influence circuit function. The integration of barcoding and ExM approaches towards creating a comprehensive ExM-derived molecularly annotated connectome is nascent, with some pioneering work at Janelia.</li><li>**Inter-individual Variability:** While there are indications of stereotypy in the fly brain's neural circuits, systematic understanding remains limited as the community has reconstructed little more than "one and a half" connectomes to date. Moreover, it does not yet account for sex differences.</li></ul> | <ul><li>**Aligned neuronal recordings and connectomics:** Ideally, both males and females will reconstruct whole CNS connectomes, combined with extensive calcium imaging prior to reconstruction.</li><li>**Whole Body EM:** Scan the whole organism and trace nerves with more detail across the organism.</li><li>**Multiplexed Molecular Mapping Protocol:** Develop a protocol for iterative antibody labeling and imaging compatible with high-expansion (>40×) ExM and enable reliable detection of 15+ proteins through serial rounds of staining.</li><li>**Synaptic-Scale X-ray Microscopy:** Given that Drosophila has some of the finest neurites known to exist, demonstrating sufficient resolution imaging of *Drosophila* brain tissue using X-ray ptychography and validating against electron microscopy ground truth could pave the way for future X-ray-based whole-brain imaging applications.</li></ul> |
| **Computational Neuroscience** | <ul><li>**Missing Molecular Foundation:** While the FlyWire connectome provides complete structural connectivity, current models must bridge a significant structure-to-function gap through multiple assumptions, including neurotransmitter identities, synaptic strengths, and more (Eckstein et al., 2024).</li><li>**Integrated Neuromechanical Framework:** Current tools typically separately simulate neural activity and biomechanics. To integrate these components, unified simulation platforms are needed.</li><li>**Limited Experimental Electrophysiological Data:** Electrophysiological data (e.g., membrane potentials or spiking frequency) is often the most useful inputs</li></ul> | <ul><li>**Resting state Drosophila brain model:** There still has not been a full *Drosophila* brain simulation in which all neurons would be active at least at their spontaneous levels across the whole brain. One needs a spontaneously active brain network within which some sensory or other signals can be instantiated.</li><li>**"Emulate the fly" roadmap:** An end-to-end plan for collecting all components to create a compelling emulation of the fly.</li><li>**OpenFly framework:** An effort that aggregates all the computational neuroscience data and provides computational tools for everyone.</li></ul> |

| | | and outputs for computational modeling, yet this data is challenging to collect in *Drosophila*. | |

# Mouse

## Anatomy & Behavior

The house mouse (*Mus musculus*) is one of the most widely used model organisms in neuroscience. At birth, the typical mouse weighs approximately 1 gram, and males reach around 36 grams while females reach around 27 grams ([JAX](JAX)). In laboratory conditions, mice typically live 2-3 years, with some individuals reaching up to 5 years under optimal conditions, though wild specimens rarely survive beyond 1-2 years due to predation and environmental pressures.

The mouse brain reaches approximately 90% of its adult size by two weeks of age ([Orr et al., 2016](Orr et al., 2016)). Neuron counts increase from approximately 57 million at 4 weeks to a peak of 69 million by 15 weeks before stabilizing at 63-70 million in adulthood ([Fu et al., 2012](Fu et al., 2012); [Herculano-Houzel et al., 2006](Herculano-Houzel et al., 2006)). These neurons are packed into a volume of 420-460 mm$^3$ ([Vincent et al., 2010](Vincent et al., 2010)) – ~5,000x larger than the brains of *Drosophila* and larval zebrafish (~0.04-0.08 mm$^3$), and roughly 3,000x smaller than the human brain (1,200 cm$^3$). The neurons are distributed across major regions, including the cerebellum (~52 million neurons), neocortex (~8.6 million neurons), and olfactory bulb (~7.2 million neurons). Outside the brain, the spinal cord contains an estimated 8 million neurons ([Fu et al., 2012](Fu et al., 2012)). Typical firing rates in vivo are estimated to fall broadly within a range from below 0.001 Hz to 50 Hz or more, varying significantly with cell type and behavioral state. Energy budget models for the rodent cortex use an estimated average firing rate of 4 Hz ([Attwell & Laughlin, 2001](Attwell & Laughlin, 2001); [Howarth et al., 2012](Howarth et al., 2012)), a value derived from earlier in vivo studies in rats where population averages ranged from 1.5-4 Hz (and individual neurons from 0.15-16 Hz). The electrophysiological properties of neurons and synapses in the mouse brain, particularly within the cortex, are relatively well characterized. Combined with the extensive knowledge of cell types derived from transcriptomics, the mouse emerges as an attractive model system for whole-brain emulation efforts.

> **Videos of rodent behavior (rat, as no similar mouse videos were available)**
>
> [Various complex motor and learning behaviors](Various complex motor and learning behaviors)

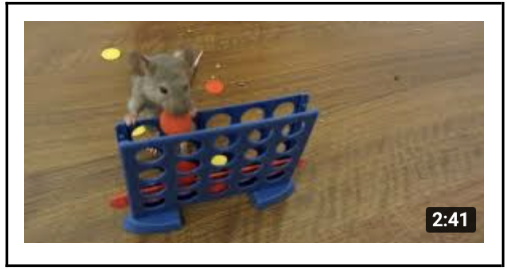

Even before birth, the maternal environment and hormonal factors mold behavioral tendencies. Individual differences in traits like anxiety and exploration emerge and gradually stabilize throughout development. However, major life events such as social stress or environmental challenges can still significantly influence the adult behavioral phenotype ([Brust et al., 2015](#)).

Adult mice exhibit a sophisticated behavioral repertoire encompassing complex social, cognitive, and reproductive domains. Their social organization features intricate dominance hierarchies maintained through scent marking, aggressive displays, and vocal and chemical communication. Cognitively, they demonstrate remarkable capabilities in spatial navigation, associative learning, and behavioral flexibility, readily adapting to environmental changes and remembering both positive and aversive experiences. Even among genetically identical individuals, mice show stable personality differences in traits like anxiety, exploration, and sociability - variations that persist despite standardized laboratory conditions. Their reproductive behavior involves elaborate courtship rituals, and females display comprehensive maternal care, including nest building, pup retrieval, and nursing. These core behavioral patterns remain relatively stable throughout adulthood, though aging gradually diminishes exploratory drive, learning speed, and overall activity levels ([Brust et al., 2015](#)).

## Neural Dynamics

Neural activity recording

In recent years, we have seen remarkable progress in our ability to record neural activity in behaving mice. Key approaches include optical techniques, such as calcium and voltage imaging, and electrophysiological recordings, notably with high-density probes like Neuropixels. Nevertheless, fundamental tradeoffs remain between the number of neurons that can be recorded simultaneously, temporal resolution, and the animal's freedom of movement. While electrophysiology offers unparalleled temporal resolution for individual spikes of up to a few thousand neurons (as discussed later), current approaches to calcium imaging in mice have also seen tremendous advances and can be broadly divided into head-fixed and freely moving preparations, each with distinct advantages and limitations.

In awake, head-fixed preparations, mice are typically positioned on treadmills that allow some degree of movement while maintaining the stability needed for high-quality imaging. This approach has enabled increasingly comprehensive recordings of neural activity. The MICrONS consortium, for example, recently demonstrated simultaneous recording of approximately 75,000 excitatory neurons across layers 2-5 of visual cortex at 6-10 Hz, spanning multiple visual areas during 14 80-minute sessions over 6 days, close to 20h in total ([MICrONS Consortium, 2024](#)). Additionally, the Allen Institute for Brain Science has produced extensive open-access calcium imaging datasets under standardized conditions, often targeting specific cell types. Notable examples include their 'Visual Coding 2P' dataset, featuring recordings from nearly 60,000 neurons during passive sensing ([de Vries et al., 2019](#)), and their 'Visual Behavior 2P' dataset, with data from over 50,000 neurons collected during active behavioral tasks ([Piet et al., 2024](#)), provide deep insights into cortical function.

**Figure Overview of the optical neural recording landscape in Mouse:** Radar plots based on the optical recording literature cited in the report. We plot the following dimensions of brain recordings: spatial resolution, brain volume, temporal resolution, and (estimated) individual and cumulative recording duration. The plots split recordings from fixated (A) and freely moving experiments (B). The outer ring is normalized to the maximum known values. Each ring represents one order of magnitude. The data for this figure is available in the linked data repository.

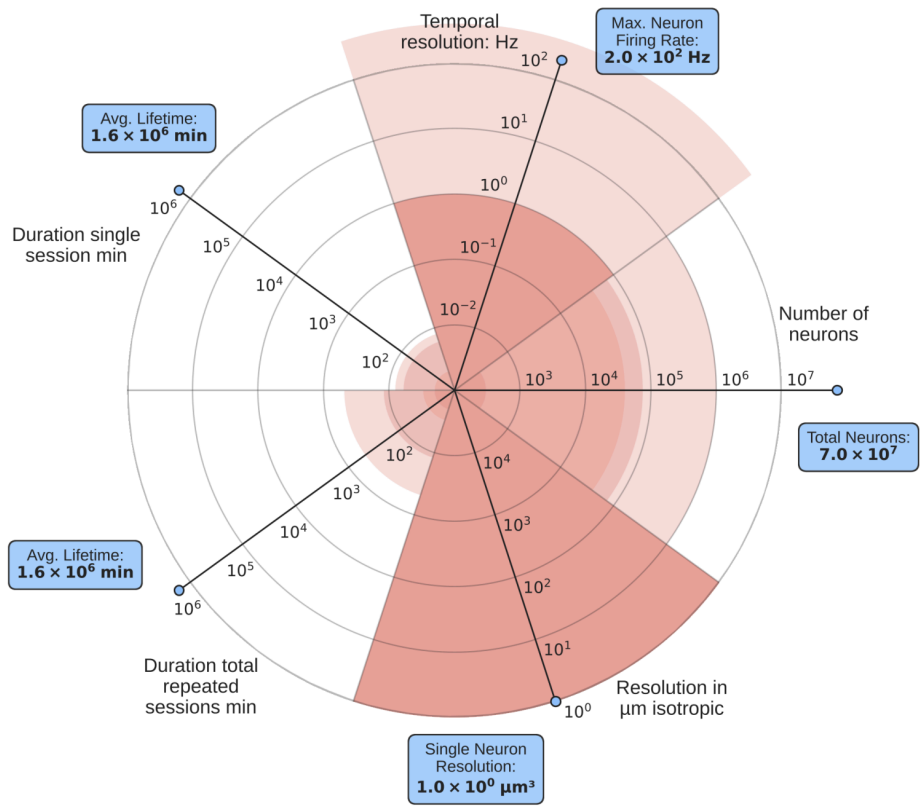

A) fixated

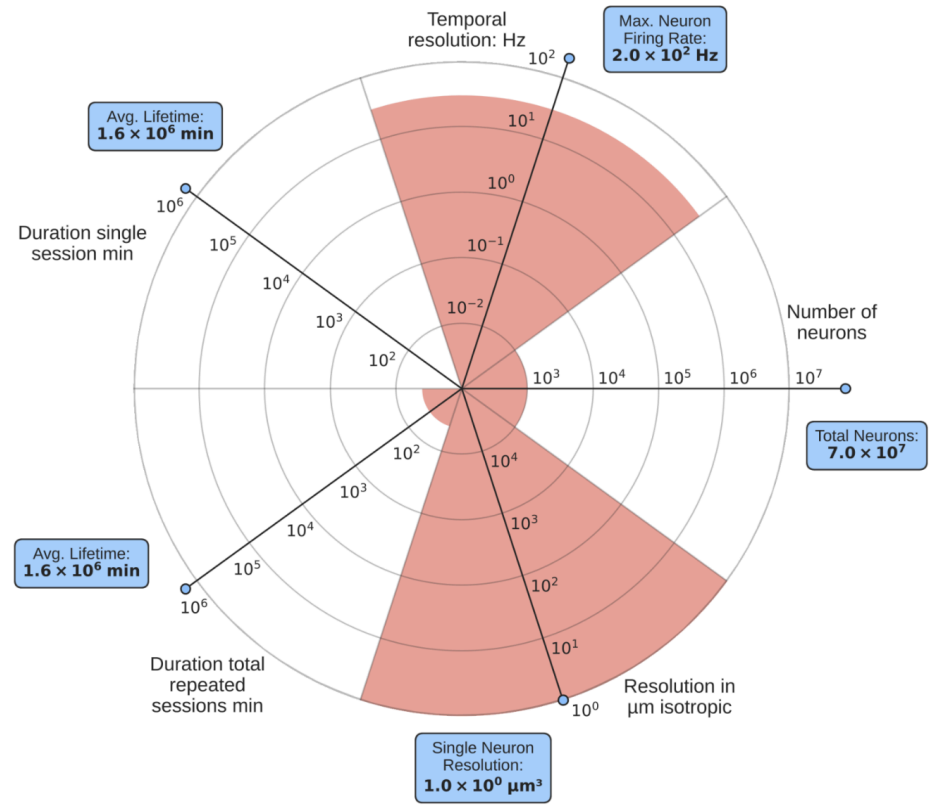

B) moving

Even more expansive capabilities have been shown using another 2P-imaging approach (light beads microscopy), which has enabled simultaneous imaging of up to 1 million neurons across the dorsal cortex while maintaining cellular resolution at 2 Hz ([Manley et al., 2024](#)). In parallel, developments in surgical approaches have enhanced optical access – curved glass windows ("crystal skull") that replace the dorsal cranium – remain viable for imaging at least 11 weeks post-surgery, providing access to an estimated 800,000-1,100,000 neurons spanning over 30 neocortical areas, though the maximum sustainable recording durations possible with this approach have not been systematically characterized ([Kim et al., 2016](#)). However, despite these technological advances, head-fixed preparations fundamentally limit the range of natural behaviors that can be studied. To mitigate this limitation and expand the repertoire of investigable behaviors, sophisticated virtual reality (VR) environments have been integrated with head-fixed preparations. These setups allow mice to perform complex tasks, such as spatial navigation or decision-making, while neural activity is monitored with cellular resolution using 2P microscopy ([Dombeck et al., 2010](#); [Harvey et al., 2012](#)).

Advances in freely moving imaging have also been substantial, driven particularly by the development of lightweight microscopes. The MINI2P microscope weighs under 3g and uses a flexible 0.7mm fiber bundle cable, enabling two-photon imaging at 15-40 Hz from 300-600 neurons per plane across up to four imaging planes (approximately 1000 neurons total) within a 420 × 420 μm² field of view. When stitching across multiple fields of view to cover a 2.2 × 2.2 mm² area, over 10,000 neurons can be recorded, though at reduced temporal resolution ([Zong et al., 2022](#)). Even lighter microscopes have since been developed, such as the 0.43g TINIscope ([Xue et al., 2023](#)). However, imaging systems compatible with freely moving mice face particularly harsh tradeoffs between the number of neurons recorded, temporal resolution, and signal quality. Even with optimal recording conditions, calcium indicators remain too slow to capture individual action potentials for many neurons in mice. Recent advances in voltage imaging offer promise for tracking individual spikes, with new approaches enabling simultaneous recording from over 300 spiking neurons in the mouse cortex at 400 Hz for over 20 minutes ([Bai et al., 2024](#)).

An alternative to either calcium or voltage imaging is electrophysiology. A large-scale example of such an effort is the International Brain Laboratory ([The International Brain Laboratory, 2017](#)). This initiative was able to collect a "comprehensive set of recordings from 115 mice in 11 labs performing a decision-making task with sensory, motor, and cognitive components, obtained with 547 Neuropixels probe insertions covering 267 brain areas in the left forebrain and midbrain and the right hindbrain and cerebellum". Mice were trained to turn a wheel in response to the position of a visual grating. This generated roughly 60 GB of electrophysiology data and 1 TB of video data, of roughly 150 hours in total. The Allen Institute has also generated extensive Neuropixels datasets. For example, their 'Visual

Coding Neuropixels' dataset, derived from mice passively viewing visual stimuli, comprises recordings from over 40,000 quality-controlled units cumulatively across many experiments, typically using up to six probes per mouse to target visual cortex, hippocampus, and thalamus (de Vries et al., 2023). An even larger dataset, the 'Visual Behavior Neuropixels' project, was acquired while mice performed a change detection task, yielding over 200,000 units recorded from 81 mice using a similar multi-probe strategy that included visual cortical, thalamic, hippocampal, and midbrain structures (Allen Institute for Brain Science, 2022). It is important to note that these large unit counts are aggregates from numerous experiments; individual multi-probe recordings simultaneously capture activity from hundreds to potentially a few thousand units, depending on the number of probes and the density of recorded regions.

## Neurotransmitters and Neuromodulators

Several genetically encoded indicators now enable monitoring of key neuromodulatory systems in the mouse brain, including sensors for acetylcholine, dopamine, histamine, norepinephrine, and serotonin (Muir et al., 2024). Additionally, sensors have been developed for select neuropeptides, including CRF, dynorphins, enkephalins, GRP, orexins, oxytocin, and somatostatin. However, these tools cover only a small portion of known neuromodulatory systems. Mass spectrometry studies have identified between 500-850 peptides in the mouse brain, with roughly half being classical neuropeptides derived from about 40-45 neuropeptide precursor genes, and half being peptides derived from intracellular proteins (Fricker et al., 2010; Zhang et al., 2012). Developing tools to monitor this broader range of signaling molecules remains an important challenge for understanding neuromodulation in vivo.

## Perturbation

Optogenetic techniques enable precise spatiotemporal control of neural activity. Advanced approaches like two-photon optogenetics aim for cellular-scale targeting, while three-photon excitation can offer access to deeper brain regions, typically 1-1.3 mm into brain tissue, allowing targeting of structures like layer 5 of the cortex (Xu et al., 2024; Lee et al., 2020; Adesnik and Abdeladim, 2021). Nonetheless, the capabilities offered by optogenetics, including both one-photon and multi-photon methods, have enabled several discoveries, including neural populations and activity patterns responsible for diverse behaviors (Piatkevich and Boyden, 2023). These behaviors include parental care (Kohl et al., 2018), spatial object recognition (Kempadoo et al., 2016), aggression against intruders (Lin et al., 2011), breathing rhythm (Sherman et al., 2015), social memory (Oliva et al., 2020) and social-spatial association formation (Murugan et al., 2017), visual perception (Lee et al., 2012), wakefulness (Cho et al., 2017), locomotion (Hagglund et al., 2013), sleep (Kitamura et al., 2017), face gender discrimination (Afraz et al., 2015), water (Zimmerman et al., 2016) and food consumption (Nectow et al., 2017), responses to odors (Root et al., 2014), movement (Gritton et al., 2019) and aversion or preference of a place (Kim et al., 2019), reward-seeking (Otis et al., 2017), parental behavior (Stagkourakis et al., 2020) and encoding of places (Zhang et al., 2013). Suthard et al. demonstrated that optogenetic stimulation of hippocampal engram cells can recapitulate the cellular activity patterns seen during natural fear memory recall, revealing coordinated neuron-astrocyte dynamics that underlie both naturally and artificially induced fear states (Suthard et al., 2024). Importantly, one- and two-photon optogenetic perturbations in the mouse have helped uncover circuit mechanisms underlying a variety of neural computations involved in perception and action (e.g., Reinhold et al., 2015; Li et al., 2015; Lien and Scanziani, 2018; Carrillo-Reid et al., 2019; Marshel et al., 2019; Chettih and Harvey, 2019; Keller et al., 2020; Daie et al., 2021; Green et al., 2023; Vinograd et al., 2024) and even enabled comparative studies with human subjects, such as shedding light on the mechanisms of dissociative first-person experience (Vesuna et al., 2020).

## Connectomics

Establishing the IARPA MICrONS consortium in 2016 marked an important development in mouse connectomics. The consortium's work progressed in two major phases – first imaging a focused volume in cortical layer 2/3 (250 x 140 x 90 μm), then expanding to image, segment and partially reconstruct an entire cubic millimeter spanning six cortical layers and three higher visual areas. MICrONS combined in vivo calcium imaging with electron microscopy, generating over a petabyte of imaging data (MICrONS Consortium et al., 2025). During this period, a parallel effort by Motta et al. mapped a 90 x 90 x 60 μm volume of somatosensory cortex within just 4,000 person-hours (Motta et al., 2019), largely comprising targeted manual work to correct the automated reconstruction (focusing on resolving algorithm-identified axon splits and mergers, and completing dendritic spine attachments).

In 2023, the NIH launched the $150 million BRAIN Initiative Connectivity Across Scales (BRAIN CONNECTS) program, funding 11 projects over 5 years to develop tools for brain-wide connectivity mapping. One prominent project within this initiative, a $33 million Harvard-led effort (in collaboration with Google Research, Allen Institute, MIT, Cambridge University, Princeton University, Johns Hopkins University) aiming to reconstruct and proofread 10 mm³ of the mouse hippocampus using high-throughput electron microscopy (Januszewski, 2023), was seemingly impacted by changes in NIH funding priorities in early 2025 (Markowitz, 2025), though the technical work appears to be continuing independently of NIH support. However, the BRAIN CONNECTS program includes other significant synaptic connectomics efforts. For instance, a project led by the Allen Institute received approximately $6.1 million in its first year to image up to 10 mm³ of the mouse cortico-basal ganglia-thalamo-cortical loop at synaptic resolution ([NIH, 2023](#)). This project uses serial section tilt TEM tomography and aims to develop a pipeline for high-throughput integrated volumetric electron microscopy for whole mouse brain connectomics, including linking to cell types via gene expression data. While focusing on just a fraction (roughly 2-3%) of the mouse brain, this volume represents a notable scaling challenge, approximately 10 times larger than the previous MICrONS dataset. The five-year project will use two 91-beam electron microscopes operating in parallel. Success would demonstrate whether current electron microscopy and automated reconstruction technologies can scale sufficiently to tackle the complete mouse connectome eventually. Notably, maintaining a similar 5-year timeline for imaging an entire mouse brain would require scaling to approximately 40-50 such microscopes working in parallel, highlighting both the technical challenges and infrastructure requirements for mapping complete mammalian brains at synaptic resolution.

Regarding expansion microscopy in the mouse brain, the availability of sophisticated genetic tools combined with the prohibitive scale of EM-based reconstruction has accelerated the development of alternative approaches. E11 Bio, one of the first FROs launched by Convergent Research, recently detailed its PRISM (Protein-barcode Reconstruction via Iterative Staining with Molecular annotations) platform, which addresses key bottlenecks in light-microscopy connectomics by providing neurons with unique molecular signatures for self-correcting reconstruction ([Park et al., 2025](#)). The platform achieves this by combinatorially expressing 18 antigenically distinct, cell-filling proteins via AAVs, which are then visualized in 5x expanded tissue through iterative immunostaining. This approach enables automated proofreading across spatial gaps of hundreds of microns and was shown to increase automatic tracing accuracy by 8-fold over conventional single-color methods. In a demonstration on a ~10 million μm³ volume of the mouse hippocampus, the technique also enabled detailed molecular mapping of synapses, revealing that large, complex synaptic structures known as 'thorny excrescences' tend to have similar sizes when they are clustered closely together on the same dendrite.

Among other ExM developments, ExA-SPIM ([Glaser et al., 2024](#)) demonstrated unprecedented imaging throughput, achieving 946 megavoxels per second with an effective resolution of 250x250x750 nm³ after 4x expansion. This effective resolution is calculated by dividing the microscope's native optical resolution by the tissue's linear expansion factor, indicating the resolving power relative to the sample's original, unexpanded dimensions. However, this resolution is far from sufficient for dense connectomic reconstruction. The team successfully imaged entire 3x expanded mouse brains in just 24 hours per channel, tracking sparsely labeled subcortical projection neurons and their axonal projections across the brain. This combination of high throughput, multi-color capability, and relatively low system cost – approximately $175,000 to $250,000, depending on the laser configuration (A. Glaser, 2024, personal communication) – makes it particularly promising for whole-brain mapping efforts. This potential has motivated the development of new optical systems, including a custom lens that, when combined with higher expansion factors of 3-12×, will enable effective lateral resolutions of 50-150nm, potentially bringing the system not far from the resolution required for dense reconstruction. Meanwhile, Tavakoli et al with their LICONN (light-microscopy based connectomics) approach demonstrated dense reconstruction in mouse cortex, imaging a ~1 million cubic micrometer volume spanning cortical layers II/III-IV at effective resolutions of ~20nm laterally and ~50 nm axially through ~16-fold expansion, with the 0.47 teravoxel dataset acquired in just 6.5 hours at an effective voxel rate of 17 MHz (Tavakoli et al., 2025). In parallel, Kang et al. developed multiplexed expansion revealing (multiExR), achieving visualization of more than 20 distinct proteins within the same mouse brain specimen through sequential rounds of antibody staining and imaging while achieving median registration precision of 25-39 nm (Kang et al., 2024).

X-ray microscopy efforts also show promise. Early demonstrations using X-ray holographic nano-tomography (XNH) achieved sub-100 nm resolution across 300x200x1000 μm tissue volumes (Kuan et al., 2020). More recently, Bosch et al. developed a correlative workflow combining in vivo calcium imaging, synchrotron X-ray tomography, and volume electron microscopy to investigate both function and structure within the same tissue (Bosch et al., 2022). In follow-up work, Bosch et al. also demonstrated another key advance: using X-ray ptychographic tomography under cryogenic conditions with specialized radiation-resistant resins to achieve sub-40nm resolution capable of resolving individual synapses (Bosch et al., 2023), an important achievement for x-ray connectomics. While whole-brain X-ray imaging has been demonstrated at cellular resolution (Humbel et al., 2024), achieving synaptic resolution across large volumes remains an active area of development.

The mouse hippocampus connectome is expected to generate an estimated 25 petabytes of data (Google Research, 2023). Imaging a whole mouse brain (approximately 500 mm$^3$) at 10 nm isotropic resolution would theoretically generate approximately 5 x $10^{17}$ voxels (or 3.2 x $10^{16}$ voxels at 25 nm isotropic resolution). At 2 bytes per voxel for a single channel, this would thus require approximately 1 exabyte (or 64 petabytes at 25nm isotropic resolution). The significant storage requirements of connectomes at the scale of the whole mouse brain (and beyond) have motivated the development of compression techniques capable of alleviating data management constraints: EM-compressor, for example, can decrease the storage needed to store raw EM data by as much as 128x without compromising subsequent neuron reconstruction (Li et al., 2024). Meanwhile, neuron reconstruction efforts are ongoing for the MICrONS cubic millimeter, containing an estimated 120,000 neurons. Automatic synapse detection has identified over 523 million synapses within the volume, and, as of January 2025 (v1300), over 1,700 axons have been manually proofread and cleaned, establishing over 500,000 verified connections to somas within the dataset ([MICrONS Consortium, 2024](#)).

## Computational Modeling

### The Blue Brain Project

The Blue Brain Project (BBP) was a pioneering effort to construct large-scale, biophysically detailed in silico models of rat cortical microcircuitry. Although BBP focused on rat rather than mouse cortex, we include it for historical context, as its biophysically detailed reconstruction methods and data-integration pipelines have substantially informed subsequent mouse modeling. Having been started before large-scale connectomic datasets were available, the BBP developed various tools to integrate disparate datasets describing cell types and connectivity in the rodent cortex. Cortical anatomy of their most significant recent model release ([Reimann et al., 2024](#)) was based on a three-dimensional cell atlas, in which cell bodies of 60 morphologically distinct cell types were placed based on experimentally reported densities. From there, detailed multi-compartment models of cells were expanded based on known shape constraints and populated with ion channels based on parameter-tuning to recreate *in vitro* recordings. Connectivity is inferred based on the principle of axonal-dendritic overlap: computationally generated neuronal morphologies, cloned from available reconstructions, are placed in the model volume, and connections are formed where their processes are sufficiently close, with subsequent extensive pruning to match experimentally observed synaptic densities. While the resulting synaptic density statistics are compared against experimental data for validation, this approach assumes that geometric proximity is the primary determinant of local connectivity and does not directly incorporate more recently available, detailed maps of specific circuit wiring, potentially missing more complex organizational rules. Synapses are modelled on a detailed level, including a pool of available neurotransmitter vesicles and short-term plasticity. In total, they developed a detailed model of 36 mm³ of rat somatosensory cortex (a process they term 'reconstruction', which, it is important to note, involves extensive model building with numerous assumptions, distinct from data-driven connectomic reconstructions), comprising 4.2 million neurons with 14.2 billion synapses between them. They had access to the [Blue Brain 5](#) supercomputer providing 0.8 TFlops of computational power.

Electrical properties of single neurons were validated by optimising ion channel densities in different neuronal compartments to give rise to firing properties, action potential waveforms, and passive properties observed in vivo (Reva et al., 2023). Synaptic parameters were fitted similarly (Ecker et al., 2020). The size of the model allowed modelling not only of local connections within a cortical column, but also mid-range connections between close brain regions (Isbister et al., 2024), and spontaneous activity was comparable to that observed *in vivo*. Long-range connections, such as sensory input, still had to be approximated. As an example of external sensory input, the movement of whiskers was modelled by directly injecting current into the soma of neurons projecting from the thalamus to the cortex, mimicking the flow of information in real brains. Notably, while the BBP model is explicitly positioned to represent a non-barrel somatosensory cortex, this validation approach relied on whisker stimulation, which primarily engages the barrel cortex. Similarly, other validation efforts utilized visual-like stimuli characteristic of the visual cortex, indicating that some key physiological validation data were drawn from cortical areas or sensory modalities different from the model's specified domain. This produced activity congruent with that observed *in vivo* on the millisecond scale in model parameterisations representing awake and anesthetized animals. Additionally, by artificially hyperpolarizing selected neurons in their simulation, they could reproduce results produced by optogenetically inactivating neurons in animals.

However, the project's reliance on algorithmic inference for connectivity, a necessity given the lack of comprehensive, experimentally-derived connectomes at the time (Reimann et al., 2015), remains a significant limitation. Critics such as Frégnac have critically observed that this method relies on a 'bootstrap' logic to generate 'realistic instantiations of possible connectomes,' aiming for a brain 'realistically connected in the statistical sense' rather than one based on direct, comprehensive empirical mapping (Frégnac, 2021). Furthermore, even when model outputs, such as activity patterns, resemble experimental data, the challenge of parameter degeneracy (Marder, 2015) makes it difficult to ascertain whether the model truly captures the correct underlying biological mechanisms. Different configurations of neuronal and synaptic parameters could potentially produce similar macroscopic outputs, meaning that a match to some experimental data does not, by itself, confirm the model's biological accuracy or its generalization abilities as to predict novel neural phenomena.

**Replication from Figure 1 in Isbister et al., 2024:** Overview of the physiology and simulation workflow. 1. Anatomical model: Summary of the anatomical nbS1 model described in the companion paper. 2. Neuron physiology: Neurons were modeled as multi-compartment models with ion channel densities optimised using previously established methods and data from somatic and dendritic recordings of membrane potentials in vitro. 3. Synaptic physiology: Models of synapses were built using previously established methods and data from paired recordings in vitro. 4. Compensation for missing synapses: Excitatory synapses originating from outside nbS1 were compensated with noisy somatic conductance injection, parameterized by a novel algorithm. 5. In vivo-like activity: They calibrated an in silico activity regime compatible with in vivo spontaneous and stimulus-evoked activity. 6. In silico experimentation: Five laboratory experiments were recreated. Two were used for calibration, and three of them were extended beyond their original scope. 7. Open Source: Simulation software and a seven column subvolume of the model are available on Zenodo (see data availability statement). Data generalisations: Three data generalisation strategies were employed to obtain the required data. Left: Mouse to rat, middle: Adult to juvenile (P14) rat, right: Hindlimb (S1HL) and barrel field (S1BF) subregions to the whole nbS1. Throughout the figure, the corresponding purple icons show where these strategies were used.

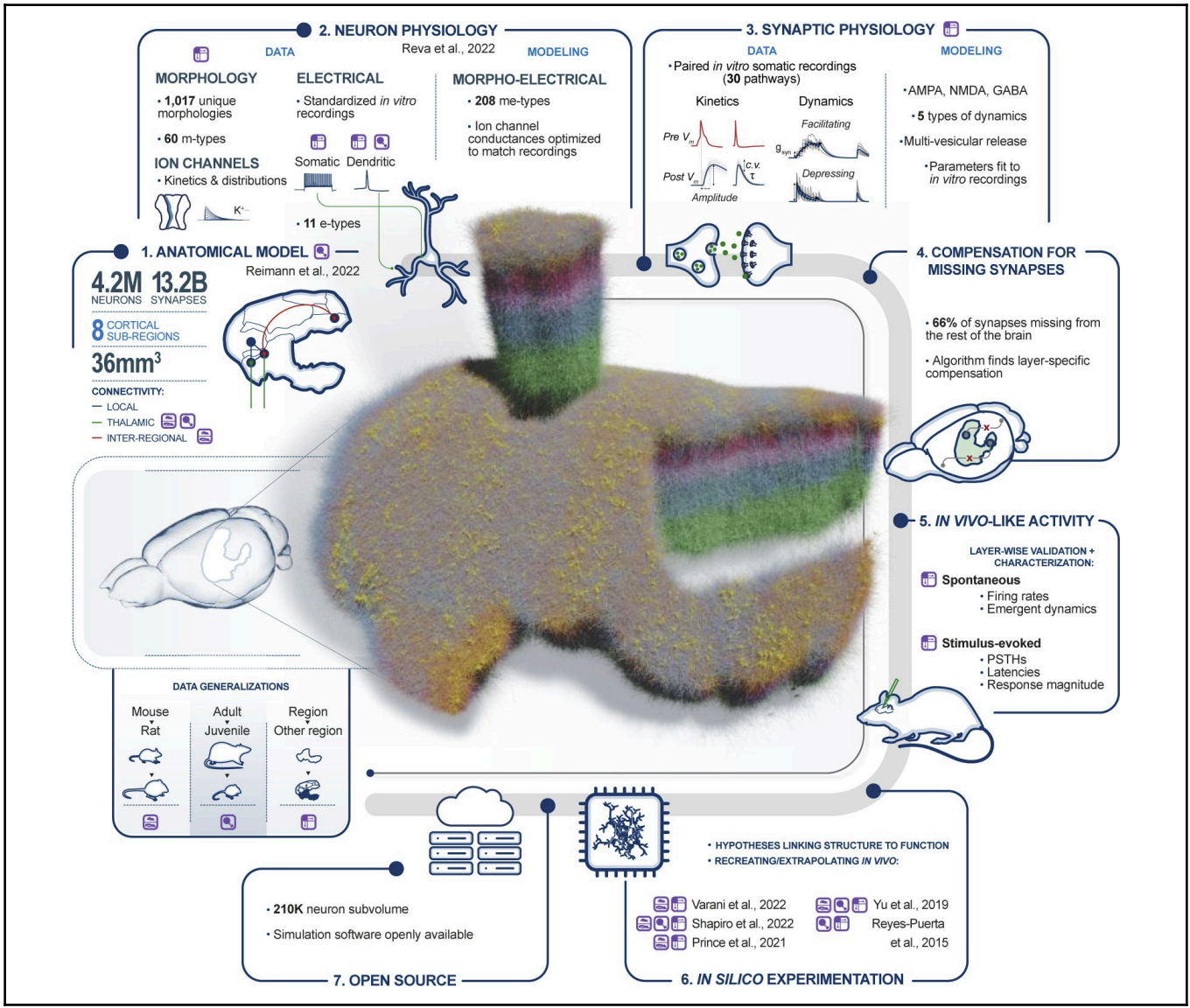

Billeh et al., 2020

While sharing the goal of large-scale, biophysically detailed cortical modeling with projects like the Blue Brain Project, the approach taken by the Allen Institute for Brain Science (Billeh et al., 2020) placed a distinct emphasis on the deep integration of concurrently acquired experimental data specific to the mouse visual cortex circuit they were modeling. By estimating cell densities and types from anatomical data, they modeled a 400μm-radius column of visual cortex containing ~52,000 neurons with high biological fidelity, surrounded by 179,000 point-neurons to avoid boundary artifacts. Connection parameters were inferred from various physiological datasets, and connections were optimised layer-wise to produce firing dynamics that matched *in vivo* Neuropixels recordings (Siegle et al., 2021).

Input was modelled via a module that mimics thalamic input to the visual cortex. This module comprised spatial-temporal filters following experimentally reported distributions, which were then connected to cortical neurons to give rise to established tunings. Connection strengths were set to match experimentally reported current strengths. This allows encoding arbitrary visual spatiotemporal stimuli (movies) via simulated thalamocortical spike trains.

The neural dynamics exhibited by the model were verified by comparing the activity of simulated neurons in response to drifting gratings with *in vivo* Neuropixels recordings, focusing on such response features as average firing rates, direction selectivity, and orientation selectivity for different neuron classes and cortical layers. Comparing these responses between the data and the model, including model versions with altered circuit architecture, resulted in several predictions regarding the organization of the visual cortical connectivity ([Billeh et al., 2020](#)), with some of the predictions confirmed by independent experimental studies ([Rossi et al., 2020](#)). A later study ([Rimehaug et al., 2023](#)) found that model responses on the level of current source density were inconsistent with results from Neuropixels recordings, and improved it by adjusting connection weights and including feedback connections.

### Wang et al., 2025

Wang et al. based their work on cortical activity rather than on explicit connectivity information: Using a dataset containing 900 minutes of two-photon recordings of calcium traces in 67000 visual cortex neurons in response to natural movies, as well as recordings of behavioural variables such as pupil position and movement speed, they trained a recurrently connected deep neural network to predict the activity of individual neurons [(Wang et al., 2025)](#). The whole network consists of a perspective network, which uses ray-tracing to infer the retinal activation of the mouse based on pupil position and stimulus, the modulation network, an LSTM network that encodes behavioural variables, a recurrent foundation core that captures abstract aspects of brain processing, and a readout network, which maps activity of the core network to individual neurons. By fixing the weights of the foundation core and only retraining encoding and decoding networks, the researchers were able to combine data from 8 mice and achieve higher accuracy than with networks trained end-to-end for each mouse.

The network quality was validated by predicting the activity of neurons in relation to novel visual stimuli and calculating the normalised cross-correlation with held-out *in vivo* recordings. Additionally, the properties of *in silico* neurons were compared to those of *in vivo* neurons, showing that they had developed the same parametric tuning properties concerning orientation and spatial position.

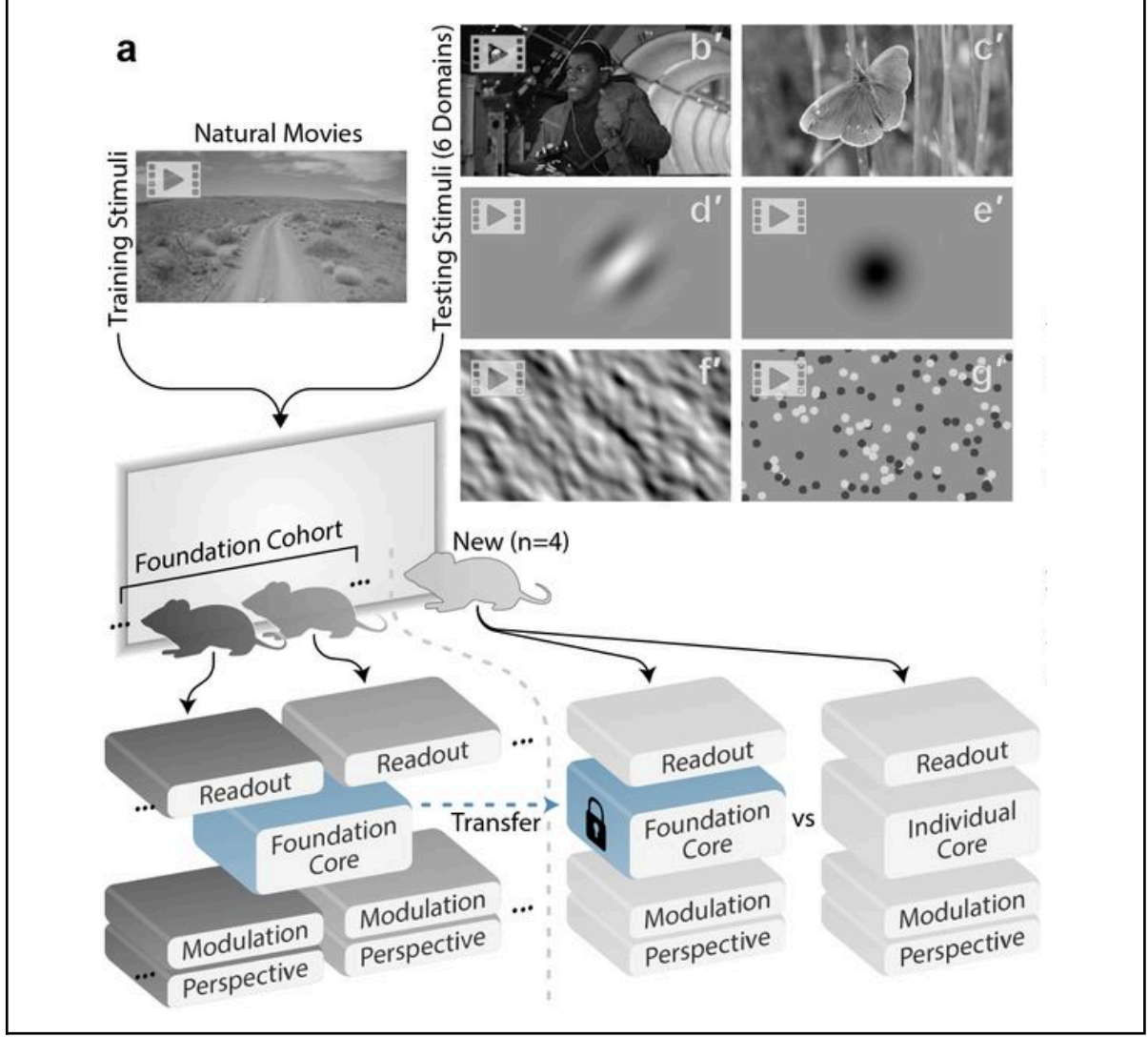

**Replication from Figure 3a from Wang et al. (2023) Predictive accuracy of foundation models. a. Schematic of the training and testing paradigm.** Natural movie data were used to train: 1) a combined model of the foundation cohort of mice with a single foundation core, and 2) foundation models vs. individual models of new mice.

# Gap Analysis

The mouse is the most commonly used model organism in medicine. Accordingly, there is vast experience with it as a model organism, and it is highly relevant to vast parts of the life sciences.

But unlike smaller organisms discussed so far, the scale of data acquisition is herculean at minimum. The mouse brain represents a fundamental transition point, in particular for brain-wide functional recording. With ~70 million neurons spread across its sizable brain, no current or easily foreseeable technology – barring, speculatively, neural dust – will enable simultaneous recording from all neurons at physiologically relevant timescales. Light scattering restricts cellular-resolution imaging to superficial layers (~1-1.5 mm depth), making subcortical and deep cortical circuits inaccessible ([Marblestone et al., 2013](#)). While novel approaches like tissue transparency agents ([Ou et al., 2024](#)) show promise, whole-brain imaging at cellular resolution remains physically infeasible due to unresolved scattering and absorption in deeper brain regions. This reality forces a pivot in approach: rather than pursuing exhaustive functional characterization, successful mouse brain emulation will require learning to predict neural dynamics from structural and molecular properties. The mouse thus serves as the critical test case for whether we can bridge the structure-to-function gap. This challenge will only become more pressing as we move toward larger mammalian brains.

| Model Organism Overview: Mouse | Pros | Cons |
|---|---|---|
| Anticipated Scientific Insights | <ul><li>**Mammalian brain:** Mammalian brains exhibit structural and functional similarities to the human brain, with thoroughly characterized homologues.</li><li>**Structure to function:** For smaller volumes, like MICrONS, complete reconstruction of structure and function for every neuron is possible, which offers wide-ranging opportunities for understanding the structure-function relationships.</li><li>**Rich social and cognitive behavior:** Verify whether simulations can achieve complex social and cognitive behavior patterns.</li><li>**Critical milestone:** achieving brain emulation in mice is likely the milestone that will trigger massive investments in human-scale emulation efforts.</li><li>**Various applications in biomedical research**: Given the mouse's role in biomedical research, in-silico simulations might replace some in vivo experiments.</li></ul> | <ul><li>**Not miniature humans**. Although mice and humans are both mammals, their brains are still reasonably different. For example, humans rely on vision much more than mice; mice would be classified as legally blind.</li></ul> |
| Experimental Tractability | <ul><li>**Strong validation with major research initiatives:** BRAINS CONNECTS is evaluating the feasibility of reconstructing a whole mouse brain. MICrONS demonstrated the collection of aligned structure and function datasets. .</li><li>**Extensive electrophysiological and morphological data exist** for [many cell types in the cortex](#) and their [synaptic connections](#).</li></ul> | <ul><li>**Very Large investments:** Comprehensive mouse programs will require investments in the 100M to billion-dollar range</li><li>**Centralization**: The scale of the mouse connectome likely necessitates moving away from a decentralized academic model towards a central, large-scale facility, which might reduce the number of actors who can contribute.</li></ul> |

| Gaps and opportunities: Mouse | Gaps (non-exhaustive selection) | Illustrative Project Opportunities |
|---|---|---|
| **Neural dynamics** | - **Harsh Tradeoffs Between Coverage, Resolution, and Noise.** Current methods force extreme compromises. Head-fixed systems achieve high coverage (~1M neurons at 6-10 Hz) but restrict natural behavior. Systems compatible with freely moving behavior enable studying naturalistic behavior, but they can only record about a thousand neurons. Voltage imaging resolves spikes at 400 Hz but is limited in coverage and recording duration ([Bai et al., 2024](#)). In general, increasing coverage degrades signal-to-noise ratio (SNR) due to shot noise ([Rupprecht, 2021](#)), with setups compatible with free behavior suffering the steepest penalties.<br>- **Limited Functional Data for Refinement and Validation.** Achieving the scale of functional data needed for comprehensive model refinement and validation remains a significant challenge, as current recording approaches capture only a fraction of the brain's neurons and attempts to increase scale face inherent trade-offs with signal quality and noise levels ([Rupprecht, 2021](#)). | - **Integrated Barcoding-Function Atlas**. Combine E11 Bio's PRISM with in vivo calcium/voltage imaging to generate a multi-modal dataset linking neural activity patterns to post-mortem structural connectivity.<br>- **Inverse Scattering Depth Benchmark**. Test the limits of deep-brain imaging by integrating Kang et al.'s inverse scattering algorithms ([Kang et al., 2024](#)) with three-photon microscopy. Quantify maximum imaging depth in the mouse brain.<br>- **Chronic Crystal Skull Endurance Study.** Systematically quantify the maximum viable recording duration using "crystal skull" glass windows in mice. Track signal quality over 6–12 months post-implantation. |
| **Connectomics** | - **EM-pipelines remain bottlenecked by proofreading requirements**. Electron microscopy (EM), the most mature imaging approach, faces prohibitive costs due to manual proofreading, which accounts for >95% of total project costs ([Jefferis et al., 2023](#)). Even with AI-assisted segmentation, proofreading a whole mouse connectome would require prohibitive amounts of human labor. Alternative approaches like E11 Bio's PRISM (genetic barcoding + ExM) aim to bypass this bottleneck but remain unproven at scale.<br>- **Molecularly Annotated Connectomes**. EM provides only structural ("naked") connectomes, lacking synaptic-level molecular information. While ExM protocols now resolve ~20 proteins in expanded tissue ([Tian et al., 2024](#)), scaling this to whole-brain volumes requires orders-of-magnitude improvements in staining throughput, antibody compatibility, and automated analysis.<br>- **Inter-Individual Variability Unquantified.** Synaptic wiring varies across mice due to experience-dependent plasticity. Current efforts focus on single specimens, but modeling learning/memory requires mapping multiple connectomes – a cost-prohibitive task even for small volumes ([Abbott et al., 2020](#)). | - **Synaptic Receptor Necessity Study**: Perform paired in vivo electrophysiology and post-mortem expansion microscopy (ExM) to map synaptic receptor distributions (AMPA, NMDA, GABA, etc.) in the same neurons. Determine the minimal set of molecular markers required to predict synaptic properties. Publish open datasets linking receptor density, synapse size, and functional measurements.<br>- **AI Proofreading Scaling Laws**: Quantify how automated proofreading accuracy (c.f, RoboEM) scales with training data volume and model size. Identify computationally optimal frontiers.<br>- **Ultra-Expansion Protocol Development**: Adapt re-PKA protocols to achieve 40–50× isotropic expansion in adult mouse brain tissue. Solve distortion challenges. |
|  | - **Exascale Data Demands.** A whole-brain EM dataset (~1 exabyte) strains storage and analysis pipelines. While tools like EM-compressor reduce raw data needs by 128× ([Li et al., 2024](#)), they do not address the computational challenges of querying or analyzing petascale connectomes.<br>- **Alternatives to EM remain underdeveloped**. X-ray ptychography has demonstrated synaptic resolution ([Bosch et al., 2023](#)) but currently lacks throughput for whole-brain imaging. ExM has demonstrated molecular profiling but struggles with isotropic expansion and tissue distortion at scale. Correlative workflows (e.g., XRM-to-EM) remain experimental and labor-intensive ([Jefferis et al., 2023](#)). |  |
| **Computational Neuroscience** | - **Models remain circuit- or region-specific**. Scaling simulations to the whole mouse brain poses substantial computational challenges. Further, the field is bottlenecked by the absence of a whole-brain connectome able to constrain model architecture and parameters ([Igarashi, 2024](#)).<br>- **Structure-to-Function Translation Challenge**. The mouse represents the first organism where inferring function from structure becomes essential for brain simulation. Even with a complete connectome, translating structural connectivity into functional circuit dynamics poses fundamental challenges that will require the development of novel approaches ([Abbott et al., 2020](#)).<br>- **Lack of Standardized Benchmarks.** While theoretical evaluation frameworks like the embodied Turing test have been proposed ([Zador et al., 2023](#)), and initial practical benchmarks for specific sensory systems are emerging ([Turishscheva et al., 2023](#)), comprehensive practical implementation remains challenging. Even if whole-brain mouse simulations were achieved, comparing results across different modeling approaches would remain difficult without clear quantitative benchmarks, creating a barrier for systematic progress. | - **Virtual Reality Benchmarking Suite**: Develop standardized VR environments to test mouse neuromechanical models (e.g., DeepMind's biomechanical simulators) in naturalistic tasks like foraging, social interaction, and predator evasion.<br>- **Whole-Brain Simulation Performance Comparison**: Systematically analyze the computational costs of simulating mouse-scale networks (70M neurons) across frameworks (NEST, NEURON, ARBOR, Jaxley, BrainPy, NeuronGPU) at varying biophysical resolutions (LIF vs. HH models, different models of synaptic transmission, etc.).<br>- **Neuromechanical Integration Standards**: Develop open APIs and data formats to unify neural simulators with biomechanical engines (MuJoCo, PyBullet). Enable bidirectional sensory-motor integration for embodied tasks like locomotion. |

# Humans

## Anatomy & Behavior

The human brain reaches its peak volume in the third to fourth decade of life, typically measuring around 1000-1400 cm³ and weighing approximately 1.3-1.4 kg, or roughly 2% of total body weight ([Steen et al., 2007](), [Raichle et al., 2002]()). The adult brain measures approximately 140 mm in width, 167 mm in length, and 93 mm in height, though considerable individual variation exists. The brain contains an estimated 86 billion neurons, with distinct regional distributions: approximately 16 billion neurons in the cerebral cortex, 69 billion in the cerebellum (which comprises only 10% of brain volume), and less than 1 billion distributed throughout other brain regions ([Azevedo et al., 2009]()). While estimates for total brain synapses range between 100-1000 trillion, the most precise ones exist for the neocortex, containing an estimated 164 trillion synapses ±17% ([Tang et al., 2001]()). Typical baseline firing rates observed in human cortical and hippocampal recordings (often derived from clinical patients) frequently average 0.5-4 Hz ([Aghajan et al., 2023]()), with firing rates increasing in response to salient stimuli or during tasks, potentially reaching averages of 10-20 Hz or higher in some cortical neurons.

The breadth of human behavior patterns is not treated here, as we take reader familiarity as given.

## Neural Dynamics

Ethical considerations and the extensive genetic toolkit requirements make optical methods for single-neuron resolution recording currently non-viable in humans. Consequently, invasive electrophysiological techniques represent the only modality capable of achieving such resolution in the human brain. Typically employed within clinical research (e.g., as part of brain-computer interface trials) or during neurosurgical procedures, these methods provide valuable albeit highly localized data on individual neuron activity.

Pioneering efforts in chronic human intracortical recording have heavily relied on Utah Electrode Arrays (UEAs). These devices, often comprising a 96-channel grid of stiff silicon electrodes, are FDA-cleared for investigational BCI studies ([Sponheim et al., 2021]()). UEAs have demonstrated impressive longevity in human participants, with successful recordings maintained for multiple years. While UEAs have enabled significant BCI achievement, they sample neurons primarily within a 2D cortical plane and typically yield a modest number of separable single units (often well below 150) (Chung et al., 2022; Sponheim et al., 2021).

High-density silicon probes, such as Neuropixels, have recently allowed for a significant increase in the scale and resolution of acute single-neuron recordings in humans during intraoperative settings (Chung et al., 2022). For example, a recent study used Neuropixels probes (10 mm shank; 384 selectable channels out of 960 contacts/recording sites) in 8 neurosurgical participants, isolating 596 neurons in total and up to ~100 neurons simultaneously from a single insertion, with recordings typically lasting 10–20 minutes (Chung et al., 2022). Another study employing the thicker Neuropixels 1.0-ST similarly obtained hundreds of spike-sorting clusters in three participants, with well-isolated neurons identified from those clusters after curation (Paulk et al., 2022). An important challenge in these open craniotomy settings is the substantial brain motion due to cardiac and respiratory pulsations, which is negatively correlated with unit yield and necessitates sophisticated motion correction algorithms.

Further applications of Neuropixels have provided detailed accounts of single-neuron activity during human language processing. Recordings from 685 neurons across cortical layers in the superior temporal gyrus (STG) of 8 participants listening to speech (Leonard et al., 2024) indicated that individual neurons encoded various speech sounds (e.g., consonants, vowels, pitch), with neurons at different cortical depths tuned to different speech features. In a study using Neuropixels in the language-dominant prefrontal cortex of 5 participants (Khanna et al., 2024), activity from 272 neurons, even before speech, outlined the structure of upcoming words. This information about the word's structure was encoded in a specific, timed order. Similarly, Neuropixels recordings from 3 participants (Jamali et al., 2024) identified single neurons in the prefrontal cortex that selectively represented word meanings, with activity also reflecting sentence context and semantic relationships. Such studies demonstrate the use of high-density probes to examine how individual human neurons process complex language elements, revealing specific encoding patterns and their organization.

Efforts towards next-generation, fully implantable, high-channel-count BCIs include Neuralink's N1 implant. This system uses 1,024 electrodes distributed across 64 flexible, robotically inserted "threads," designed for wireless data transmission and inductive charging (Neuralink, 2024). As part of their PRIME clinical trial, three participants with quadriplegia have received the N1 implant. These participants have had their implants for over 670 days and used the BCI system for over 4,900 hours, with recent independent daily use averaging 6.5 hours (Neuralink, 2025). Initial BCI performance for cursor control has been reported at up to 8.0 bits-per-second (BPS). Wireless implants like Neuralink's N1 allow neural recording during more naturalistic daily activities than traditional wired systems. Challenges reported include the retraction of some electrode threads post-surgery in one participant, which required algorithmic adjustments to maintain performance. While still in early clinical stages, these systems aim to substantially increase the scale and practicality of human chronic neural recording.

While the aforementioned electrophysiological methods provide increasingly detailed recordings of neural activity, the ability to perturb specific human neurons at single-cell resolution in vivo to map function is severely limited. Optogenetic perturbation, a powerful tool in model organisms, is not currently viable for targeted modulation within the living human brain due to ethical barriers, challenges in precise gene delivery for opsin expression, and difficulty delivering light safely to deep brain structures.

Patch-clamp electrophysiology, however, serves as a robust method for perturbative studies on *ex vivo* human brain slices. Tissue resected during neurosurgery can be kept viable, allowing researchers to perform whole-cell patch-clamping. This enables precise current or voltage injections to characterize neuronal firing properties, pharmacological manipulations to block specific ion channels or activate receptors, and paired recordings with synaptic stimulation to investigate microcircuit connectivity ([Menéndez de la Prida et al., 2002](); [Peng et al., 2019]()). Nevertheless, performing true whole-cell patch-clamp inside the intact, living human brain remains technically and ethically unfeasible. Other in-vivo human perturbation techniques, such as Transcranial Magnetic Stimulation (TMS) or Deep Brain Stimulation (DBS), affect larger neuronal populations and lack single-cell specificity.

# Connectomics

Despite the considerable technical challenges of imaging such large volumes of nervous tissue at synaptic resolution, electron microscopy efforts in primate and human brains have achieved several notable milestones. Wildenberg et al. imaged a 0.8 mm × 2.4 mm × 40 nm block of rhesus macaque cortex, revealing that primate neurons receive 2-5 times fewer synaptic inputs than their mouse counterparts ([Wildenberg et al., 2020](#)). Loomba et al. conducted comparative connectomic analysis across species by imaging multiple cortical samples, including ~175 × 220 × 100 μm³ volumes of layer 2/3 from macaque and human, and a larger 1.7 × 2.1 × 0.03 mm³ volume spanning all cortical layers in human temporal cortex ([Loomba et al., 2022](#)). A landmark achievement came in 2024 when Shapson-Coe et al. demonstrated high-throughput serial section electron microscopy of a 170-μm-thick slab of human temporal cortex at nanoscale resolution. This volume, obtained during epilepsy surgery, spans approximately 1.05 mm³ of tissue (accounting for sectioning-induced compression) and was imaged at a resolution sufficient to resolve individual synapses and subcellular structures. Thus, it represents the most significant volume of human brain tissue imaged at a resolution sufficient for dense connectomics to date ([Shapson-Coe et al., 2024](#)).

The most significant X-ray imaging effort for human connectomics is the SYNAPSE (Synchrotron for Neuroscience – an Asia-Pacific Strategic Enterprise) collaboration, which aims to map an entire human brain at 0.3 μm resolution ([Stampfl et al., 2023](#)) in 2-3 mm-thick sections ([Chen et al., 2021](#)). The collaboration has made remarkable technical progress, achieving imaging speeds of 1 mm³ per minute and coordinating roughly 10 synchrotron facilities, including the first beamline entirely dedicated to connectomics at the Taiwan Photon Source ([Chen et al., 2021](#)). Their next phase, SYNAPSE 2.0, aims to achieve another 10-100x speed increase to 0.2 mm³/second ([Stampfl et al., 2025](#)). Despite this, it is important to note that the project's target resolution is designed for mapping cellular distributions and long-range projections, not for resolving individual synapses, and the SYNAPSE roadmap does not currently include plans to reach synaptic-level detail. However, separate theoretical proposals suggest that a similar high-throughput synchrotron pipeline could potentially reach synaptic resolution by integrating expansion microscopy (ExxRM), thus bridging the current resolution gap ([Collins, 2023](#)).

The reconstruction of primate connectomes at synaptic resolution remains a massive undertaking. Storage requirements alone would be significant: assuming two bytes per voxel, a typical marmoset brain would require 11.6 exabytes at 10 nm isotropic resolution (0.74 exabytes at 25 nm); a typical macaque brain 208 exabytes (13.3 exabytes at 25nm); and a human brain roughly 2-2.8 zettabytes

(128-180 exabytes at 25nm). The volume imaged by Shapson-Coe et al., representing an extremely minute part of a whole human brain, or approximately 0.00007% of a whole human brain, produced over 1.4 petabytes of data ([Shapson-Coe et al., 2024](#)), motivating the development of dedicated infrastructure ([Maitin-Shepard and Leavitt, 2022](#)). Given the scale of this challenge, some whole-brain mapping efforts opt for lower-resolution imaging approaches. Even so, storage requirements remain substantial - the SYNAPSE collaboration, which aims to map the human brain at 0.3 μm isotropic resolution using X-ray imaging, estimates they will need exabyte-level storage for a single brain dataset and approximately 100 exabytes for their complete connectome mapping goals. Reconstruction would also represent a daunting challenge. The H01 dataset released by Shapson-Coe et al., even with thousands of Google-developed Tensor Processing Units (TPUs) for automated neuron segmentation and synapse detection methods ([Blakely and Januszewski, 2021](#)), still required extensive manual proofreading. While the computational pipeline successfully identified about 16,000 cells and 150 million synapses, ensuring accuracy still demanded intensive human effort, with expert reviewers spending over 3.5 hours per neuron to correct remaining errors. Proofreading efforts continue to this day, with the original release including only 104 proofread cells.

## Computational Modeling

Unlike organisms like *Drosophila* or larval zebrafish, computational neuroscience of the human brain operates under severe data constraints. There is no synapse-level connectome, and functional recordings either cover only hundreds of cells or provide only indirect measures of neural activity at low spatial and temporal resolution (in the case of non-invasive methods like fMRI). This fundamental limitation has shaped a landscape dominated by feasibility studies - efforts focused primarily on demonstrating the possibility of simulating human brain-scale networks on current supercomputing infrastructure, rather than attempting to replicate specific circuits or behaviors with high biological fidelity.

Yamazaki et al., 2021

Yamazaki and colleagues developed one of the first human-scale simulations of the cerebellum using their MONET simulator on Japan's K supercomputer, capable of 11.3 PFLOPS ([Yamaura et al., 2021](#)). Their model leveraged anatomical studies providing cerebellar layer thicknesses and cell density measurements across different regions. This structural data informed the spatial organization and the number of neurons of each type in their model.

They constructed a network of 68 billion neurons and 5.4 trillion synapses from this anatomical data. Neurons were modeled as conductance-based leaky integrate-and-fire units with α-function synapses implementing four receptor types (AMPA, NMDA, GABAA, GABAB). Connection patterns between neurons were defined using two-dimensional Gaussian distributions, with parameters like connection probabilities and spatial extents derived from anatomical studies.

Model validation occurred in two stages. First, they examined resting state activity, comparing baseline firing rates across different cell types with experimental data. Second, they tested the model's ability to reproduce the optokinetic response (OKR), a reflexive eye movement controlled by the cerebellum. For OKR, they found Purkinje cells modulated their firing rates between 50-80 Hz, within the range observed across various animal studies (20-100 Hz). Running on 82,944 nodes of the K computer, the simulation achieved speeds 578 times slower than real-time.

Lu et al., 2023

Lu and colleagues developed the "Digital Twin Brain" (DTB) platform, implementing a whole-brain simulation of 86 billion neurons and 47.8 trillion synapses on a GPU-based supercomputing cluster ([Lu et al., 2024](#)). Their model's structure was informed by three types of macroscopic brain imaging data: structural MRI provided regional neuron densities at 3x3x3mm³ voxel resolution, diffusion tensor imaging (DTI) determined voxel-to-voxel connection probabilities, and PET data established what proportion of connections remained local vs projecting to other regions (varying from ~29% external connections in cortex to ~13% in cerebellum). Of all possible connections between voxels, only 0.72% were realized, reflecting the brain's sparse connectivity.

They developed a "hierarchical mesoscopic data assimilation" (HMDA) approach to tuning model parameters. This involved first training a smaller network (0.2 billion neurons) to estimate hyperparameters for synaptic conductances and then using these to initialize the full model. Parameters were iteratively refined by comparing simulated BOLD signals to real fMRI data during a visual evaluation task. Each neuron was modeled as a conductance-based LIF unit with four receptor types (AMPA, NMDA, GABAa, GABAb). Average input synapses per neuron were set to 1000 for cortical/subcortical regions and 100 for brainstem/cerebellum. Long-range connections were restricted to excitatory neurons only. To manage computational demands on their 14,012 GPU cluster (each node contained four GPUs with 16GB of memory each), they developed a partitioning algorithm that balanced computational load and minimized inter-GPU communication. The fitted model achieved correlations above 0.98 with experimental BOLD signals in input regions (with a 2-timepoint lag) and 0.75 across the cortex. When trained to predict participants' emotional image ratings, the model significantly correlated with actual ratings (r=0.655, p=0.006). Running at different firing rates (7Hz, 15Hz, and 30Hz), the simulation achieved speeds between 65-118.8 times slower than real-time.

## Gap Analysis

The human brain sits in a league of its own, being more than three orders of magnitude larger than the mouse brain. This extraordinary scale creates a fundamental divide in feasibility: while whole-brain recording at cellular resolution is achievable in zebrafish and likely feasible in *Drosophila*, and whole-cortex recording remains at least imaginable for the mouse, comprehensive neural recording in humans faces insurmountable physical barriers with current or near-future technologies, even setting aside regulatory and ethical challenges. Most models are based either on low-resolution functional data (fMRI) or anatomical observations, requiring significant extrapolation from these sparse constraints. Unlike more approachable organisms where neural recordings can constrain brain simulations, human-scale models will likely need to find different approaches, potentially relying on infering parameters primarily from structural datasets. Given current technology, human brain connectome efforts would require industrial-scale operations that dwarf any existing neuroscience facility. Facilities the size of modern semiconductor manufacturing plants: thousands of electron or optical microscopes running in parallel, or entire synchrotron facilities with multiple beamlines dedicated to brain mapping. Such facilities would demand large-scale storage and computing infrastructure to process the multi-exabyte datasets generated daily.

| Model Organism Overview: Human | Pros | Cons |
|---|---|---|
| **Anticipated Scientific Insights** | <ul><li>**Fundamental insights into Neuroscience:** It seems plausible that advanced brain models will transform our understanding of how the brain works and how consciousness and personality traits arise.</li><li>**Applications for human health**</li><li>**Feasibility of personality-preserving brain emulations**</li></ul> | <ul><li>**Miscellaneous ethical consideratioins:** Potential risks to individuals or even whole society have to be carefully evaluated.</li></ul> |
| **Experimental Tractability** | <ul><li>**Computational models approaching human scale:** Recent modelling on high-performance computing datacenters is getting close to managing the loads necessary to model whole human brains. Additionally, progress towards more efficient neuromorphic hardware is ongoing.</li><li>**First connectome reconstructions of human tissue:** First reconstructions of human brain tissue can rely on the same methods as smaller organisms.</li></ul> | <ul><li>**Methodological limitations in humans:** Many technologies require genetic engineering or highly invasive brain surgery.</li><li>**Resolution vs Coverage Trade-off.** Non-invasive methods provide whole-brain coverage but at the cost of spatial and temporal resolution. fMRI measures slow hemodynamic responses averaged across hundreds of thousands of neurons, while EEG and MEG provide faster temporal sampling but even coarser spatial resolution. Conversely, invasive recordings can resolve individual neurons but are restricted to tiny fractions of brain tissue, capturing thousands of neurons at best compared to the brain's ~86 billion. This trade-off between resolution and coverage appears fundamental rather than technological, making comprehensive recordings with full human brain coverage at single-neuron resolution permanently out of reach.</li><li>**Scale of human brain:** At 1000x the scale relative to the mouse brain, the operational challenges mentioned in the section about mouse are magnified substantially. For instance, achieving synaptic resolution in a volume of ~1,200 cm³ (vs. ~0.5 cm³ for the mouse) demands capacity and throughput far beyond the capabilities of current laboratories. Meeting this challenge will require near-industrial levels of coordination, equipment scale-up, and data infrastructure, elevating human connectomics from a traditional scientific project to a massive engineering and logistics undertaking.</li><li>**Likely strong interindividual and intercultural differences:** Given human diversity differences in brain connectivity are expected.</li></ul> |

| Gaps and opportunities: Human | Gaps (non-exhaustive selection) | Illustrative Project Opportunities |
|---|---|---|
| **Neural dynamics** | - **Structure-to-Function Dependency.** A human connectome's ultimate value for brain emulation depends critically on bridging the structure-to-function gap. Even with molecular annotations, a fully reconstructed wiring diagram is only the first step. | - **Extensive "Human in a dish recording":** Instead of gathering data in vivo, human neuronal cell lines could be used to collect large-scale data sets and facilitate modelling. |
| **Connectomics** | - **Electron Microscopy.** At current effective scan speeds, achieving a complete human EM dataset within a decade would require approximately 30,000 parallel electron microscopes (Jefferis et al., 2023). Beyond microscope availability, sample preparation, automated image segmentation, and proofreading each introduce substantial computational and manual labor demands, though recent advances in machine learning have improved automation capabilities (Januszewski et al., 2020; Schmidt et al., 2021).<br>- **More scalable methodologies: Expansion Microscopy (ExM) and X-ray Microscopy (XRM):** ExM has never been attempted at anything close to the scale of a human brain. Tissue anisotropies, distortion, and batch-to-batch consistency remain unresolved, particularly at the extremely high expansion factors needed to achieve synapse-level resolution in large volumes. XRM promises high-speed volumetric imaging of thick specimens, thanks to high-brightness synchrotron beams, and is not constrained by the diffraction limit. However, XRM's practical viability for a complete human connectome remains untested, with significant challenges in achieving dense reconstructions at synapse-level resolution.<br>- **Lack of baselines:** In humans we have a surprisingly poor understanding of variables such as number of synapses per neuron in different areas of the brain or the ratio of local and distant connections for neurons. | - **High-throughput EMs:** Substantial optimization of beamlines, automated sample handling, etc., could increase effective data acquisition speeds by an order of magnitude, potentially more.<br>- **Prototypes of advanced imaging modalities:** Establishing protocols and scalability tests for methodologies like ExM and XRM.<br>- **Small connectomics studies across many areas of the brain:** Sample tiny areas from many areas of the brain from multiple individuals to determine variables for estimating the total computational demands. |
| **Computational Neuroscience** | - **Data Scarcity**: The most fundamental limitation is the lack of adequate functional and structural data to constrain human brain models. While future technologies may eventually provide detailed structural data through connectomics, functional data at cellular resolution will likely remain permanently out of reach due to physical and ethical constraints. This forces models to rely on massive extrapolations from animal studies or indirect measurements, severely limiting their biological validity.<br>- **Extreme Resource Demands**: Even with optimal implementation, biophysically detailed models at accurate human scale (86B neurons with realistic synaptic connectivity) would require exascale computing systems for | - **Human-scale neuromorphic computing experiments:** Stress testing the currently existing human-scale neuromorphic computing systems and identifying and iterating on their strengths and weaknesses. |

real-time simulation. The "memory wall" and interconnect bandwidth limitations pose particular challenges for efficiently simulating such massive, densely connected networks.
- **Model Validation Challenges**: A key issue is the limited direct data on human brain activity. In animal models (e.g., mice), scientists can gather extensive and controlled neural recordings. However, data from human brains is significantly less comprehensive, and researchers have less control over recording locations. Further, relying solely on behavioral validation is insufficient to confirm that the emulation truly replicates the brain's underlying neural mechanisms.

# Methods for Brain Emulation

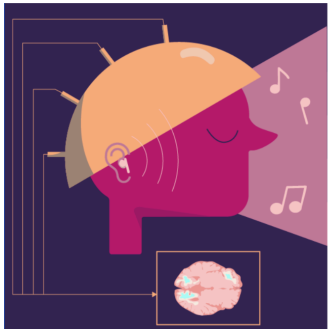
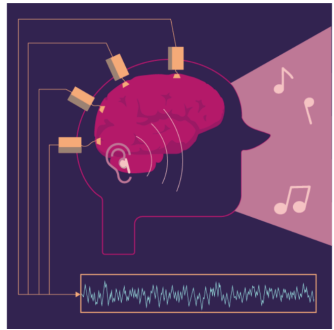
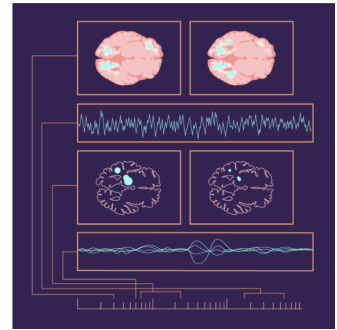
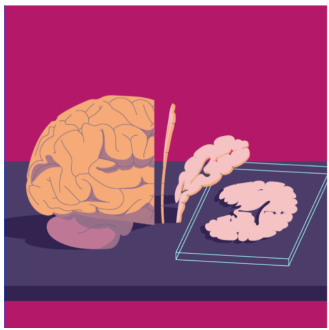
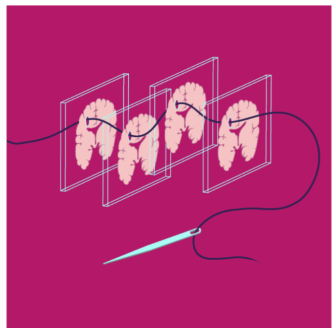
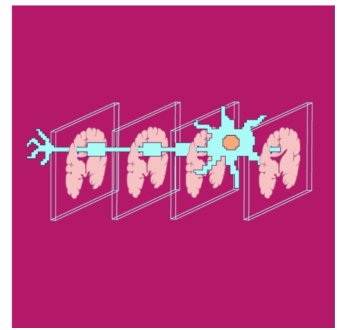

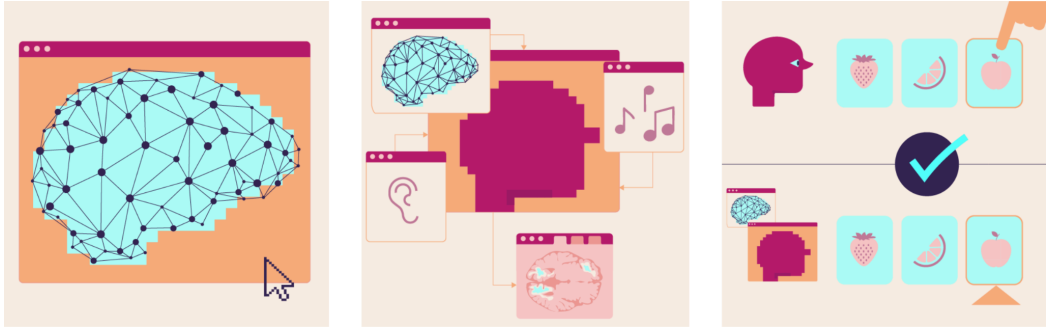

# Neural Dynamics: Brain Function & Activity

In the 1920s, English physiologist Edgar Adrian wondered if it might be possible to record the electrical activity of a single neuron as it fired, an idea that seemed almost impossible given the technology of the time. Most scientists believed neural signals were too fast and too small to measure. Using a primitive amplifier and an electrode thinner than a human hair, Adrian managed to record the electrical pulses from a single sensory nerve fiber. His 1928 paper *The Basis of Sensation* ([Adrian, 1928](#)) revealed for the first time that neurons communicate through discrete electrical impulses, or action potentials, and that the frequency of these impulses encodes the intensity of the stimulus. This discovery earned Adrian the Nobel Prize ([Adrian, 1932](#)).

Close to a century and many Nobel Prizes later, we now appreciate that neuronal activity is primarily shaped by fast synaptic inputs through chemical synapses and electrical coupling via gap junctions. These signals are modulated by slower-acting neuromodulators, hormones, and, to some extent, by surrounding glial cells. The structural features of neurons, particularly their dendritic/synaptic architecture, membrane properties, and ion channels, influence how these signals are integrated. Activity-dependent plasticity mechanisms allow neurons to adjust their properties based on experience. Note that the brain operates over time scales covering at least twelve orders of magnitude; electrical activity happens at the millisecond timescale, whereas the brain changes physically over decades.

Many modalities have been developed over the past century to measure neuronal activity. They differ substantially, by multiple orders of magnitude, in their temporal (sampling rate and recording duration) and spatial (neuron resolution and brain coverage) properties. The following figure illustrates this by comparing five different modalities across those axes. The figure highlights the benefits and drawbacks of each recording modalities, as well as the amount of information captured by the respective method. The ideal neuroscientific recording achieves single-neuron, single spike resolution across the whole brain and permits chronic recording.

> **Comparison of different recording modalities:**
> Comparison across four dimensions. Resolution as number of individual cells recorded. Speed as temporal resolution in frames per second. Maximum recording duration per session and total volume recorded. An ideal recording method for whole-brain human recording would rank at the top of each bar.. Calculations are in the [data repository](data repository).

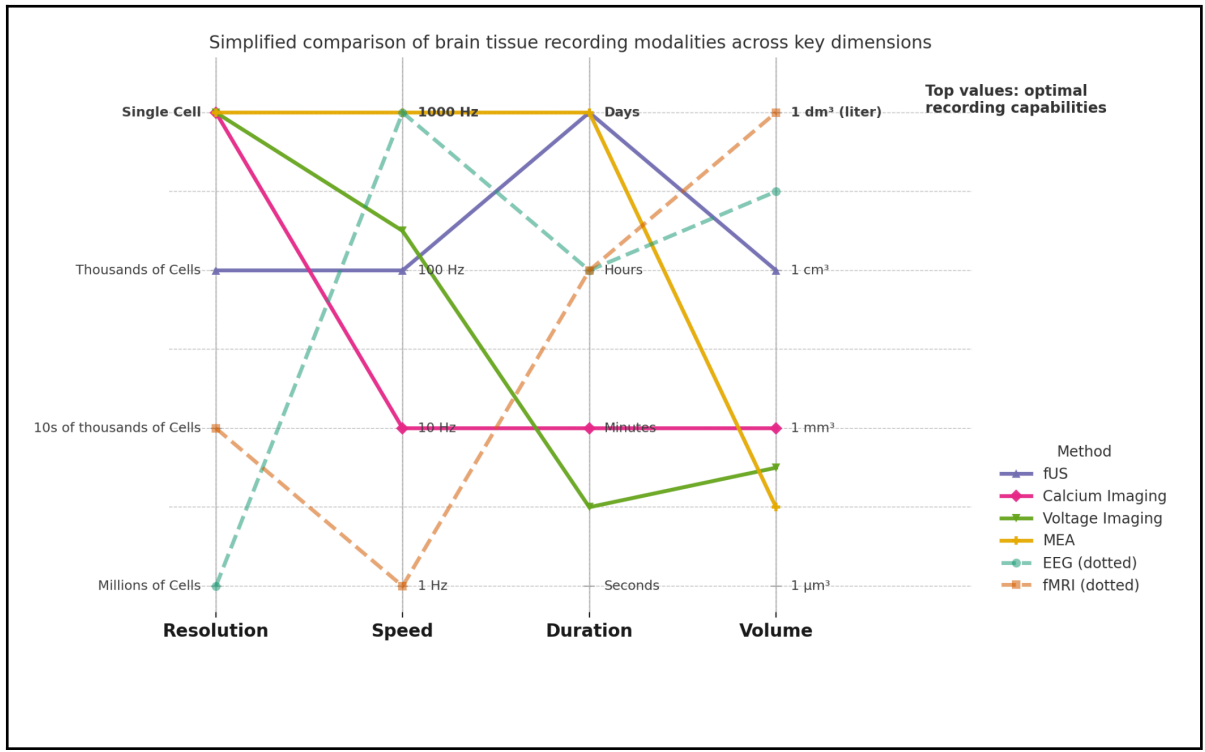

Generally speaking, non-invasive recording methods such as fMRI, EEG, fNIRS, or MEG are incapable of single-neuron resolution. While non-invasive methods tend to have a comparably high penetration depth, i.e., recording across deeper areas of the brain (which matters in particular for large brains) and can record over hours, they lack spatial resolution and only provide aggregate (indirect) information of the neuronal activity. Invasive methods, however, can provide spatial resolution of individual neurons and extremely high sampling rates, i.e., recordings at the millisecond level. Since the first single-cell recordings in the 1950s, these methods have evolved into sophisticated patch clamping, high-density microelectrode arrays (MEAs), and calcium or voltage imaging systems that measure electrical activity of individual neurons. The number of neurons being simultaneously recorded – with electrophysiological methods like MEA and patch clamping or imaging modalities such as calcium imaging – has roughly doubled every 5 years (Stevenson & Kording, 2011; Urai et al., 2022; Mineault et al., 2024), but has rapidly accelerated with the rise of optical techniques. Parallel recording of individual neurons over time is now possible for up to one million cells simultaneously, though at low temporal resolutions (sampling rate of about 1 Hz). Invasive methods not only require surgery with direct access to the brain, but also often genetic engineering to express reporters of neural activity. Electrophysiological arrays are often limited to the more superficial parts of the brain.

Another dimension in which recording modalities differ is their ability to capture the full scope of an organism's behavioral repertoire. Most modalities work exclusively or best in non-moving or fixated organisms, substantially limiting the range of behaviors. Many non-invasive methods are prone to movement artifacts and are almost unusable beyond resting or head-fixed organisms. Some methods, for example, calcium imaging, can work in both head-fixed and freely behaving animals; however, head-fixing animals typically reduces motion artifacts and permits higher spatial and temporal resolution. Miniaturization of microscopes permits calcium imaging in behaving mice, however, the full behavior repertoire is still substantially limited, given the equipment attached to the organism.

While acknowledging the importance and briefly characterizing non-invasive recording modalities as well as electrophysiology studies in the following box, the method section as well as organism-specific reviews focus on modalities that – at least in theory – allow for whole brain recordings with single neuron resolution: Calcium imaging, voltage imaging and the nascent field of ultrasound currently in development.

While electric coupling via neurotransmitters is the primary way of communication for neurons, neuropeptides modulate neuronal activity and regulate long-term physiological processes, such as structural changes. The scale of signaling peptides is far from millions or billions of neurons; hundreds of neuropeptides could interact with their environment in different permutations. This adds an entirely new dimension to what needs to be captured to understand information processing in the brain. The totality of neuropeptides interacting with each other is sometimes called "chemical / peptidergic connectome" on top of the electric connectome, which is highly conserved across evolution. ([Jekley & Yuste 2024](#)). Unlike classical neurotransmitters, neuropeptides lack specific clearance mechanisms, allowing for sustained signaling that matches behaviorally-relevant timescales ([Guillaumin and Burdakov, 2021](#)).

At this point, it is not well understood what the relative contributions of neuropeptides to electrical activity and the overall information processing of the brain are. In *C. elegans*, for instance, where neuropeptide effects have been studied extensively ([Bhat et al., 2021](#)), a long list of behavioral defects can be demonstrated in mutants without the respective neuropeptide. A large variety of behaviors are regulated by neuropeptides in *Drosophila*,, including feeding, aggression, and sleep ([Nassel and Zandawala, 2022](#)). One prominent example for humans is oxytocin, which has been convincingly demonstrated to influence complex bonding behaviors, e.g., in interactions of mothers and their newborn children ([Scatliffe et al., 2019](#)).

Lastly, combining recording modalities with methods that allow a deliberate and controlled manipulation or alteration of a specific part of a neural circuit (perturbation) is of exceptional value to understanding brain dynamics. By selectively activating or inhibiting specific neurons, researchers can directly test hypotheses and collect less correlated datasets for later computational modelling. Early studies focused on manipulating broad neural populations; modern approaches aim for increasingly precise control, from targeting anatomically defined regions to manipulating individual neurons.

In the following chapter, we survey major neural recording modalities for electrical activity and neuromodulators, including both invasive and non-invasive approaches, then cover perturbation methods, provide representative cost ranges, and discuss data management, standardization, and analysis.

# Electroencephalography (EEG)

When performed on humans, EEG records the brain's electrical activity by placing electrodes on the scalp to detect voltage fluctuations produced by large populations of neurons, predominantly cortical pyramidal cells. The scalp-recorded signals in EEG are blurred by the skull and scalp, which limits spatial precision to roughly one centimeter or more. This resolution aggregates the activity of millions of neurons. EEG also predominantly captures signals from superficial cortical layers, as signals from deeper or subcortical structures attenuate significantly before reaching the scalp. Researchers can enhance spatial precision somewhat using high-density electrode arrays (64, 128, or more electrodes) and computational source localization methods. However, the reliance on scalp measurements makes it challenging to pinpoint activity at the resolution of individual cortical layers. Despite the poor spatial resolution, these signals allow researchers to sample data at rates of 250 Hz to over 1 kHz, i.e., well above the neuronal firing rates. In clinical or research settings, EEG can be collected continuously for hours or even days, enabling extended monitoring of conditions such as epilepsy.

Because EEG relies on electrodes attached to the scalp, participants can be seated upright or lying down, and some mobile systems allow moderate movement and study of real-world tasks. Despite this relative flexibility, excessive motion degrades data quality, and researchers must constrain movement. The most common disruptions come from eye movements and blinks (which can be 10-100 times larger than brain signals), muscle activity from jaw clenching or forehead movements, and electrical interference from nearby devices or power lines (showing up as a 50/60 Hz hum in the data). EEG systems are highly accessible, likely with tens of thousands of devices worldwide across research labs, hospitals, and increasingly, consumer applications. A research-grade system typically costs between $20,000 and $200,000 ([Ledwidge et al., 2018](#)). Operating costs are relatively modest, mainly involving electrode gel, cap maintenance, and technician time, typically ranging from $50-200 per hour and

relatively modest data sizes compared to imaging modalities, though high-density systems recording continuously can generate substantial datasets.

# Functional Magnetic Resonance Imaging (fMRI)

fMRIs use powerful magnets and radio waves to measure changes in blood oxygen levels (the BOLD signal) throughout the brain. This technique identifies regions with heightened neuronal activity by detecting differences in oxygen-rich and oxygen-poor blood: as neurons become more active, they consume more oxygen, and the subsequent increased blood flow to replenish this oxygen is what fMRI visualizes to map brain function. A key strength is non-invasive, whole-brain imaging. While typical fMRI studies often use voxel sizes on the order of a few millimeters, high-resolution protocols can achieve 1 mm³ whole-brain coverage ([de Martino et al., 2011](); [Heidemann et al., 2012]()), though this remains relatively uncommon. At 1 mm³, fMRI aggregates signals from tens of thousands of neurons, with the BOLD signal's spatial point-spread being a limiting factor at this scale ([Marblestone et al., 2013]()). State-of-the-art scanners push spatial resolution further, with custom 7T systems achieving 0.56 mm isotropic whole-brain BOLD imaging ([Feinberg et al., 2023]()), and the 11.7T "Iseult" scanner reaching sub-0.5 mm "mesoscale resolution" for functional imaging ([Boulant et al., 2024]()). This approaches the spatial (though not temporal) resolution of functional Ultrasound Imaging (fUSI) at ~0.1 mm³. Even at these advanced fMRI resolutions, BOLD signals represent hundreds to thousands of aggregated neurons; single-neuron MRI remains far away without new contrast mechanisms ([Marblestone et al., 2013]()). Temporally, the BOLD signal's intrinsic ~10-second rise and fall limits the effective sampling rate to ~0.1 Hz, despite scanners technically acquiring images every 1-3 seconds. As a consequence, to put this temporal resolution in context, a single brain image can be influenced by over 20 spoken words ([Tang et al., 2023]()).

Human fMRI across data repositories like OpenNeuro, DANDI, and Brain Image Library is typically on the order of 10-20h max with sessions of 1h each at ~1 mm³ resolution, with the most "intensive" datasets reaching 200 hours per subject ([Kupers et al., 2024](), [Boyle et al., 2020]()). Different groups generated multimodal datasets, such as fMRI and EEG ([Mayhew et al., 2013](); [Pisauro et al., 2017]()), or fMRI-fNIRS-EEG ([Scarapicchia et al., 2017]()). The key to good fMRI pictures is compliant participants. Motion correction and headcases compensate for subtle head movements during scanning, as even millimeter-scale shifts can contaminate the signal. Accordingly, fMRI is also limited in that it is restricted to analysis of supine, generally non-moving participants, meaning that movement and the complete behavior repertoire of organisms are challenging to study. Currently, thousands of these devices are used globally for research purposes. A state-of-the-art 7T machine can cost over $10M ([Balchandani and Naidich, 2015](#)). Operational costs are often quoted at approximately $1000/hour ([Marek et al., 2022](#)).

# Electrophysiology

Although invasive electrophysiological systems demonstrate remarkable and rapidly advancing capabilities for BCIs and fundamental neuroscience, their utility for whole-brain emulation is inherently limited. They provide exceptionally high-quality, high-temporal-resolution passive recordings of neuronal spiking activity. However, electrophysiological recordings are, and will likely remain, highly localized to the small brain volumes immediately surrounding the implanted electrodes. Indeed, scaling current electrophysiological approaches to achieve significant coverage of the human brain faces fundamental physical, biological and ethical hurdles. The sheer number of discrete implants required would be infeasible, and the associated tissue response to such widespread penetration would be prohibitive. Thus, these methods primarily offer sparse, high-fidelity data points rather than a pathway to comprehensive functional mapping of the entire human brain.

## Patch Clamp

Patch clamp recordings involve attaching a glass micropipette to a neuron's membrane to measure electrical currents directly through individual ion channels. This technique achieves the highest possible spatial resolution by recording from single neurons or even isolated membrane patches containing individual ion channels. The temporal resolution is exceptional, capturing signals at microsecond timescales (>10 kHz sampling rates), allowing researchers to observe individual ion channels' opening and closing kinetics. However, patch clamp recordings, depending on the exact technique, are typically limited to minutes or hours for a single cell due to membrane destabilization and cell deterioration. In slice preparations or cell cultures, experienced researchers might maintain stable recordings from multiple sequential cells over an 8-12 hour experiment ([Mayer Jr and Brown, 1998](#)).

Patch clamp is highly invasive and requires precise micromanipulation under microscopic visualization, limiting its use mostly to in vitro preparations (brain slices, cell cultures). Patch clamp experiments have been done in vivo, but involve highly laborious and relatively low-throughput procedures. We estimate that a basic setup costs between $50,000 and $150,000 (e.g., ScienceDirect Topics, *Patch-Clamp Technique* reports 'almost $100 000 at current list prices' ), with operational costs of approximately $200-500 per day (see e.g. the Ohio State University whole cell patch clamping service at $375/day) in consumables (micropipettes, solutions) and requiring highly skilled operators.

## Microelectrode Arrays (MEAs)

Microelectrode Arrays (MEAs) consist of recording electrodes embedded in a substrate to simultaneously record from multiple individual neurons. MEAs bridge the gap between single-cell and population-level recordings, typically capturing activity from dozens to thousands of neurons, depending on electrode density. Modern high-density MEAs can contain thousands of electrodes within a 1mm² area, approaching single-neuron resolution within local networks. MEAs sample neural activity at 10-50 kHz, capturing action potentials and local field potentials with millisecond precision. Unlike patch clamp, MEAs can maintain stable recordings for days to weeks in vitro, and implanted chronic arrays can sometimes function for months to years in vivo, enabling long-term studies of neural network dynamics and plasticity. Common challenges include signal quality decreases with motion, electrode impedance changes over time, and tissue inflammatory responses around electrodes. MEA systems are increasingly available in research settings, with perhaps thousands of systems globally. We estimate that equipment costs range from $50,000 for basic in vitro systems to over $250,000 for high-density or wireless in vivo systems. Operational costs include replacement arrays ($500-2,000 each), surgical procedures for implantation ($1,000-3,000 per animal), and data storage for the substantial volumes generated.

# Ultrasound

Ultrasound-based neuroimaging is distinct from electromagnetic approaches in that it uses mechanical waves, brief "pings" of sound, to probe tissues. These waves travel through soft tissue at around 1,540 m/s and get partially scattered back, carrying information about local density and compressibility. Frequencies typically range from about 3 MHz (wavelength ~500 μm) to 25 MHz (wavelength ~60 μm), with attenuation in water-rich tissue scaling linearly (Marblestone et al., 2013).

Bone attenuates ultrasound more severely, though this can be addressed through minimally invasive preparations such as acoustically transparent cranial windows.

Currently, most neuroscientific research involving ultrasound revolves around hemodynamic functional ultrasound imaging (fUSI), which measures changes in blood flow rather than neural activity (Rabut et al., 2020). Hemodynamic fUSI involves transmitting plane waves at multiple angles and detecting Doppler shifts from moving red blood cells, producing images that quantify local cerebral blood volume with spatial resolution around 100μm and temporal resolution of 400 ms, all while enjoying significant coverage. State-of-the-art probes are reportedly capable of imaging over 30% of the human brain by volume (Forest Neurotech, 2025). However, as discussed above with fMRI, the hemodynamic signal is ultimately a slow, indirect correlate of neural activity: the blood-flow response emerges with a latency of around one second and is integrated over multiple neural events (Aydin et al., 2020). Norman et al. achieved a 0.4 mm$^3$ resolution with fUS for an 8 cm$^3$ volume in monkeys (Norman et al., 2021).

A key development toward more direct ultrasonic measures of neural activity involves gas vesicles (GVs): air-filled protein nanostructures that scatter ultrasound waves. These 2-nm-thick protein shells enclosing gas compartments can be genetically introduced into mammalian cells and engineered to respond to biological signals, making them promising "reporter genes" for functional ultrasound (Shapiro et al., 2014). Recent developments include the engineering of GVs to respond dynamically to calcium transients, resulting in ultrasonic reporters of calcium (URoCs) – the first genetically encodable calcium indicators for ultrasound imaging (Jin et al., 2023). Significant challenges remain, including slow sensor kinetics (requiring 3.5 seconds to reach half-maximum signal upon calcium binding) and achieving robust expression in neurons. Further, research has only demonstrated resolutions of about 100 μm, insufficient to capture single neuron activity.

While the approaches described above leverage ultrasound as an imaging modality, another proposed ultrasound-based technique, "neural dust", uses ultrasound for power delivery and communication with implanted electrodes for direct neural recording (Seo, 2018). These miniature devices ("motes") leverage piezoelectric materials so that when ultrasound waves from an external transducer reach these implants, some of the ultrasound's energy powers the device while changes in electrical impedance caused by neural activity modulate how ultrasound is reflected back, creating a backscatter communication channel. This approach has been successfully demonstrated in peripheral nerves and muscles in rodents, achieving wireless recordings comparable to conventional wired systems (Seo et al., 2016). In general, although still in its early stages, neural dust offers significant potential: its ultrasonic powering can enable extremely small (50-100 μm) implants, promising high-density, minimally invasive neural recording. However, many significant challenges remain, including delivering these microscopic motes to target tissues, achieving sufficient backscatter sensitivity for reliable signal detection, developing robust biocompatible encapsulation, and employing the sophisticated beamforming required to isolate signals from large numbers of motes.

In conclusion, ultrasound represents a modality with significant potential. For imaging approaches like fUSI and gas vesicles, at least theoretically, higher frequencies (60-100 MHz) could achieve single-neuron resolution (~15-25 μm) with significantly higher attenuation. Meanwhile, neural dust offers a complementary approach through distributed, miniature implantable sensors that can directly record neural activity with potentially less tissue displacement than traditional electrodes. Together, these ultrasound-based technologies could advance the frontier of neural recording by potentially enabling whole-brain recordings at cellular resolution in organisms as large as mice, while offering broader coverage in larger brains than is currently achievable with other techniques. However, substantial engineering challenges remain across all ultrasound approaches, and realizing the full potential of ultrasonic neural interfaces will require significant further research and development.

# Optical methods

Before we dive into two of the most prominent recording modalities (calcium and voltage imaging) we want to provide a quick primer on photon microscopy. Readers familiar with the systems can skip this section.

Traditional single-photon fluorescence microscopy relies on a fundamental quantum mechanical process. When a fluorescent molecule absorbs a photon of the appropriate energy (typically in the visible light range), an electron in this molecule is excited from its ground state to a higher energy level. After a brief period (nanoseconds), the electron returns to its ground state, releasing a photon with slightly less energy (and thus longer wavelength) than the excitation photon. This difference between excitation and emission wavelengths, known as the Stokes shift, allows us to separate the emitted fluorescence signal from the excitation light using optical filters. In the context of brain imaging, this process faces significant limitations. When visible light enters brain tissue, it encounters numerous cell membranes, protein structures, and other cellular components that can either absorb the light or change its direction (scattering). Both effects reduce the number of photons that reach the focal point and make it harder to collect the emitted fluorescence. Moreover, because single-photon excitation occurs wherever an excitation photon encounters a fluorescent molecule, out-of-focus fluorescence creates a background signal that reduces contrast and spatial resolution. While techniques like confocal microscopy can reduce this out-of-focus fluorescence using a pinhole, they cannot overcome the fundamental depth limitation imposed by tissue scattering, typically restricting 1P imaging to depths of 200-300 micrometers of tissue ([Xu et al., 2024](#)).

Two-photon microscopy, first demonstrated by Denk and colleagues in 1990 ([Denk et al., 1990](#)), revolutionized deep tissue imaging through an elegant quantum mechanical principle: instead of using one high-energy photon to excite a fluorescent molecule, it uses two lower-energy photons that arrive nearly simultaneously (within about $10^{-16}$ seconds). These lower-energy photons, typically in the infrared range (700-1040 nm), can penetrate tissue much more deeply than visible light because they experience less scattering. Critically, the probability of two photons arriving simultaneously is only high at the focal point of the microscope, creating natural optical sectioning and reducing photobleaching and photodamage in surrounding tissue. This allows imaging up to about 600-800 micrometers deep ([Xu et al., 2024](#)) in the cortex.

Three-photon microscopy extends these principles further by using three even lower-energy photons (typically >1300 nm) to achieve excitation. This technique offers several advantages for deep imaging: the longer wavelengths experience even less scattering, and the requirement for three coincident photons provides better background rejection. Three-photon microscopy can reach depths of 1-1.3 millimeters ([Xu et al., 2024](#)), accessing structures like the hippocampus in mice that are largely inaccessible to two-photon imaging. However, three-photon systems require more expensive laser sources and typically operate at slower speeds due to the need for higher pulse energies; further, another consideration with three-photon systems is that three-photon imaging heats brain tissue more readily than two-photon systems.

Because two-photon and three-photon microscopy rely on point-scanning, increasing the number of recorded neurons requires distributing the available laser power across more points and reducing the dwell time per neuron to maintain temporal resolution. Analysis of publicly released datasets demonstrates this relationship clearly - as the number of simultaneously recorded neurons increases from hundreds to thousands, there is a systematic increase in noise levels unless temporal resolution is sacrificed ([Rupprecht, 2021](#)). This relationship holds across different imaging configurations and preparations, reflecting the inherent physical limits of current optical recording approaches.

# Calcium imaging

Calcium imaging uses genetically encoded proteins that light up when they bind to calcium. These fluorescent proteins, most commonly variants of GCaMP, are introduced into neurons through genetic engineering. When neurons fire, calcium rushes into the cell; GCaMP binds to calcium, resulting in a conformational change that permits fluorescence.

The preparation approach for introducing the GCaMP transgene varies significantly. Viral injection methods generally result in higher and more rapid expression, though this expression can be quite variable. For viral approaches, researchers perform surgery to inject viral vectors carrying the calcium indicator genes. After this injection surgery, there is a waiting period of days to multiple weeks for proper protein expression, with expression levels changing over time. In contrast, generating transgenic lines provides reliable, heritable expression but may require months to years to develop and validate. For in vivo imaging with either method, a surgical procedure to implant a clear window in the skull is also necessary to provide optical access to the brain.

The technique requires substantial infrastructure – a typical multiphoton microscope costs more than $500,000 ([Holmes et al., 2022](#)) and comes with significant service and maintenance costs. Globally, we estimate the number of two-photon microscopes in active use for neuroscience research to likely be between a few hundreds and a few thousands. Additional costs include surgical supplies, viral vectors, and specialized habituation equipment. The price per hour falls into the range of $20-$100 ([ref](#), [ref](#)).

Calcium imaging operates in an interesting middle ground between fMRI and electrophysiology. The calcium indicator GCaMP7f, for example, has a half-rise time of about 60 milliseconds and half-decay time of about 150 milliseconds ([Dana et al., 2019](#)), while the newer jGCaMP8f reaches half-rise times of about 2–7 ms and half-decay times of about 40 ms ([Zhang et al., 2023](#)). Typical two-photon systems can image at 30 Hz for a single plane, though this drops to 1-5 Hz when imaging multiple planes to capture a volume. Recordings can last from hours in acute preparations to months in chronic experiments. Key factors limiting continuous imaging are photobleaching and photodamage (phototoxicity). Photobleaching, the irreversible loss of indicator fluorescence due to light exposure, is counteracted by cellular synthesis of new indicator proteins. Given typical fluorescent protein half-lives of approximately 24 hours ([Snapp, 2009](#)), significant fluorescence recovery through this replenishment can occur over hours to days. Separate from signal loss, photodamage refers to light-induced cellular injury and physiological disruption, often mediated by reactive oxygen species or thermal effects, particularly with two-photon microscopy ([Grienberger et al., 2022](#); [Icha et al., 2017](#)). This damage can be subtle, affecting cellular processes before morphological changes are evident ([Icha et al., 2017](#)). Since both photobleaching and photodamage restrict imaging duration and can compromise data integrity, careful optimization of light exposure is critical.

The spatial resolution of calcium imaging is sufficient for individual neurons and even their dendrites at approximately 0.5-1 μm resolution. A typical field of view might be 500x500 μm, containing hundreds of neurons. Standard two-photon microscopes can image up to 600-800 μm deep in the cortex, while three-photon systems can reach 1-1.3 mm ([Xu et al., 2024](#)). For advanced systems, this results in up to 1 mm$^3$ volumes that can be imaged at single-cell resolution. While these capabilities allow for recording large neuronal populations, it's important to recall the inherent trade-offs in point-scanning systems between neuron count, signal quality, and temporal resolution, as discussed previously.

## Voltage Imaging

While calcium imaging indirectly measures neural activity through calcium transients, voltage imaging directly detects changes in membrane potential. This allows voltage imaging to capture the rapid dynamics of action potentials, spike timing, and subthreshold synaptic events, all of which remain invisible to calcium imaging ([Peterka et al., 2011](#)). These capabilities are also valuable for mapping dendritic computation, axonal propagation, and inhibitory/excitatory balance ([Kulkarni & Miller, 2017](#)), and position voltage imaging as complementary to both calcium imaging and electrophysiology. Similar to calcium imaging, both viral and transgenic approaches exist.

Voltage imaging relies on indicators that transduce membrane potential changes into optical signals through two primary approaches: voltage-sensitive dyes (VSDs) and genetically-encoded voltage indicators (GEVIs). VSDs are small organic molecules that often employ mechanisms like electrochromism, where the electric field directly shifts the dye's absorption or emission spectrum, and that exhibit fast kinetics (<1 ms) and high sensitivity but lack cell-type specificity ([Aseyev et al., 2023](#)). GEVIs, in contrast, are fluorescent proteins engineered to be voltage-sensitive, typically using voltage-sensing domains from proteins like Ci-VSP or microbial rhodopsins, enabling genetic targeting to specific neuronal populations. However, GEVIs typically trade speed for sensitivity: most have response times of 2–10 ms, limiting single-trial action potential detection in vivo ([Kulkarni & Miller, 2017](#)). Both methods face challenges in signal-to-noise ratio (SNR) due to low photon counts at high acquisition rates (>1 kHz) and phototoxicity.

Despite significant advances, voltage imaging still faces substantial challenges. The signal-to-noise ratio of voltage indicators generally remains problematic, with fluorescence changes of only 2-50% per 100 mV compared to calcium indicators' signals that can exceed 1000%. Moreover, voltage indicators are restricted to the membrane (since they necessarily must detect the membrane potential), which is a much smaller volume than the cytosol, and consequently, voltage imaging is often much dimmer than calcium imaging. These challenges create a tradeoff where faster indicators often require substantial excitation light, causing photodamage that restricts recording sessions to typically just minutes rather than the hours possible with calcium imaging. Furthermore, the membrane localization requirement makes cell segmentation difficult, as adjacent labeled neurons create overlapping "chicken wire" patterns instead of easily distinguishable volumes ([Kulkarni & Miller, 2017](#)). Generally, voltage imaging to whole brain coverage in mammals would require orders-of-magnitude improvements in sensor brightness, photostability, and multiplexed imaging systems ([Aseyev et al., 2023](#)). Nonetheless, voltage imaging has already proven invaluable for studying aspects of neural computation inaccessible to other techniques, and its continued development holds significant promise, particularly in organisms with relatively optically accessible brains.

## Neurotransmitter imaging

While the methods discussed above capture electrical activity, understanding brain function also requires monitoring the dynamics of chemical signaling. Neurotransmitters and neuromodulators orchestrate neural communication, from rapid synaptic transmission to slower, widespread modulatory effects. Genetically encoded fluorescent biosensors have emerged as powerful tools for visualizing these chemical signals in real-time, offering insights that complement neural recordings. These biosensors generally operate by fusing a specific ligand-binding domain – which recognizes a particular neurotransmitter or neuromodulator – to a fluorescent protein, often a circularly permuted fluorescent protein (cpFP). The binding of the target molecule induces a conformational change in the sensor, which in turn alters the fluorescence properties of the cpFP. This change can then be detected using standard microscopy techniques similar to those employed for calcium or voltage imaging. A key advantage of this approach is the ability for targeted expression in specific cell types or even subcellular compartments, typically achieved via viral vectors (such as AAVs) or through the creation of transgenic lines.

The detection of fast excitatory neurotransmission, particularly glutamate signaling, has been a significant focus. Early FRET-based sensors for glutamate, such as FLIPE, had limitations in signal-to-noise ratio (SNR) and kinetic performance, which restricted their application for studying rapid synaptic events ([Hao and Plested, 2022](#)). An important development in this area was the introduction of iGluSnFR (intensity-based glutamate-sensing fluorescent reporter) by Marvin and colleagues ([Marvin et al., 2013](#)). This sensor, which utilizes a glutamate-binding protein from *E. coli* (GltI) fused with a cpGFP, offered improved dynamic range and SNR, enabling the detection of glutamate transients with a temporal resolution of approximately 100 ms in various experimental settings. iGluSnFR has since been widely adopted for studying glutamate dynamics in contexts ranging from sensory processing and synaptic plasticity to investigations of pathological states ([Hao and Plested, 2022](#)).

The initial utility of iGluSnFR prompted the engineering of numerous variants with refined characteristics. These include versions with faster off-kinetics (e.g., iGluf, SF-iGluSnFR.S72A) to better resolve successive release events, improved brightness and stability through the use of superfolder fluorescent proteins (e.g., SF-iGluSnFR), and altered emission spectra (e.g., R-iGluSnFR1) to facilitate multicolor imaging ([Helassa et al., 2018](#); [Marvin et al., 2018](#); [Wu et al., 2018](#)). To enhance spatial precision, iGluSnFR has also been fused to synaptic proteins like Neurexin 1 for presynaptic targeting or Stargazin for postsynaptic localization ([Kim et al., 2020](#); [Hao et al., 2021](#)). These ongoing improvements have broadened the applicability of glutamate sensors, for example, in quantal analysis and disease modeling. Similar design principles have also been applied to create sensors for other fast neurotransmitters, including GABA (e.g., iGABASnFR) and acetylcholine (e.g., iAchSnFR) ([Marvin et al., 2019](#); [Jing et al., 2018](#)).

For monitoring neuromodulators and neuropeptides, many of which signal through G-Protein Coupled Receptors (GPCRs), a common sensor design strategy involves engineering the native GPCRs themselves. This is typically achieved by inserting a cpFP into an intracellular loop of the target GPCR. The design aims to preserve the receptor's natural ligand affinity while blocking endogenous G-protein coupling, thereby preventing downstream signaling but allowing ligand binding to be transduced into a fluorescence change ([Girven et al., 2022](#)). This approach has yielded sensor families such as "GRAB" (GPCR Activation-Based) sensors and others like dLight for dopamine. Currently, validated sensors are available for approximately 12-15 different neuromodulators and neuropeptides, including dopamine (e.g., dLight, GRABDA), norepinephrine (e.g., GRABNE), serotonin (e.g., GRAB5-HT), acetylcholine (acting on muscarinic receptors, e.g., GRABACh), and various opioid peptides ([Muir et al., 2024](#); [Sun et al., 2018](#); [Feng et al., 2019](#)).

These GPCR-based sensors generally provide temporal resolution (1-10 Hz frame rates) and allow recording durations (hours to months) comparable to calcium imaging systems. For example, the dopamine sensor dLight1.2 has reported response times of approximately 9.5 ms for binding and 90 ms for unbinding ([Patriarchi et al., 2018](#)). Spectrally distinct sensors permit simultaneous monitoring of multiple neuromodulators, such as green dLight1.3b for dopamine alongside red GRABNE2m for norepinephrine ([Muir et al., 2024](#)). Such tools have enabled new observations, for instance, tracking dopamine transients with ~100 ms precision during reward-related behaviors ([Mohebi et al., 2019](#)) and identifying selective opioid peptide release during specific physiological states ([Castro et al., 2022](#)). The imaging setups, data handling, and physical limitations for these biosensors are largely analogous to those for calcium imaging. While the range of available sensors continues to grow, expanding this toolkit to cover a broader array of neurochemicals and further enhancing sensor performance remain active areas of research.

# Invasive perturbation experiments

Under one of the most common evaluation criteria for brain emulations, the emulation should match the internal dynamics of the target brain, likely at least at the level of neural activity. Meeting this benchmark requires accurately modeling each neuron's input-output function – how its outputs depend on the various inputs it receives. While passive recordings constrain possible input-output functions, standard inputs and the fact that different circuit configurations can produce identical activity patterns make it challenging to gather enough data to uniquely identify each neuron's parameters ([Haspel et al., 2023](#)). This fundamental limitation persists even with extensive recordings under diverse conditions, as the system's inherent complexity and feedback loops mean that many different parameter sets remain consistent with observed activity patterns ([Pospisil et al., 2024](#)). This challenge of parameter identification from passive recordings alone, which exists even when the connectome is known, means that experimental, perturbation methods will also be key to generating accurate brain emulations.

In this discussion, we focus on perturbation techniques like optogenetic activation and silencing, as well as patch-clamp electrophysiology, since these permit temporally and spatially precise changes to neural activity. Other forms of perturbing neural activation, such as ablation experiments, are generally not discussed, since these techniques are often less useful for downstream computational modeling, as their exact influence on circuit activity is typically less precise and often irreversible, complicating the modeling of dynamic input-output functions.

Invasive perturbation methods provide a solution by breaking natural correlations between neurons and allowing direct manipulation of specific circuit elements. As established in causal inference theory, determining genuine causal relationships generally requires perturbations ([Woodward, 2004](#); [Pearl, 2009](#)). In neural circuits specifically, establishing the causal influence of one neuron on another requires stimulating the former while recording from the latter ([Haspel et al., 2023](#)). By precisely controlling individual neurons or groups of neurons while recording from others, experimenters can drive circuits into novel states that would not arise naturally. Data from such causal perturbations enables dramatically more sample-efficient fitting of neuronal input-output functions than passive recordings alone ([LaFosse et al., 2024](); [Wagenmaker et al., 2024]()). These targeted interventions thus provide crucial constraints for reverse engineering neural computation.

Patch clamp recording, one of the oldest of these interventional techniques, provides exact control and measurement of individual neurons' electrical activity. By forming a tight seal with the cell membrane it enables direct manipulation of membrane potential while simultaneously recording with sufficient temporal resolution to detect individual action potentials and minute features of the membrane voltage dynamics. Dual patch recordings are required to establish causal relationships between neurons, stimulating one neuron while recording from another. This method produces the highest-quality data for determining input-output functions, with minimal off-target effects and exceptional signal fidelity. However, patch clamp recordings are technically demanding, especially in vivo, labor-intensive, and typically limited to short durations. Combined with the need for dual recordings, this makes it impractical for mapping large circuits.

Optogenetics enables precise temporal control of genetically defined neuron populations using light-sensitive ion channels or pumps ([Rost et al., 2022]()). This approach allows simultaneous manipulation of many neurons and can be combined with large-scale recording methods. Optogenetic approaches can be used to both activate and silence sets of neurons. Standard optogenetic approaches using fiber optics suffer from light scattering in tissue, limiting spatial resolution. However, methods like two-photon holography combined with soma-targeted opsins can achieve single-cell resolution in vivo, though spatial precision varies with the specific tools and light delivery systems used ([Adesnik and Abdeladim, 2021]()). While optogenetic stimulation offers millisecond-precise control, recording neural activity in response to these perturbations presents additional challenges. The most common recording method, calcium imaging, struggles to reliably detect individual spikes, particularly in fast-spiking neurons, despite ongoing improvements in indicators. Furthermore, optical crosstalk between stimulation and imaging wavelengths can interfere with simultaneous recording and perturbation.

Beyond detailed circuit mapping, it is important to note that optogenetic perturbation experiments linking neural population activity to behavior have the potential to contribute significantly towards embodied emulations, possibly more so than whole-brain imaging alone for certain objectives ([Cowley et al., 2024]()). Optogenetic activation behavioral data can be relatively cheap and fast to collect compared to some functional recording datasets, and can generate very useful, quantitative data ([Cande et al., 2018]()). Furthermore, such perturbation experiments directly test causal relationships, avoiding the interpretational problems of messy correlations often found in neural recording data, and directly link neural activity to behavior.

Chemogenetics complements these approaches by offering sustained but less temporally precise control through engineered receptors (DREADDs) that respond to specific synthetic compounds ([Roth, 2016](); [Minamimoto et al., 2024]()). After viral delivery of these receptors to targeted neuron populations, systemic drug administration can modulate neural activity over hours to days. While this approach enables manipulation of specific cell types and even specific neural projections through retrograde viral vectors, its utility for reverse engineering precise input-output functions is more limited. The slow kinetics of drug action and intracellular signaling cascades, combined with the sustained nature of the manipulation, make it impossible to probe the precise temporal interactions needed to characterize neuronal computation. Instead, chemogenetics is most valuable for behavioral studies and understanding the role of broader cell populations over longer timescales.

# Data management: storage, standardization, analysis

The advent of high-density neural recording technologies routinely generates terabyte datasets that were unimaginable a decade ago. This scale presents both opportunities and fundamental challenges for data management and analysis.

To illustrate – high-resolution scanning at 1 mm³ can generate about 60 GB of raw image data per hour of continuous scanning (brain dimensions: ~150mm × 180mm × 150mm = 4,050,000 voxels at 4 bytes over 1 hour at 1Hz is 16.2 MB × 3600 = 58.32 GB). High-density arrays recording at high sampling rates can produce 10-100 GB per hour of continuous recording. A typical two-hour calcium imaging session at 30 Hz can generate 500 GB to 2 TB of raw data, depending on the field of view and resolution ([Stringer et al, 2024]()). Raw data includes both the brain data and associated behavioral measurements.

Nowadays, functional recordings in neuroscience are often shared with the community via dedicated data repositories. The appendix provides a list of over 50 such repositories.

Because large scale neuroscience projects can generate huge data volumes, individual academic labs often do not make their raw data easily accessible, and instead state that the data is "available upon request" ([Tedersoo et al., 2021](#)). Public, easily accessible data sharing has not been the default over the past two decades (although it has long been pioneered by dedicated organizations like the [Allen Institute](#)) and is only slowly finding adoption, as some funders make it a requirement. Only recently have efforts like Neurodata Without Borders ([Rübel et al., 2022](#)) started to standardize neurophysiology data, making them more comparable and interoperable. From private conversations with multiple experts, they assume that less than 5% of all the existing data is publicly available (in any form) at this point. In conversations with computational neuroscientists, the anecdotally reported access to multiple consistent and well-cleaned datasets is one of the main limiting factors to computational models making use of said data. This data barrier also prevents non-neuro-specialist engineers from other fields from getting involved and making meaningful contributions.

We can think about data repositories in 4 categories.
1. **Preferred repositories:** these comprise the majority of the well-formatted, easily accessible data: OpenNeuro, DANDI, Brain Image Library, EBrains, FigShare, Allen Institute, or Zenodo.
2. **Single datasets:** Study Forrest and FlyWire are examples of large datasets hosted on their dedicated websites.
3. **Other repositories, such as** CRCNS, Neurovault, and others, are often much less well-maintained and/or easily accessible.
4. **Meta-repositories, registries, etc.:** neuinfo does not host data and just has metadata scraped from elsewhere. OpenEphys predominantly lists other repositories, and the data it does link might just be repeats of previously mentioned repositories.

While some data repositories offer APIs facilitating mass data download, others may be impossible to scrape due to absent or scattered metadata; despite this, a repository's user-friendliness correlates positively with data volume, allowing access to significant portions of data across various modalities. It is unclear how much duplication exists across repositories. Heterogeneity of the data quality is another issue, with image data particularly vulnerable to quality issues like poor resolution.

The standardization of neurodynamic datasets presents unique challenges due to the extraordinary complexity of neural recordings across different temporal and spatial scales across different organisms. These technical challenges are compounded by sometimes significant methodological variations across laboratories and institutions. Different research groups often employ distinct preprocessing pipelines, quality control procedures, and analysis methods, making direct comparisons between datasets difficult. Even within the same recording modality, variations in experimental protocols, equipment configurations, and environmental conditions can introduce systematic differences that complicate data integration. The situation is further complicated by inconsistent documentation practices, with many datasets lacking crucial metadata about preprocessing steps, quality control measures, or experimental conditions.

The raw data distributed needs various post-processing steps. For calcium imaging, for instance following motion correction, cell identification and segmentation algorithms can process hours of recording in parallel. Real-time processing of 10,000 neurons requires a modern workstation and a high-end GPU. Scaling to 100,000 neurons demands at least 128GB RAM and multiple GPUs, with processing taking 3-5x real-time. At the extreme end, analyzing 1,000,000 neurons requires distributed computing systems with at least 1TB of RAM, processing at 10-20x real-time. ([Stringer et al 2024](#)) The primary bottlenecks emerge in population analyses that require pairwise computations between neurons – for instance, a correlation matrix for 100,000 neurons requires 80GB of RAM just for storage. Scaling these analyses to even larger populations (millions to billions of neurons) would require fundamental algorithmic innovations. Current approaches often scale quadratically with neuron count, making them impractical for large populations.

Recently, Mineault et al. published an analysis of the "contents of DANDI, OpenNeuro, iEEG.org, as well as large-scale individual datasets" as part of their NeuroAI for AI Safety roadmap ([Mineault et al., 2024](#)). We will quote them verbatim here.

**Availability of neural data to train a large-scale model (from [Mineault et al, 2024](#))**

"The past decade has seen an explosion in the quantity of neural data freely available online. These public datasets represent a unique opportunity to learn good representations of neural data for a variety of downstream tasks, including brain-computer interfaces, clinical diagnoses for computational psychiatry and sleep disorders, and basic neuroscience.

Here we present a breakdown of the available data sources from an analysis of the contents of DANDI, OpenNeuro, iEEG.org, as well as large-scale individual datasets. Some of the highlights from this analysis include:

- There are around 100,000 hours of neural data available in freely accessible archives.
- There are roughly 3.3 million neuron-hours of single-neuron recordings from animals.
- The most abundant data type in terms of number of hours is intracortical EEG in humans–an invasive modality generated from the typically continuous, week-long recordings performed during epilepsy monitoring.
- Single neuron data is concentrated in a few datasets; the top 10 largest datasets in terms of neuron-hours account for more than 94% of total neuron-hours across all of DANDI. These come mostly from zebrafish and mouse, with one dataset from macaques.
- Large fMRI recordings are split into two categories: broad neuroimaging surveys,

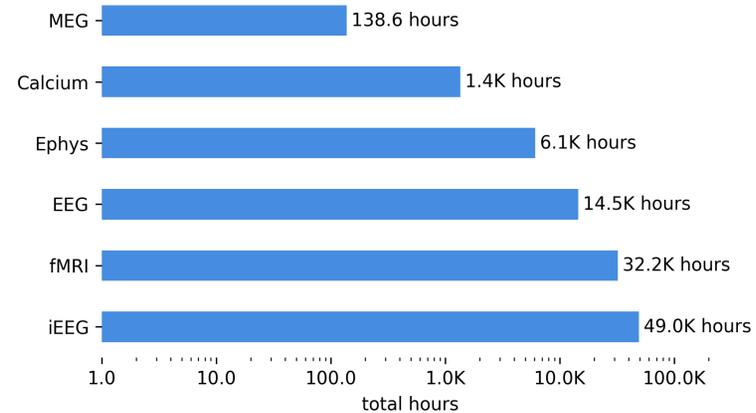

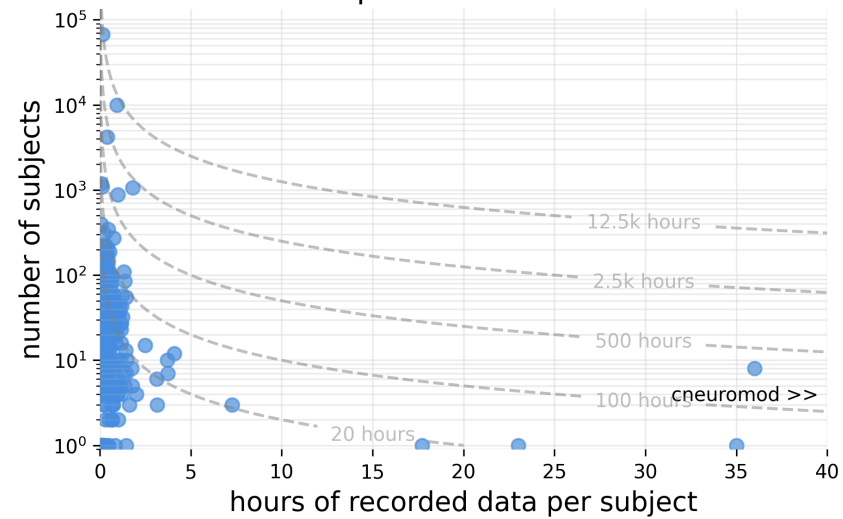

including HCP and UK Biobank, which scan many people for a short time; and intensive neuroimaging datasets, including Courtois Neuromod and the Natural Scenes Dataset, which scan few people for a very long time."

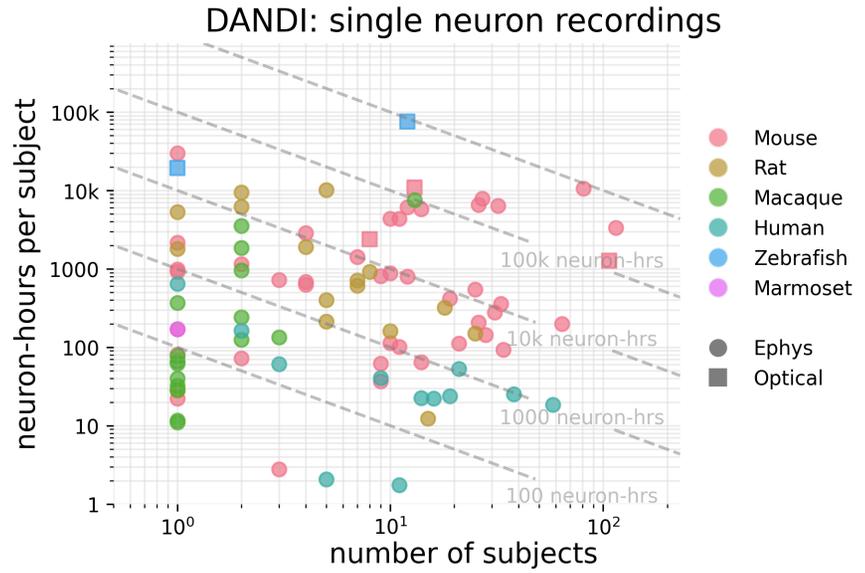

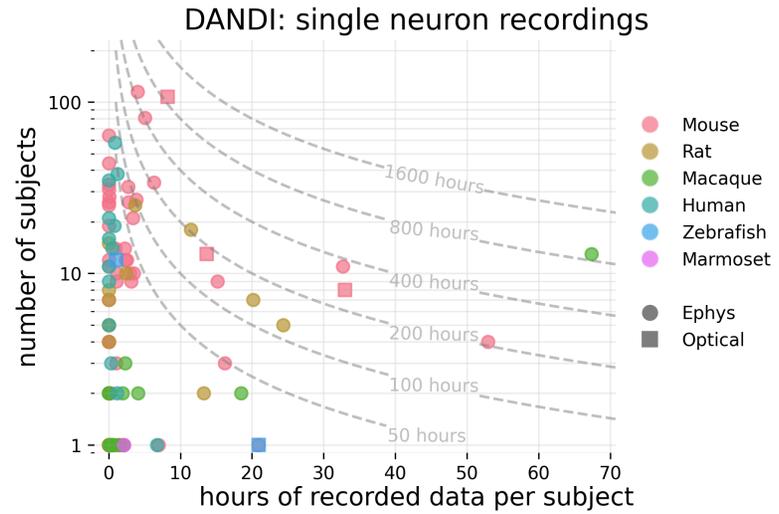

# Connectomics: Brain structure reconstruction

Connectomics, the endeavor to reconstruct the brain's intricate structural architecture, fundamentally depends on imaging technologies capable of resolving the smallest elements of neural circuitry. The smallest unit here is either the synapse, between 200 and 800 nm wide ([Sheng and Kim, 2011](#)), the synaptic cleft between two synapses, about 20-30 nm wide ([Yang & Annaert, 2024](#)), or gap junctions between neurons, which are approximately 2 nm wide. The smallest (unmyelinated) axons are around 50 nm ([Helmstaedter et al., 2013](#)) in width and can reach from the motor cortex to the lower parts of the spine. This necessitates voxel sizes at the single-digit to low double-digit nm scale. The connectomics community typically cites resolution requirements between 10-20 nm per voxel to resolve synapses properly (that is, each pixel in an image should represent a 3D cube of the brain of the dimensions of say 10x10x10 nm) ([Jefferis et al., 2023](#)). Dense connectomics refers to the analysis of all neurons and their connections within an analyzed volume. This is often accompanied by non-neural cells such as astrocytes, microglia, oligodendrocytes, and vascular cells. Staining for these approaches often utilizes chemicals that broadly bind to biomolecules within the samples (such as lipids) to produce recognizable subcellular divisions. Sparse connectomics, on the other hand, captures a subset of the neurons.

> For an excellent illustration of the sizes of neurons and synapses, see Figure 1 in [Iascone, 2020](#) where a pyramidal neuron of mouse primary somatosensory cortex is traced and individual synapses identified. Due to copyright restrictions, we cannot replicate this image here.

Santiago Ramón y Cajal's pioneering work established synapses as the cornerstone of neuroscience - a view later confirmed by electron microscopy in 1956 ([Jekely and Yuste, 2024](#)). Electron microscopy development reached resolutions less than ten nanometers in the 1930s and, in 1944, broke 2 nm resolution ([Haguenau et al., 2003](#)), far surpassing light microscopy constrained to a lateral resolution of ~250 nm and an axial resolution of ~550 nm by the diffraction limit ([Huang et al., 2010](#)).

The first connectome, a complete map of all neurons and their synaptic connections to each other, was imaged by John White, Sydney Brenner and colleagues in the 1980s. It was Brenner who first thought to do this "radical experiment," according to a 2020 article in *Cell* ([Abbott et al., 2020](#)):

> "…might it be possible to obtain the complete wiring diagram of an animal's nervous system by serially sectioning it into many exceedingly thin slices, imaging each of these sections at high resolution with an electron microscope (EM), and painstakingly tracing each neuron's branches and synaptic connections with other neurons? This audacious idea became reality in 1986 when Brenner, John White, and several other extraordinary scientists produced a 340-page magnum opus, "The Structure of the Nervous System of the Nematode *Caenorhabditis elegans*" (with the running head "[The Mind of a Worm](#)")…"

Brenner's work was decades ahead of its time. The worm connectome had to be reconstructed by hand because digital image processing tools at that time were inadequate. Brenner and his colleagues at Cambridge's Laboratory for Molecular Biology imaged and reconstructed all 300 neurons of the highly stereotyped worm neurons, combining eight different individuals, tracing and connecting each neuron's spindly branches by hand. An intuitive way of illustrating this process is to compare it to satellite images of the earth (see figure below): Not only do you need to take images at extremely high resolution, you also need to determine (encircle individual areas and colorize) and annotate objects (save meta information) in order to make the helpful map. The first worm connectome paper [has been cited thousands of times](#), and almost every year since its publication more than three decades ago, the rate of citations has increased.Still, it took neuroscientists nearly 40 years of work to go from a 300 neuron worm connectome to a 140,000 neuron fruit fly connectome.

Neurons make new connections and abandon unused synapses. Some areas even withstand the trend of declining neuron numbers over life, and we see new neurons emerge. Neuronal activity and other variables impose changes on the connectome, which influence long-term information processing and thus represent the structural equivalent of learning over time. Modern studies have revealed multiple timescales of structural plasticity, from rapid synaptic modifications occurring within minutes to slower axonal and dendritic remodeling processes that can take days or weeks. Understanding the rules governing these structural modifications may be as crucial as mapping the connections themselves, as these rules determine how the network can reconfigure itself in response to experience and environmental demands. To lean into our satellite image of the world metaphor again, we want to have a dynamic video of the world, rather than a single static picture. Despite progress in methods to study neuroplasticity ([Velicky et al., 2023](#)), reliably tracking morphological dynamics and, through that, understanding neuroplasticity largely remains beyond what the field is capable of.

**Figure Comparison of a map of the globe and connectomics** A) Current Google Maps maximum zoom is approximately 1 pixel ≈ 15cm on Earth (Google, 2014). Intelligence experts estimate that the most advanced US military satellites can achieve resolutions of up to 6 cm per pixel. (Richardson, 2024). An illustration in the image below (source). Scanning the human brain at the resolution necessary to reconstruct a connectome (~20nm sized voxels necessary to trace synapses) is the same as having a map of the world 10x sharper than the current best satellites (with 6371 km radius, earth's total surface area is ~510 million km², which at 1 mm² resolution / pixel = $5.1 \times 10^{20}$ pixels. 1200 cm³ average human brain volume at 20nm isotropic voxels size is $1.5 \times 10^{20}$ voxels.). B) Figure illustrating various stages of image processing in comparison to raw satellite data: following registration (left), following segmentation and tracing (center), and following annotation (right). Source: Schlegel, 2024

A
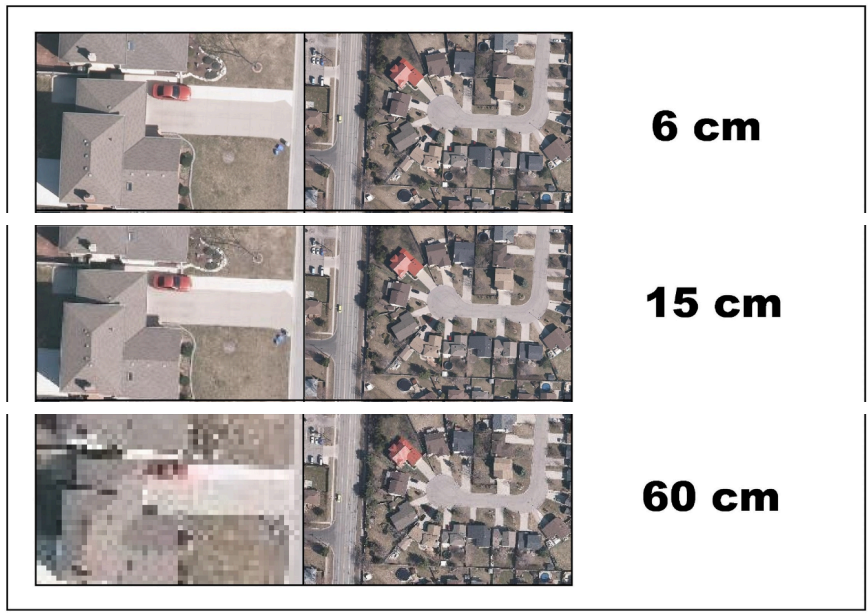

B
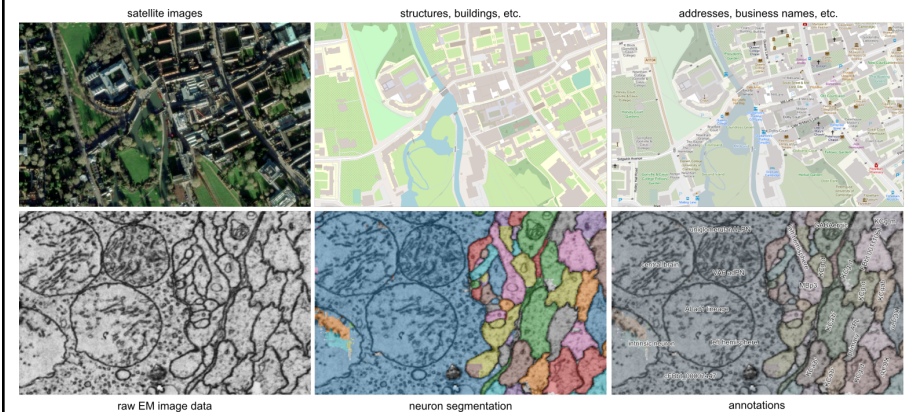

# Methods of reconstructing the brain

Reconstructing brain structure at resolutions necessary to derive the connectome pushes the boundaries of imaging techniques. On a high level, various imaging approaches use electron beams and various forms of electromagnetic waves, ranging from hard X-rays to visible light, and use one or multiple tissue processing techniques such as staining, tissue expansion, or molecular annotations. All static brain structure imaging methods perfuse the brain with chemicals to stabilize its structure during euthanization, after which it is carefully extracted and undergoes subsequent tissue preparation steps. The purpose of such tissue processing varies from increasing visibility of important structures to compensating for the resolution limitations of imaging techniques like X-ray or standard light microscopy.

The figure below demonstrates how quickly even rare errors tracing neurons across the vast number of images can interfere with the successful reconstruction of a neuron (a 1mm long axon imaged with the above-mentioned axial resolution of 20 nm comprises 50,000 images). Tissue loss and damage increase the difficulty of following neurons across different sections during final reconstruction. Even the brains of relatively small organisms must be sliced to arrive at processable sizes that match equipment capacity, diffusion speeds, and parallelization across multiple machines. So far, 1 mm-thick samples have routinely demonstrated robust results. However, cutting brain tissue into ~1 mm slices with "ultra-smooth" vibratomy becomes increasingly difficult with bigger brains.

As we will see in the following chapter, tracing neurons via morphology relies on the high resolution and subsequent time-consuming reconstruction and error correction. If it were possible to uniquely identify the same neuron at its soma and far end, that would loosen the hampering constraint of requiring virtually error-free tracing. This is the idea of barcoding: place a uniquely identifiable molecule in each neuron that can unambiguously determine the identity of each cell. Then, even if it were impossible to trace a neuron continuously through tissue, the expressed barcode would allow correct identification of distal synapses. Loss of an entire section or slicing errors would be much less catastrophic than in any morphologic approach, as barcoding would not rely on perfect traceability.

Barcodes can come in many forms, and scientists are becoming increasingly creative, generating more ambitious variants.

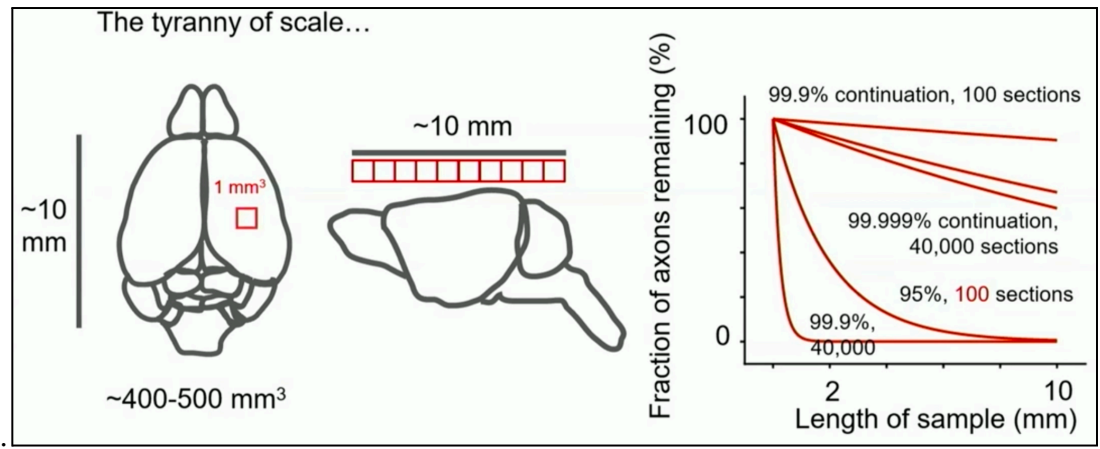

**Replication of the Tyranny of Scale Figure from the Brain Connects Workshop Series.** Even at relatively low error rates, tracing accuracy deteriorates given high numbers of imaging sections with large brain volumes.

# Structural imaging approaches

Morphology staining approaches are the most common method in connectomics. Tissue sections are completely stained to highlight morphological features, particularly cell boundaries (membranes) and protein densities. The advantage is that "everything" is visualized, and the technique is general to any tissue type. The disadvantage is that imaging at a high enough resolution to resolve desired features is time-consuming and data-intensive. Electron microscopy, connectomics and X-ray approaches typically use osmium or other heavy metal stains, whereas antibody- or lipid-based morphology staining is used in light microscopy.

## Electron microscopy

Electron microscopes use electron beams that are accelerated through a high voltage and are precisely guided by electromagnetic lenses to focus on a sample with nanometer precision. As these electrons interact with the sample's atoms, various signals are generated – such as secondary electrons, backscattered electrons, or transmitted electrons – depending on the type of microscope. Specialized sensors detect these signals, which are then processed and transformed into highly detailed images revealing the sample's structure, composition, and even atomic arrangements. For Volume Electron Microscopy (vEM), series of such 2D images are first acquired and then computationally reconstructed into 3D volumes. The imaging techniques for acquiring these series primarily fall into two categories:

Transmission Electron Microscopy (TEM)-based and Scanning Electron Microscopy (SEM)-based approaches.

TEM-based vEM techniques, such as serial section TEM (ssTEM), involve physically cutting the biological sample into a series of ultra-thin sections (typically 30-100 nm thick). Each section is collected on a support grid and imaged individually in a TEM, where the electron beam passes through it to form a 2D projection image. These serial 2D images are subsequently aligned and stacked to reconstruct the 3D volume. TEM generally offers excellent lateral resolution due to the thinness of the sections and the physics of electron transmission, but its axial resolution is inherently limited by the physical thickness of each section. Advances like GridTape TEM ([Phelps et al., 2021](#)) further automate section handling and imaging for ssTEM, significantly increasing throughput ([Peddie et al., 2022](#)).

SEM-based vEM techniques typically involve iteratively imaging the surface (or "block-face") of a sample within the SEM, then removing a thin layer to expose a new surface for the next image. Several approaches exist. Focused Ion Beam SEM (FIB-SEM) uses a focused ion beam (e.g., gallium) to ablate or "mill" away very thin layers (typically 5-20 nm) from the sample surface. After each layer's removal, the newly exposed block-face is imaged by the SEM using backscattered or secondary electrons. This process allows for very high axial resolution, making isotropic voxels achievable. For large-volume acquisitions, where milling proceeds over significant depths (e.g., hundreds of micrometers), the milling front can become uneven; periodic replanarization steps are often necessary to re-flatten the sample surface and maintain a consistent cutting plane. Serial Block-Face SEM (SBF-SEM) employs an ultramicrotome (a diamond knife) integrated directly inside the SEM chamber. The knife cuts a thin section (typically 25-100 nm) from the block-face, which is discarded, and the newly exposed surface is then imaged. SBF-SEM is generally faster for very large fields of view compared to FIB-SEM but offers lower axial resolution. Array Tomography (AT) involves first cutting an entire series of ultra-thin sections, collecting them in an ordered array (e.g., on a silicon wafer or specialized tape), and then imaging them sequentially in an SEM. While its z-resolution is limited by section thickness (e.g., 30-50 nm), AT uniquely allows for post-section staining, re-imaging of regions of interest, and is well-suited for correlative light and electron microscopy. For very large tissue samples, they may first be subdivided into more manageable "slabs" or ribbons, for instance using a hot knife, before serial sectioning for AT ([Peddie et al., 2022](#)).

To accelerate the inherently slow process of SEM imaging, multi-beam SEM (mSEM) systems have been developed. As early as 2015, Zeiss produced 61-beam SEMs capable of approaching GHz data acquisition speeds ([Zeidler et al, 2015](#)). Current state-of-the-art devices feature 91 beams ([Riedesel et al., 2019). In mSEM, the primary electron beam is split into multiple sub-beams, each scanning a small region of the sample in parallel to simultaneously generate images of the underlying tissue. These individually captured high-resolution images are then computationally stitched together to form a larger composite image. While mSEMs can achieve burst imaging speeds of up to 3.6 GHz (e.g., ~40 MHz per beam for a 91-beam system), the effective rate, when projected over 24/7 scanning operations considering factors like downtime and sample exchange, typically falls to 100-200 MHz. In this context, 1 Hz is equivalent to imaging one voxel per second. To put this in perspective: scanning a 500 mm³ volume of mouse brain (approximately 500 petavoxels at 10nm isotropic resolution) would take about twenty 91-beam EM systems roughly four years, assuming continuous 24/7 scanning at an effective imaging rate of 200 MHz per system.

Only a handful of facilities globally possess the multiple, high-throughput vEM systems required for such large-scale endeavors. Setup costs for a single advanced vEM instrument range between $0.5 million and $10 million, with similar maintenance costs over 3-5 years. Depending on uptime and specific operational context, costs per hour can range between $1,000 and $5,000.

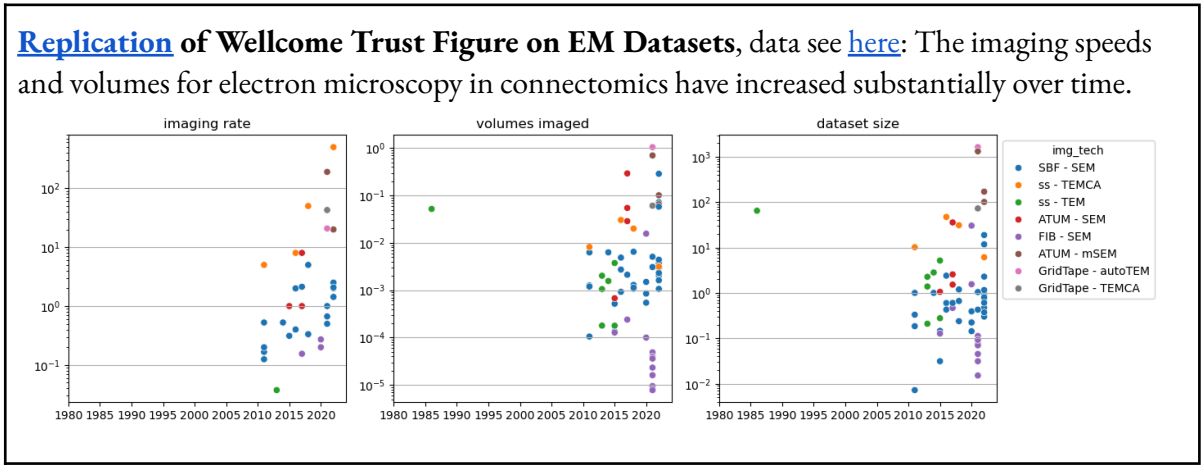

Replication of Wellcome Trust Figure on EM Datasets, data see here: The imaging speeds and volumes for electron microscopy in connectomics have increased substantially over time.

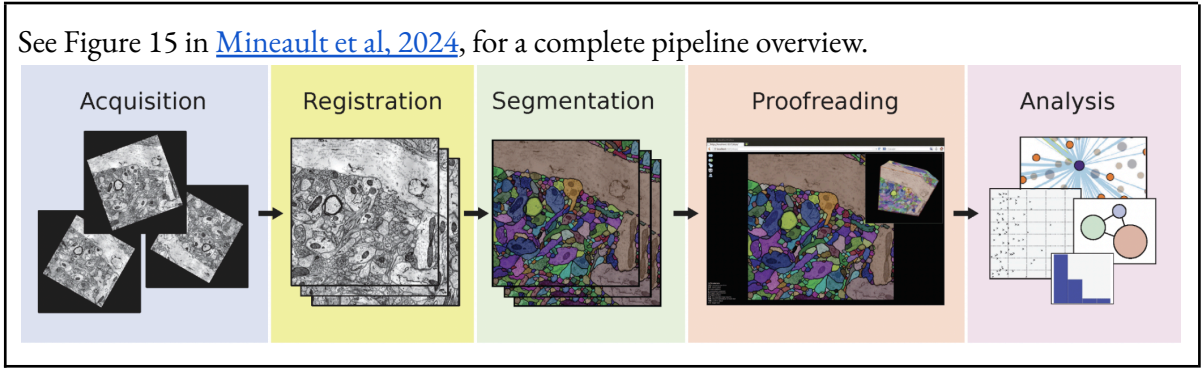

See Figure 15 in Mineault et al, 2024, for a complete pipeline overview.

## Expansion Microscopy with Dense Labeling

In 2015, Chen et al. introduced expansion microscopy (ExM), a revolutionary technique achieving initial resolutions as fine as 70 nm. Electron microscopy (EM) is constrained by physics to a resolution limit of 0.23 nm ([Penczek, 2011](#)). In contrast, traditional light microscopy is constrained to a lateral resolution of ~250 nm and an axial resolution of ~550 nm by the diffraction limit ([Huang et al., 2010](#)). Expansion Microscopy effectively circumvents the diffraction limit by isotropically physically expanding the tissue sample rather than attempting to improve the microscope's resolving power. Since then, multiple authors have demonstrated the dense labeling of protein ([M'Saad and Bewersdorf, 2020](#); [M'Saad et al., 2022](#)) and lipid components ([Karagiannis et al., 2019](#); [Shin et al, 2025](#)).

ExM pipelines physically expand biological samples by embedding them in swellable hydrogels (see figure). Routine expansion protocols can expand brains between 4- and 16-fold, while recent iterative and non-iterative expansion advances allow expansion up to 40-fold and beyond. Brain tissue sections (typically 50-70 µm thick) are first fixed, then embedded in a first hydrogel. After denaturation, the sample expands in water. The sample is then re-embedded in a second neutral gel, followed by a third expansion gel with a non-cleavable crosslinker. During this process, proteins can be labeled with antibodies and pan-stained with fluorescent dyes to reveal ultrastructure. For instance, the final expansion described by M'Saad et al. achieves approximately 24-fold total enlargement. This enables the resolution of features as small as ~15 nm in the pre-expanded sample using standard confocal microscopes, effectively bypassing the diffraction limit ([M'Saad et al., 2022](#)). The entire process takes 4-5 days from initial fixation to final imaging, with the key advantage that no specialized equipment, such as expensive electron microscopes, beyond a standard confocal microscope, is needed.

> **Figure: Expansion Microscopy Process.** Illustration of stepwise processing of brain tissue in the context of expansion microscopy. Credits to Eon Systems PBC

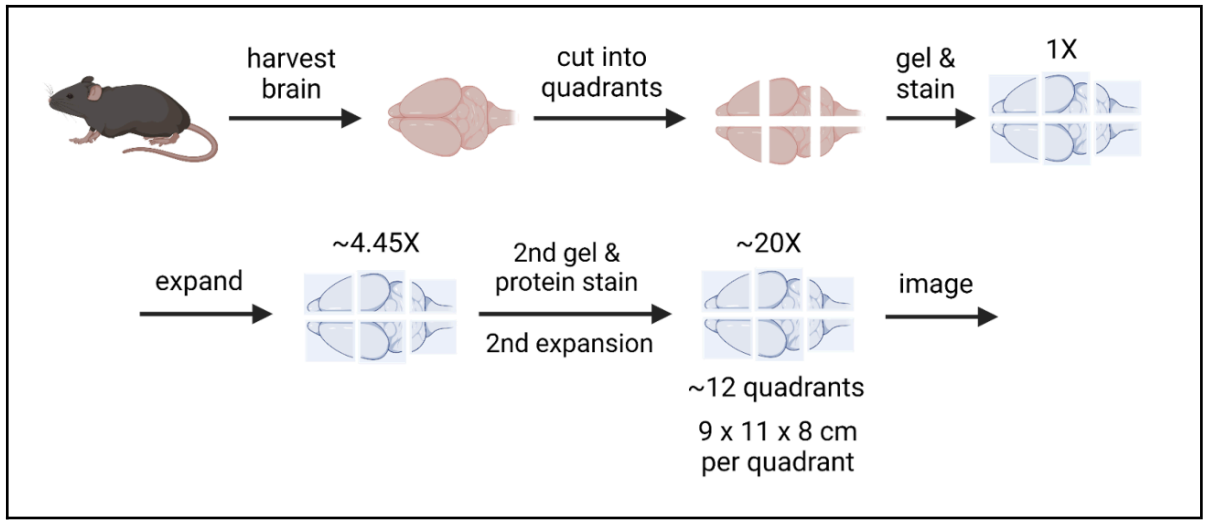

Figure by M'Saad et al., 2022: pan-ExM-t workflow for mouse brain tissue sections. (a-f) Experimental workflow. (g): Timeline summarizing the protocol. Abbreviations: FA: formaldehyde; AAm: acrylamide; NaOH: sodiumhydroxide; DHEBA: N,N′-(1,2-dihydroxyethylene) bis-acrylamide; SDS: sodium dodecyl sulfate; PBS-T:0.1% (v/v) TX-100 in PBS; ROI: region of interest.

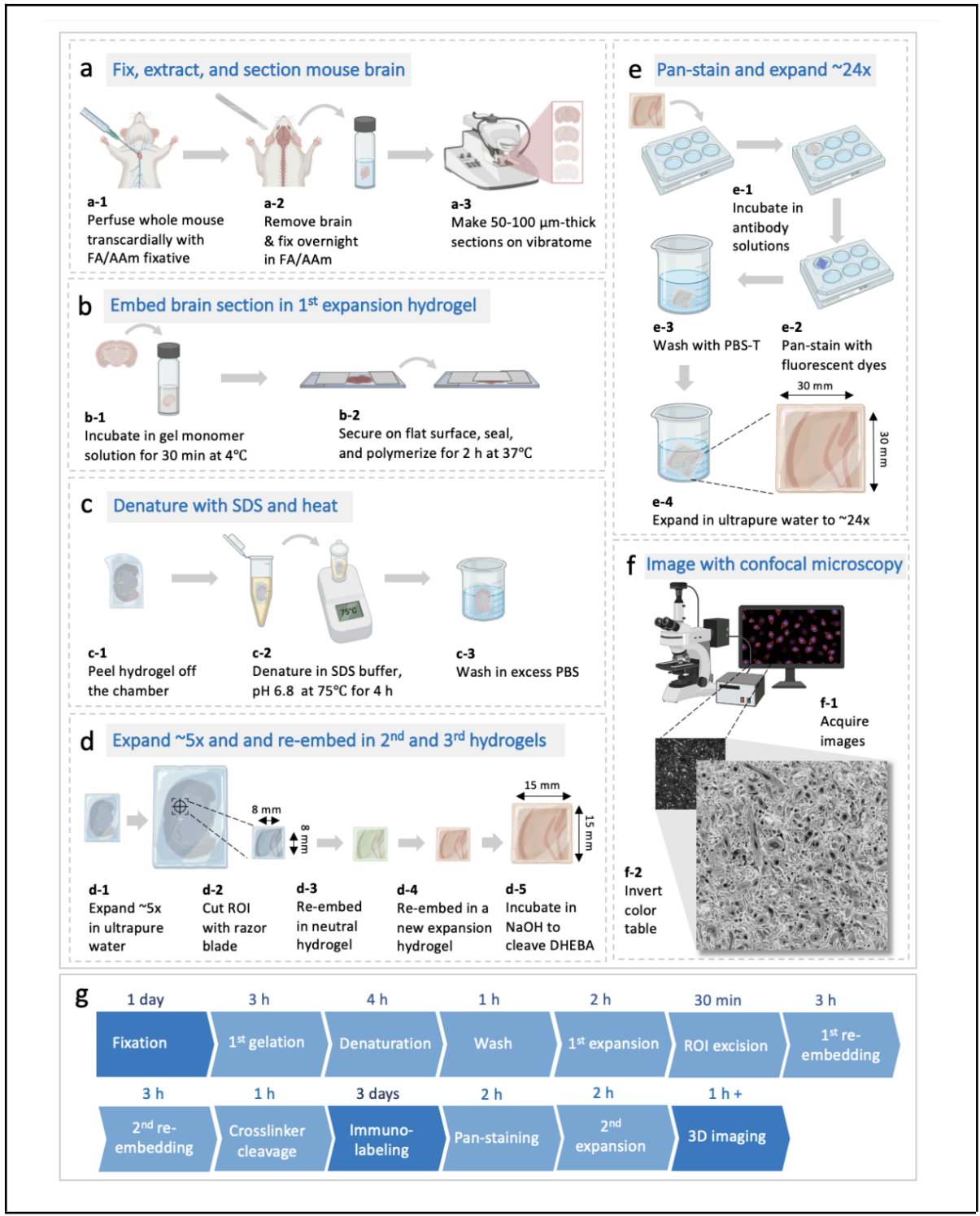

Expansion microscopy is beginning to bridge the gap between the molecular specificity of light microscopy and the synaptic resolution required for dense connectomics. Two complementary dense-labeling strategies have been demonstrated in brain tissue for ExM-based ultrastructural contrast: (i) pan-protein labeling (pan-ExM/pan-ExM-t) that produces dense, EM-like protein-density contrast and is compatible with immunostaining (M'Saad et al., 2022), and (ii) dense, continuous membrane labeling (umExM) for nanoscale visualization of membranes in intact tissues (Shin et al., 2024).

A recent proof-of-principle for dense light-microscopy connectomics is LICONN (Tavakoli et al., 2025). The method uses pan-protein labeling with iterative expansion, combined with a standard spinning-disk confocal microscope for imaging and readout. In mouse cortex and hippocampus samples, LICONN demonstrated an effective optical resolution of ~20 nm laterally and ~50 nm axially (~9.7 × 9.7 × 25.9 nm³ voxel size), and an effective imaging throughput of ~17 MHz (~0.001 mm³ imaged over 6.5 hours). For reconstruction, Tavakoli et al. adopted a pipeline directly from EM connectomics, using automated segmentation with flood-filling networks (FFNs) followed by manual proofreading, yielding reconstruction accuracy comparable to state-of-the-art EM pipelines (Tavakoli et al., 2025).

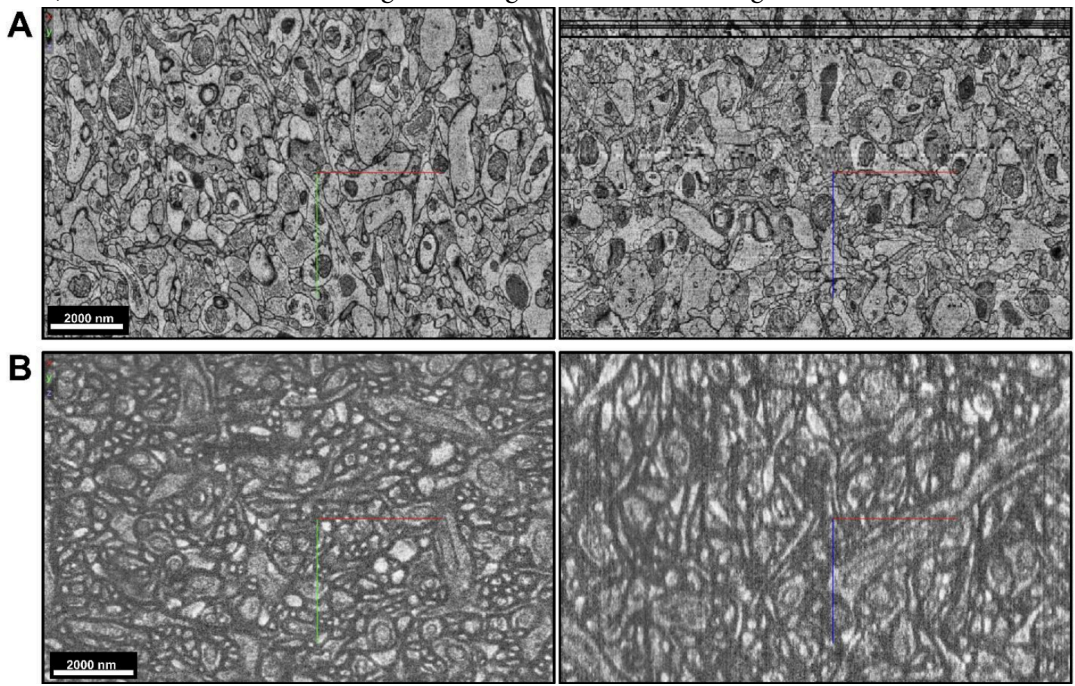

**Replication of Figure 1 in Collins et al., 2024**. Comparison of raw image data in serial section transmission electron microscopy (top) and expansion confocal light microscopy (bottom). Note the 1-2 order of magnitude higher resolution. Image from Collins et al.

Unlike EM pipelines, LICONN natively supports molecular annotation in the same high-resolution volume. Presynaptic and excitatory postsynaptic markers (e.g., bassoon and SHANK2) and inhibitory postsynaptic markers (gephyrin) are imaged together with the dense structural channel, enabling direct classification of synapses. Importantly, astrocytic connexin-43 labeling reveals gap junctions, allowing electrical coupling to be mapped at scale -- something that is typically difficult to recover with EM alone. Furthermore, because the structural and molecular channels are perfectly aligned, the data can be used to train deep learning models that predict molecular locations from the structural channel alone. The authors demonstrated this by training predictors for presynaptic and postsynaptic markers, achieving high accuracy on held-out data. This strategy can significantly reduce the number of staining rounds required when scaling up imaging to larger volumes, as the structural map can be used to infer select features.

Throughput is the main limitation at present. LICONN's ~17-MHz rate is well below multi-beam EM, but those EM throughputs depend on scarce, multi-million-dollar systems; by contrast, the LICONN result came from a standard, relatively inexpensive spinning-disk setup. The current throughput is thus a function of the chosen readout, not a fundamental limit of the sample preparation, which is not tied to any specific imaging modality. The clear path to scaling, therefore, involves pairing LICONN-like sample preparation with optics designed for high-speed volumetric imaging. Such a combination could achieve EM-competitive throughputs while preserving the molecular annotation that EM lacks. This potential to match EM speed and resolution on accessible equipment makes dense ExM a compelling route to scalable molecularly annotated connectomics.

## X-ray synchrotron approaches

X-ray synchrotron approaches use high-energy X-rays to visualize tissue architecture and cellular features. The method can leverage the natural contrast between different cellular components based on their electron density and elemental composition for minimal preparation requirements. It can also use heavy metal stains like osmium or gold to enhance the contrast of specific features, particularly membranes and synaptic proteins. These stains also increase acquisition time (Ahn et al., 2013; Depannemaecker et al., 2019). Special chemicals can be leveraged to increase radiation resistance of samples (Bosch et al., 2023). Imaging is also generally non-destructive, meaning the same sample can be imaged through other modalities.

Spatial resolutions – not limited by diffraction, but rather by the x-ray cross section of elements with low atomic numbers in unstained samples and, more generally, by the risk of damage caused by high x-ray radiation doses – down to roughly 10 nm are theoretically possible in frozen hydrated samples (Howells et al., 2009). The primary advantages are that the high penetration depth of X-rays allows for imaging of large tissue volumes (several mm³) at resolutions approaching 30 nm ([Stevens et al., 2020](#); [Du et al., 2021](#)).

The drawbacks are also multiple, however. X-ray approaches for structural brain mapping are far less developed than their EM and light-microscopy counterparts. This is partly because imaging generally requires access to a beamline at a synchrotron facility, with beamlines heavily oversubscribed. Scalability-wise, although simulations are encouraging, significant progress is needed, particularly in improved detectors, before scalability could be sufficient for imaging whole mammalian brains ([Collins, 2023](#); [Du et al., 2021](#)). Recent developments include sub-100 nm imaging and dense reconstruction of fly and mouse brain tissue using x-ray holographic nano-tomography ([Kuan et al., 2020](#)), correlative studies involving in-vivo recordings, x-ray synchrotron microtomography and electron microscopy ([Bosch et al., 2022](#)), the establishment of the SYNAPSE consortium for imaging a whole human brain at 300 nm resolution ([Stampfl et al., 2023](#)), the development of protocols for highly multiplexed x-ray fluorescence ([Strotton et al., 2023](#)) and successful imaging of individual synapses through x-ray ptychography ([Bosch et al., 2023](#)).

> For a visualization of X-ray synchrotron approaches, see Figure 1 in [Dyer et al., 2017](#). Due to copyright constraints, we cannot replicate this image here.

## Barcoding approaches

Protein barcoding for connectomics emerged from the convergence of two key technological advances: site-specific DNA recombination and fluorescent protein engineering. The watershed moment came with Brainbow in 2007, which leveraged Cre/lox recombination to stochastically express different ratios of fluorescent proteins in individual neurons. This approach achieved roughly 100 distinguishable color combinations through the differential expression of red, green, and blue fluorescent proteins. The initial proof-of-concept in mice demonstrated the potential for unique cellular labeling, though it also revealed fundamental challenges in achieving consistent expression levels and maintaining color fidelity across large tissue volumes. Expression in different cell types is uneven, trafficking of these proteins is uneven, and fluorescent staining is uneven. A distal axon may display a slightly different barcode than the soma within the same neuron, complicating correct identification. Subsequent iterations expanded the technique to other model organisms, notably *Drosophila*, while attempting to address these limitations through improved fluorescent proteins and more sophisticated genetic designs ([Livet et al., 2007](#), [Pan et al., 2011](#), [Cai et al., 2013](#), [Leiwe et al., 2024](#))

The most advanced approaches combine three key elements: genetic targeting, protein-based labeling, and multi-round imaging. In current protocols, viral vectors deliver genetic constructs encoding multiple protein markers, each under the control of an independent promoter. These markers can be fluorescent proteins, epitope tags, or engineered protein scaffolds designed for subsequent antibody labeling (see figure by [Serrano, 2022](Serrano, 2022)). The overall process requires multiple rounds of staining, washing, and imaging.

Such barcoding approaches face three distribution problems when delivery is carried out via viral vectors:
1. Distribution to all neurons.
2. Even distribution among all neurons.
3. Distribution within neurons.

> For a visualization of barcoding, see Figure 1 in [Serrano, 2022](Serrano, 2022). Due to copyright constraints, we cannot print the figure here.

In 2020, Shen et al. combined Brainbow with multi-round immunolabeling in expansion microscopy, which allowed the unique identification of neurons and reconstruction of their structure, including imaging up to 15 different stained targets ([Shen et al., 2020](Shen et al., 2020)). Notably, super-multicolor Tetbow allows reconstruction despite two sections of a neuron being separated, beginning to leverage the core strength of barcoding ([Leiwe et al. 2024](Leiwe et al. 2024)). In principle, methods like Tetbow could scale to whole-brain connectomics when combined with expansion microscopy, pan-protein, and/or lipid staining.

However, achieving whole-brain connectomics using barcoding is not currently possible. The focused research organization (FRO) [E11](E11) is actively working on advancing barcoding technologies for whole mammalian brains.

The theoretical limits of protein barcoding intersect with fundamental biological constraints and technical capabilities. In mammalian brains, where individual neurons form between 8,000 and 30,000 synapses (though individual connections between neurons can consist of up to 60 synapses or more – even if a neuron makes 10,000 synapses with other neurons, it may connect to only a few thousand unique neurons), the mathematics of unique identification become particularly challenging. With a 25-bit binary code ($2^{25}$ combinations), statistical analysis reveals that in a human brain of 80 billion neurons, approximately 10 million neurons (0.02%) would share a barcode and at least one synaptic connection. While this error rate might seem problematic, it compares favorably with current electron microscopy reconstruction error rates, particularly considering that spatial information can help resolve ambiguous cases. It is important to note that whether state-of-the-art barcoding methodology can capture most neurons and especially most synapses will need to be verified in statistical comparisons against "ground-truth" electron microscope data in small, similar reconstructed samples.

Reaching almost all neurons with the viral vectors carrying the barcodes is difficult. This will require further breakthroughs in delivery, such as the intravenous bCap1 AAV capable of reaching 5-20% of cells in the brain. ([ref](#)) Various virus-based platforms (lyssavirus, Sindbis, HSV, etc.), better intravenous delivery, high-density intra-CNS injections, and creative approaches to multi-site injection throughout the brain will be required – or perhaps all four simultaneously.

Notably, nucleic acid barcoding approaches exist, where unique RNA/DNA barcodes (typically 15-30 nucleotides) are introduced into neurons using viral vectors or genetic engineering and then use sequencing as a readout. The field emerged from spatial transcriptomics methods developed in the 2010s. The transition to connectomics applications began with MAPseq ([Kebschull et al., 2016](#)), which introduced high-throughput projection mapping, tracing where neurons from one brain region send their axons, without detailing individual synaptic connections. For connectivity attempts (BRICseq): Although not yet working at scale, there are current efforts to barcode neurons and transform connectomics from an imaging problem to a sequencing and analysis problem ([Huang, 2020](#)). Most recently, ExBarSeq ([Goodwin, 2022](#)) combined barcoding with expansion microscopy to improve spatial resolution to ~20 nm. However, these methods remain more powerful for projection mapping than detailed connectivity analysis. Current throughput for projection mapping reaches 100,000+ neurons per experiment. BARseq, for instance, employs in situ sequencing of both endogenous mRNAs and synthetic RNA barcodes to infer long-range projections of neurons across whole mouse forebrain hemispheres, analyzing millions of cells ([Chen et al, 2024](#)). In contrast, detailed connectivity mapping remains limited to hundreds of neurons due to challenges in reliably getting barcodes to and reading them from synapses.

# Data storage and processing

## Storage & Bandwidth

While high-throughput imaging of mammalian brains demands extraordinarily effective bandwidths, existing technology can largely meet these requirements, especially when combined with modern compression methods.

A human brain is roughly 1000-1400 cm³, and such a volume, if imaged at 10 nm isotropic resolution, represents roughly 1.2 zettavoxels. Assuming 2 bytes per voxel and a total acquisition time of about a year, grayscale imaging thus requires a total effective bandwidth of about 600 terabits/s. Multiplexed imaging scales bandwidth requirements linearly with the number of colors; thus, streaming 30-color imaging data requires a total bandwidth of roughly 18 petabits/s. However, state-of-the-art compression methods already demonstrate a 128x reduction in data size without compromising subsequent reconstruction (Li et al., 2024). Thus, data compression at the source could see practical bandwidth requirements decrease to about 140 terabits/s for 30-color imaging and about 5 terabits/s for grayscale imaging. Beyond these initial compression approaches, as our understanding of neural tissue matures, additional representations such as connectivity graphs and neuron skeletons could complement the detailed data, potentially offering further significant reductions in storage requirements. However, for early whole-brain reconstructions, preserving sufficient raw data to resolve subcellular structures, including synaptic vesicles and cellular organelles, will likely remain essential, as it is not yet clear which biological details are essential for faithful emulation.

Bandwidth requirements thus represent an important challenge, particularly if no compression is used. However, they are not far beyond what existing technology enables in fields such as high-energy physics. Indeed, experiments within the LHC at CERN can produce roughly a petabyte (that is, eight petabits) of raw data per second, with various filtering and compression algorithms within the sensor decreasing the effective data rate to a more manageable 50 terabytes (400 terabits) per second (Radovic et al., 2018). Such high data rates, of course, require highly performant networking infrastructure, but that too is not far beyond current capabilities: Google data centers, for example, have demonstrated within-datacenter bandwidths of up to 13 petabits/s (Vahdat, 2024). As long as compression is used and processing happens mostly within a data center co-located with the imaging facility, imaging bandwidth should thus not represent a fundamental bottleneck to efforts aiming to image a whole human brain in roughly a year, provided the level of investment is sufficient.

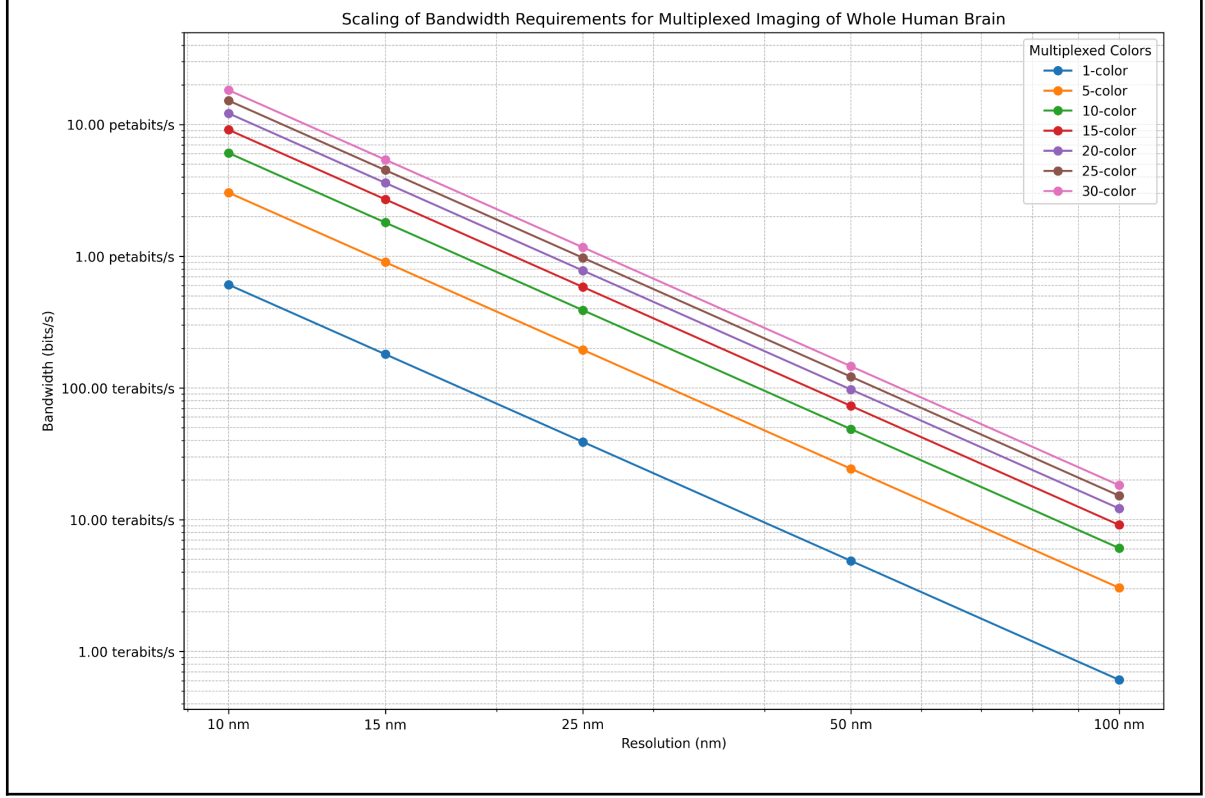

**Figure Bandwidth Requirements by Resolution and Channel Count.** Effective total bandwidth requirements for human brain imaging in less than one year at 10 nm isotropic resolution. (over one year)

Depending on the compression factor, the data storage necessary for a human connectome would be on the order of 1 Peta to 100 Exabytes, with an additional 2-3x factor to accommodate for backup infrastructure. Petabyte storage capacity is common for small to mid-sized data centers. Most large data centers operated by commercial providers do not disclose their official capacity, but can store up to several exabytes ($10^{18}$) of data ([Blackblaze, 2024](#)). The European Centre for Medium-Range Weather Forecasts (ECMWF) generates 400 terabytes of new data daily and processes this with previous dates to make predictions ([ECMWF, 2024](#)). Companies like ByteDance process over 500 PB/day ([Wu et al., 2024](#)) while the most prominent actors, such as Google, Meta, or Microsoft, do not disclose official numbers. For 2025, the total globally available digital storage capacity is estimated to be around 16 zettabytes ([Statista Inc., 2021](#)).

# Neuron Reconstruction

Neuron reconstruction aims to recreate the three-dimensional connectome from the vast quantities of microscopy data produced through imaging. This complex process generally involves three main stages: data preparation (including registration), automated segmentation and tracing, and meticulous proofreading. Each stage presents unique challenges and opportunities for automation.

The first critical step is data preparation, where raw image data is processed for downstream analysis. A key bottleneck here is registration: aligning adjacent scans to ensure spatial consistency of neural structures. Misalignments, often caused by imaging artifacts or slight sample movements, are a dominant cause of errors in subsequent automated neuron reconstructions ([Popovych et al., 2024](#)). Poor alignment can make it impossible for algorithms to correctly follow fine neurites across section boundaries. Fortunately, recent advancements in machine learning have significantly improved alignment accuracy, with algorithms now addressing artifact removal, de-warping, and noise reduction, in some cases reducing genuine misalignments to as low as 0.06% of image pairs ([Popovych et al., 2024](#); [Scheffer et al., 2020](#)).

Once the data is prepared, the process moves to segmentation of cell boundaries and tracing of neuron skeletons (axons and dendrites). This is how the initial 3D shapes of neurons and their potential connectivity are computationally derived. Historically, manual tracing and segmentation were the only options, but these are prohibitively costly for large volumes, averaging around 11.2 hours per neuron in early dense reconstructions ([Zheng et al., 2018](#)). Modern approaches heavily rely on automated techniques. Convolutional neural networks, such as U-Nets, can learn from relatively sparse training data to perform tasks like segmentation and synapse classification. For dense segmentation, a range of architectures – including U-Nets ([Lee et al., 2017](#)), flood-filling networks ([Januszewski et al., 2018](#)), and embedding-based ones ([Lee et al., 2022](#)) – have become state-of-the-art, achieving high precision in identifying which voxels belong to the same neuron. While some of these methods can be computationally intensive, ongoing architectural modifications and algorithmic refinements continue to improve efficiency and performance across the board.

Despite the capabilities of current automated segmentation algorithms, errors inevitably occur. These can be "split" errors (a single neuron incorrectly broken into multiple pieces) or "merge" errors (different neurons incorrectly joined). Correcting these errors requires proofreading, which currently represents the most significant time and cost bottleneck in EM-based connectomics, especially for dense reconstruction of complex mammalian neurons from large-scale datasets like serial section TEM (ssTEM). The human labor involved in correcting such errors is substantial. For instance, proofreading the local axonal arbor of a single mouse cortical pyramidal neuron – often considered a benchmark for the most challenging structures to trace – from a petascale EM volume (e.g., the MICrONS dataset) is estimated to take approximately 40 person-hours (T. Macrina, personal communication, 2025). This translates to direct proofreading costs of around $400 per neuron (assuming outsourced proofreader costs of $10/hr), though the experienced cost for a user engaging a reconstruction service can be higher, in the range of $500-$1000 per neuron. It's important to note these estimates often benchmark against challenging neuron types like pyramidal cells (which have extensive and fine axonal arbors) and typically focus on local arbors, with costs potentially varying for simpler neurons or different reconstruction targets. The quality of the initial imaging and automated reconstruction also heavily influences proofreading effort; imaging defects, such as missing sections, are a primary cause of errors that necessitate manual correction.

The scale of this challenge is significant: given current proofreading costs, the total proofreading expenditure for an entire mouse brain could reach upwards of $1 billion ([Jefferis et al., 2023](#)). It's also worth noting that the difficulty of proofreading also varies across species; for example, researchers report that proofreading mammalian neurons is significantly more challenging than for *Drosophila*, due to factors like larger cell sizes and more complex morphologies. Reflecting this, only about 2% (e.g., ~1700 neurons as per version v1300) of the ~100,000 neurons in the 1 mm³ MICrONs dataset has been fully proofread to date, and similarly, only a small fraction (e.g., ~100 neurons) of the about 16,000 neurons in the H01 human temporal lobe dataset has been proofread.

Given these costs, the primary focus for making large-scale EM-based connectomics feasible is to drastically improve the automation of error correction, effectively augmenting and reducing the need for manual proofreading. The field is seeing rapid improvements here. Proofreading efforts have already been significantly augmented by tools that provide machine-suggested edits, allow users to define regions of interest, and utilize morphology libraries for comparison, speeding up the process by orders of magnitude in some contexts ([Scheffer et al., 2020](#); [Plaza, 2016](#)). Experts anticipate that new AI-driven tools could further reduce proofreading costs for complex cases like mouse pyramidal neurons to below $100/neuron in the near future. While there hasn't been a recent isometric petascale dataset for direct comparison with MICrONS, improvements in imaging and reconstruction methods are evident. For example, in a 1 x 1 x 0.1 mm³ ssTEM dataset of mouse hippocampus, a 5-fold improvement in proofreading efficiency for CA3 pyramidal axons has been observed (T. Macrina, personal communication, 2025). This suggests that the cost of proofreading a cortical pyramidal axon from a new MICrONS-like dataset could potentially be reduced to around $80.

Indeed, the pace of improvement may be accelerating even faster than such estimates suggest. A recent pre-print introduced PATHFINDER, an AI system that reportedly achieves an 84-fold increase in proofreading throughput on high-quality IBEAM-mSEM imaging data (Januszewski et al., 2025). The method's strength lies in a multi-stage process that first generates numerous potential neuron assemblies and then uses a separate model with a larger field of view to evaluate their morphological plausibility. Such a leap, if validated and generalized across different datasets and imaging modalities, would represent a step-change in the economics of connectomics, moving the field significantly closer to making whole-brain projects tractable.

This trajectory of rapid improvement suggests the field may be outpacing prior forecasts. Results like those from PATHFINDER significantly reduce proofreading costs for complex cases like mouse pyramidal neurons. Besides model architecture improvements, another important component of developing these more advanced automated proofreading tools is the generation of more high-quality, human-verified ground truth data. The very act of proofreading, while currently expensive, produces the training data needed to improve the next generation of AI models for both initial segmentation and subsequent error correction. This creates a virtuous cycle where human effort refines AI, which in turn reduces future human effort.

The overarching goal remains to minimize the human burden in connectome reconstruction, rendering the mapping of entire brains, including complex mammalian ones, economically and logistically viable. Continued advancements in AI are poised to dramatically reduce, and perhaps largely automate, the intensive proofreading of EM datasets. AI's general trajectory provides a strong basis for optimism in this direction, turning the current high-cost, labor-intensive process into a more manageable one. After all, AI has approached, achieved, or surpassed human-level performance on many tasks, including chess (Campbell et al., 2002), Go (Silver et al., 2016), reasoning (OpenAI, 2024), parts of mathematics (Trinh et al., 2024), software development (Schluntz et al., 2024), natural language (Brown et al., 2020), reading comprehension (Rae et al., 2021), and visual reasoning (OpenAI, 2024). However, even if purely algorithmic solutions for EM data face persistent challenges in fully eliminating the human proofreading burden, emerging technologies might nonetheless dramatically reduce or even sidestep entirely the need for exhaustive proofreading. Indeed, as discussed above, techniques integrating expansion microscopy with high-plex barcoding aim to assign unambiguous molecular identities to neurons and their fragments. This 'self-proofreading' capability, where molecular data resolves ambiguities intractable from morphology alone, offers a powerful route to sidestep the most laborious aspects of current EM-centric pipelines. The successful maturation of either AI for EM workflows or these alternative molecular connectomic techniques thus promises to enable much more affordable and scalable neuron reconstruction.

It is important to note, however, that creating a precise historical trendline of these improvements is not currently feasible. Direct comparisons between methods published over time can be misleading due to various confounders, such as the amount of compute used, differences in the underlying data quality, and variations in reconstruction targets. To better track progress, the field would benefit from the systematization and standardization of reporting metrics in the future (e.g., errors per μm³, hours of human labor per μm³, FLOPs/μm³). A collection of relevant references can be found in the accompanying data repository.

# Computational Neuroscience: Simulating Brains

Building on very early neuronal models like Lapicque's *Recherches quantitatives sur l'excitation electrique des nerfs* ([Lapicque, 1907](#)), the quest to formally understand and potentially replicate neural computation gained significant momentum in the mid-20th century. Foundational theoretical work by McCulloch and Pitts ([McCulloch and Pitts, 1943](#)) proposed a mathematical model of simplified neurons, demonstrating how networks of such units could perform logical computations. Inspired by groundbreaking work such as Edgar Adrian's on the electrical impulses of neurons (see the Chapter

Neural dynamics), and the detailed biophysical modeling of the action potential by Hodgkin and Huxley ([Hodgkin and Huxley, 1952](#)), the idea of engineering brain-like computation emerged. Despite the limitations of the era's technology (vacuum tubes and analog circuits), Marvin Minsky and Dean Edmonds rose to the challenge by constructing the Stochastic Neural Analog Reinforcement Calculator (SNARC) in 1951. This machine, powered by vacuum tubes and motorized potentiometers, was a pioneering hardware system designed to embody a multi-neuron network capable of learning through reinforcement. Though SNARC itself was never formally published and remained relatively obscure compared to the impactful theoretical work of McCulloch and Pitts, the biophysical insights from Hodgkin and Huxley, or later influential developments like Rosenblatt's Perceptron ([Rosenblatt, 1957](#)), it nonetheless demonstrated that even a modest array of physical neuron-like units could adapt connections to solve tasks, hinting at the potential later realized in artificial neural networks.

Since then, countless researchers have followed in Minsky and Edmonds' footsteps, striving to replicate or simulate neural systems in increasingly sophisticated ways. These simulations vary widely in their scope, from the type of simulation (e.g., spiking vs. rate-based models) to the number of neurons and synapses, the hardware and software used, and even the computational cost required to simulate one biological second. However, simulating neurons and synapses is only part of the challenge. For brain models to demonstrate meaningful behavior, they must interact with their environment. This requires embodiment – the ability to encode external stimuli into neural activity and decode that activity into actions. Embodiment does not necessarily demand a physical robotic body; it can occur in a virtual environment, where sensory inputs and motor outputs are simulated alongside the brain model. This process mirrors modern brain-computer interfaces or cochlear implants, where information is encoded and decoded into neural inputs and outputs for specific tasks (e.g., seeing, hearing, memorizing, or speaking). Over time, this interaction can improve iteratively, benefiting from adaptive changes in either the emulated brain or the interface.

To explore these ideas further, we will discuss the computational architectures used to simulate neuronal networks and the tools that enable these digital neurons to interact with digital bodies.

# Methods of Emulating the Brain

Computational models of neurons and synapses have become indispensable tools for bridging the gap between partial functional measurements, structural wiring data, and the richly detailed behavior of living brains. As discussed in prior chapters, exhaustive single-neuron resolution recordings are often unattainable in large organisms, and connectomes alone are insufficient to explain circuit function. Neuron models aim to capture how each cell transforms incoming activity into an output, whether that output is a firing rate, a spike train, or a graded voltage response. By fitting a model's parameters to observed neuronal behavior, researchers gain explanatory power and a means to predict how real cells might respond under new conditions.

## Neuron Models

Neuron models aim to capture how electrical or synaptic inputs map onto changes in a neuron's membrane potential or spiking. Some frameworks treat neurons as abstract point units with minimal internal dynamics, whereas others explicitly encode multiple ion channels, compartments, or morphological details. Different modeling choices imply different storage (how many state variables must be updated) and computational costs (how many floating-point operations per millisecond). Simpler models can be integrated quickly (tens of flops/ms) but may omit key spike features. In contrast, more detailed schemes can require hundreds to thousands of flops/ms while capturing richer physiology. A variety of software frameworks exist to facilitate simulations, including NEURON ([Carnevale and Hines, 2006](#)), NEST ([Gewaltig and Diesmann, 2007](#)), Nengo ([Benkolay et al., 2014](#)), NeuCube ([Kasabov et al., 2014](#)), NeuroGrid ([Benjamin et al., 2014](#)), GeNN ([Yavuz et al., 2016](#)), Brian2 ([Stimberg et al., 2019](#)), BMTK ([Dai et al., 2020](#)), and many more.

> For a tabular comparison of different models and their respective computational demands, see Figure 2 [Izhikevich, 2004](#). Due to copyright constraints, we cannot replicate this image here.

The leaky integrate-and-fire (LIF) approach is one of the simplest spiking models. It treats the neuron membrane as a resistor-capacitor circuit, whose membrane potential $u$ integrates an incoming current while spontaneously decaying (the "leak") towards its resting potential; if the membrane potential exceeds the firing threshold, the neuron spikes. The membrane then resets to a resting voltage. This scheme introduces only one differential equation (plus a rule for thresholding and reset), and captures fundamental aspects of spiking behavior. Its computational lightness makes it a staple for large-scale network simulations, although it omits ion-channel kinetics or dendritic geometry ([Skocik and Long, 2014](#)). Because it tracks only one state variable (u) plus a fixed threshold, LIF can simulate a neuron with as few as 4–40 floating-point operations per millisecond in practice ([Izhikevich, 2004](#); [Skocik and Long, 2014](#)). This low cost, however, means less biophysical realism – one trades away phenomena like bursting or variable spike thresholds for speed and minimal memory usage.

An example of a somewhat more general and potentially more biophysically detailed model is the Izhikevich model, which is a more flexible single-compartment system with two coupled state variables, typically *v* for membrane voltage and *u* for a recovery process. Although it lacks direct biophysical details about sodium and potassium currents, it can reproduce many spike-timing patterns – regular, bursting, chattering – through suitable tuning of just four parameters. In practice, Izhikevich neurons are common in spiking network simulations that benefit from more realistic spiking patterns than LIF can provide, but do not require the fully biophysical detail of Hodgkin–Huxley ([Izhikevich, 2004](#); [Skocik and Long, 2014](#)). A caveat is that the model's properties can shift as the time step changes, so care is needed to ensure consistent solutions at different step sizes ([Skocik and Long, 2014](#)). Implementing these two state variables plus four parameters typically requires on the order of 10–40 flops/ms if run coarsely, but can climb to hundreds or thousands of flops/ms for higher accuracy or smaller time steps ([Izhikevich, 2004](#); [Skocik and Long, 2014](#)). In return, modelers gain the ability to represent bursting, rebound spikes, and other complex dynamics that simple LIF neurons cannot capture.

In contrast to these more phenomenological approaches, the Hodgkin–Huxley model preserves explicit ion-channel kinetics derived originally from voltage-clamp experiments on the squid giant axon. Its membrane potential is governed by separate sodium, potassium, and leak conductances, each described by gating variables that follow voltage- and time-dependent transition rates. Although more parameter-intensive, this scheme remains an essential tool for replicating neuronal action potentials' shapes, thresholds, and frequency responses. It can also be adapted to incorporate multiple channel subtypes or altered gating parameters ([Skocik and Long, 2014](#)). Beyond single-compartment usage, the Hodgkin–Huxley formalism commonly appears in multi-compartment neuron models, where each compartment has its own channel dynamics. Because it tracks four or more separate state variables and uses multiple exponentials or lookup tables, Hodgkin–Huxley can require anywhere from a few hundred to well over a thousand flops/ms to achieve faithful spike timing. This added cost buys far more physiological detail, enabling accurate reproduction of real spike shapes and voltage-dependent channel behaviors ([Izhikevich, 2004](#); [Skocik and Long, 2014](#)). Further biological complexity can be introduced by using multiple nonlinear conductances of the Hodgkin-Huxley type modeling a variety of distinct ion channel families with different ionic species (sodium, potassium, calcium, etc.), kinetic properties (fast, slow, transient, persistent, etc.), and gating mechanisms (voltage, ligand, etc.) ([Gouwens et al., 2018](#)).

Expanding Hodgkin–Huxley to a multi-compartment framework allows modeling of dendritic branches, axonal initial segments, and other spatial structures. These models discretize the neuron's geometry and, via the axial-current term derived from classical cable theory, couple neighbouring segments so that voltage in one piece can influence the next. Once active conductances are added to each segment, this framework can capture phenomena such as local dendritic Ca²⁺ or NMDA spikes, axonal back-propagation (Rama et al., 2018), and region-specific channel gradients. While multi-compartment models certainly can involve dozens or even thousands of compartments (Herz et al. 2006), it is thought that in many situations only a small handful, for instance, two to five, are needed to capture soma-dendrite interactions or back-propagating action potentials (Carlsmith 2020). This extra complexity enables phenomena like dendritic coincidence detection, local dendritic plateau potentials, or intricate backpropagation, which can shape the cell's coding properties in ways that single-compartment models cannot capture. Thus, large compartmental expansions represent a relatively flexible choice: increasing compartments and parameters can boost realism in dendritic computations, at the cost of heavier per-neuron memory usage and computational flops.

At the finest scale, molecular dynamics (MD) simulates individual atoms to understand ion channel mechanisms like permeation and gating (Roux, 2002; Alberini et al., 2023; Guardiani et al., 2022). This offers far better biophysical detail but comes at an extreme computational cost. This cost limits typical simulations to short timescales, often just pico- to nanoseconds, extending to only a few microseconds even on high-performance hardware (Guardiani et al., 2022). This poses a significant challenge, as critical functional processes like channel gating or ligand binding frequently occur on much longer, millisecond-to-second timescales (Guardiani et al., 2022). The extreme computational cost also severely limits the spatial scale feasible for MD simulations. Although there has been progress thanks to both GPU acceleration (Schoenholz et al, 2019; Doerr et al, 2021) and to dedicated hardware like the Anton 3 supercomputer (Shaw et al., 2021), it currently seems unrealistic to directly emulate whole brains or even a single neuron at this level of detail for functionally relevant durations. Consequently, prospective emulations will likely not employ MD directly to simulate neural activity. Instead, its primary contribution could be indirect: providing fundamental biophysical data to inform the construction and parameterization of the more computationally efficient, higher-level neuron models required for brain-scale simulation.

The choice of neuron model often reflects a specific project's target level of biological abstraction, as different models capture varying degrees of underlying function. However, perfect replication down to the atomic level at whole-brain scale will plausibly remain infeasible for the foreseeable future. Furthermore, even within the usable range of models, determining the most computationally efficient approach for a given target level of accuracy is complex; different studies reach varying conclusions based on specific benchmarks, accuracy metrics, and implementation choices (Izhikevich, 2004; Skocik and Long, 2014, Valadez-Godinez et al., 2019). Given these uncertainties and the significant cost associated with simulating unnecessary biological detail (Guardiani et al., 2022), the field must determine *empirically* what model complexity is necessary and sufficient to produce functionally faithful emulations. Resolving this challenge will not just guide model development but also define data collection and compute resource requirements, ultimately determining the viability of achieving functionally accurate brain emulations for any given level of funding.

## Synapse Models

Synapses transmit signals between neurons, exhibit short-term dynamics based on recent activity, and undergo long-term changes during learning. Computational models of synapses vary from simple rules to detailed biophysical simulations, with different models emphasizing different aspects of synaptic function.

The simplest models treat each synapse as a fixed weight: presynaptic spikes cause instantaneous jumps in postsynaptic current. While computationally efficient, this ignores that real synapses produce responses that rise and decay gradually over time. More realistic models generate a transient response for each spike, using either an alpha function or a difference of exponentials to capture this time course ([Roth and van Rossum, 2009](#)). These waveforms can be applied either to conductance (capturing voltage-dependence but requiring more computation) or directly to current. Although these temporal response functions – functions that describe how conductance or current evolves – in their basic form they still abstract away receptor kinetics and potentially ignore voltage dependence. For this reason, a common extension is to include an NMDA component with a voltage-dependent $Mg^{2+}$, typically as a multiplicative factor on the conductance. Going beyond these relatively simple kinetic schemes, more detailed models use kinetic schemes tracking multiple receptor states (closed, open, desensitized), while the most complex simulations also model neurotransmitter diffusion and geometric effects in the synaptic cleft. However, such detailed diffusion models are typically too computationally demanding for network simulations and are more commonly used to study single synapses ([Destexhe et al., 1998](#)).

The models described above capture instantaneous synaptic responses, but real synapses show activity-dependent changes over milliseconds to seconds, exhibiting either depression (decreased effect) or facilitation (increased effect) with repeated activation ([Citri and Malenka, 2007](#)). Two main phenomenological models capture these effects: the Tsodyks-Markram model treats synapses as having a pool of resources that deplete with use and recover over time, while Abbott et al's model directly modifies release probability ([Tsodyks et al., 1998](#); [Abbott et al., 1997](#)). Both models can fit experimental data well despite their simplicity. More mechanistic variants, however, can track multiple vesicle pools, presynaptic calcium, and other factors that drive more complex short-term dynamics.

Whereas short-term changes fade within seconds, synapses also undergo longer-lasting modifications that can persist for hours or days. Such long-term plasticity is thought to underlie learning and memory, and is typically modeled through Hebbian-like rules. Basic Hebbian models strengthen synapses when pre- and postsynaptic neurons are active together, but require stabilizing modifications to prevent runaway growth. Spike-timing-dependent plasticity (STDP) implements this principle at the level of individual spikes: synapses strengthen when presynaptic spikes precede postsynaptic ones by tens of milliseconds and weaken for the reverse order. Still, straightforward STDP rules cannot capture frequency-dependent or burst-dependent effects. More advanced models address these nuances through multi-spike interactions (Pfister and Gerstner, 2006) or by simulating how calcium influx drives synaptic changes (Shouval et al., 2002).

## Model Fitting and Data-Driven Approaches

Regardless of a model's complexity, its predictive usefulness depends on how effectively its parameters are constrained by experimental measurements (Almog and Korngreen, 2016). Historically, most neuron models were constrained by manually tuning a few parameters (e.g., leak conductance, threshold, or channel densities) to match qualitative observations, such as a neuron's typical firing frequency or spike waveform. Modelers would "hand-fit" the neuronal behavior by trial-and-error until the simulated voltage traces or firing rates closely resembled a reference dataset. Although workable for simple scenarios, this approach often fails to generalize and can obscure parameter degeneracies – different parameter sets may produce similar outputs without revealing which ones are biologically correct.

The effort to more rigorously constrain neuron models has been significantly advanced by detailed electrophysiological recordings, particularly from single neurons in slice preparations. Building on this, early automated optimization strategies emerged that did not require model differentiability. For instance, Druckmann et al. introduced a multi-objective framework using genetic algorithms to fit conductance-based models by comparing multiple electrophysiological features (e.g., spike rate, action potential shape) from simulations to the mean and standard deviation of those features in experimental recordings (Druckmann et al., 2007). The application of such stochastic optimization techniques, including genetic algorithms, was made more accessible and standardized by software packages like BluePyOpt (Van Geit et al., 2016), and these genetic algorithm-based approaches enabled the systematic generation of large libraries of biophysically detailed models (Gouwens et al., 2018). More recently, simulation frameworks supporting automatic differentiation have gained traction, also frequently utilizing such detailed electrophysiological data. When a model's equations are fully or partially differentiable (as Hodgkin–Huxley usually is), gradient descent, similar to training methods underlying modern AI systems, can be applied to minimize the mismatch between recorded and simulated activity systematically. This has motivated recent work on differentiable simulators that can backpropagate errors from final spike outputs or time-varying voltage traces through an entire model, yielding parameter sets consistent with optical or electrophysiological data ([Deistler et al., 2024](#)). In other cases where discontinuities or non-smoothness arise (e.g., certain spiking resets), generative methods can learn to map recordings to model parameters without needing direct backpropagation through each spike event.

For synaptic models, patch-clamp electrophysiology remains the gold standard for parameter fitting. Single-cell recordings characterize basic synaptic transmission parameters like postsynaptic current kinetics and receptor properties, while paired recordings or optogenetic stimulation reveal presynaptic release probability and short-term plasticity dynamics. Studying long-term plasticity poses additional challenges, as these changes unfold over hours to days and involve complex molecular cascades that are difficult to monitor in living tissue while maintaining sufficient temporal and spatial resolution.

While such approaches benefit from biologically informed model architectures (e.g., known ion-channel structure or a leaky-integrate-and-fire scheme), one can also relax these biophysical priors entirely and rely on purely data-driven methods. Large recurrent or feedforward neural networks – trained end-to-end on high-dimensional time-series data – can, in principle, capture dynamics that conventional equations might omit ([Wang et al., 2025](#)). However, this freedom demands enormous datasets, with recent scaling analyses suggesting that even for *C. elegans*, almost an order of magnitude more neural recordings than currently exist would be required for highly predictive models ([Simeon et al., 2024](#)). Nonetheless, in smaller, relatively stereotyped organisms where single-neuron resolution data could be collected in large quantities, purely "black-box" models with minimal biophysical assumptions could at least theoretically thrive.

## Structure to function

For larger mammalian brains, though, comprehensive, whole-brain functional data remain beyond present technology, making it impractical to learn all parameters directly from raw observations. Consequently, biophysical priors – such as those embedded in multi-compartment Hodgkin–Huxley or simpler spiking models – remain crucial to reduce parameter space and keep simulations biologically plausible. Yet even these more traditional approaches struggle to scale up to billions of cells; far more extensive measurement capabilities would be required to fit every neuron and synapse.

A promising strategy for bridging the structure-to-function gap is to develop generative models that link morphological and molecular data to neuronal function. In smaller, more experimentally tractable organisms such as *C. elegans* or larval zebrafish, one can gather both a molecularly annotated connectome and rich functional data (e.g., through optogenetic perturbations and whole-brain recordings). These models learn how specific structural attributes translate into core biophysical parameters like channel conductances or short-term plasticity kinetics by correlating neuronal responses with ultrastructural features, such as synapse size, receptor distributions, and dendritic geometry. Once validated in systems with abundant ground-truth data, the same approach can be extended to large mammalian brains, where whole-brain functional measurements remain infeasible. In this context, Holler et al stands as a landmark demonstration of how morphological information alone can inform functional predictions: by combining slice electrophysiology with electron microscopy, they showed that synapse size reliably predicts synaptic strength in mouse somatosensory cortex ([Holler et al., 2021](#)). Their work exemplifies how structural features can serve as powerful proxies for synaptic function. Incorporating molecular-level data into such pipelines would enable generative frameworks to infer parameters with even greater precision and scale, providing a viable path toward biologically grounded simulations of large neural circuits. The necessary condition for that is an extensive database of slice synaptic electrophysiology data, such as [Campagnola et al, 2022](#). In this way, integrating high-resolution structural data with targeted functional measurements can progressively narrow – and ultimately bridge – the gap between brain structure and brain function. This will require substantial investments in such aligned datasets.

## Embodiment

Brains do not function in isolation. They exist within bodies that provide sensory input and execute motor commands, creating closed sensorimotor loops that drive behavior and ground identity. This interdependence raises several questions for whole brain emulation. Firstly, is embodiment (providing the emulation with sensory inputs and motor outputs linked to an environment) a prerequisite for faithful emulation? Secondly, if embodiment is necessary, what level of fidelity is required? Is a generic, functionally adequate body sufficient, or must the emulation replicate the original body's specific physical characteristics and biochemical properties to preserve personal identity and memories?

Arguments supporting the necessity of embodiment, potentially at high fidelity, often point first to the consequences of disrupting brain-body interaction. Functional deficits following sensory or motor deprivation, for instance, are frequently cited to underscore the requirement for continuous interaction simply to preserve long-term stable brain function (Sandberg and Bostrom, 2008). Beyond this need for basic interaction, the case for replicating the specific original body posits that crucial aspects of identity are deeply intertwined with individual physical characteristics. From this perspective, learned motor skills and reflexes are often precisely tuned to the unique biomechanics of the original body (such as limb dimensions, mass distribution, and muscle properties) through adaptations occurring in both the brain and spinal cord (Shadmehr and Mussa-Ivaldi, 1994). The concern raised by this view is that transferring an emulation to a generic or different body might invalidate these learned abilities (Sandberg and Bostrom, 2008). This viewpoint further highlights that the brain's ongoing state and subjective experience are constantly modulated by the body's hormones, metabolites, and biochemistry more generally. The implication suggested is that accurately capturing an individual's baseline temperament, drives, and even subjectively essential aspects of identity requires simulating this precise internal chemical environment (Sandberg and Bostrom, 2008; McKenzie, 2022). Thus, the conclusion drawn from this rationale is that while some form of embodiment might address basic functional viability, a faithful emulation could require reconstructing the original body with sufficient accuracy (Sandberg and Bostrom, 2008).

Counterarguments emphasize the centrality of the brain for identity. This perspective often distinguishes between enduring personal identity, rooted in long-term memories and personality traits presumed encoded within the brain's structure and dynamics, and transient physiological states, such as hormonal influences or mood fluctuations, primarily mediated by the rest of the body (McKenzie, 2022). This view supports the observation that significant body portions, including limbs, organs, and even sensory apparatus (such as cochleas or retinas), can be functionally replaced or lost without fundamentally erasing personal identity or core memories (McKenzie, 2022). Further, the topographic mapping of sensory and motor functions within the brain provides a plausible pathway for interfacing an emulation with sufficiently structured sensory inputs and motor outputs without the original peripheral structures. Thus, this position suggests that an 'adequate' body simulation needs only to provide a functional feedback loop with the environment, provided that the brain emulation can adapt (for example, through emulated neuroplasticity).

Encouragingly, the development of the simulation tools needed to investigate these questions and provide embodiment for WBE is advancing. Dedicated open-source platforms like OpenSim from the biomechanics community offer increasingly sophisticated capabilities for detailed musculoskeletal modeling and analysis (Seth et al., 2018). Concurrently, the field is increasingly leveraging progress in high-performance physics engines such as MuJoCo (Todorov et al., 2012; Vaxenberg et al., 2024; Wang-Chen et al., 2024). Initially driven by robotics and reinforcement learning, these engines prioritize simulation speed and robust contact handling, while offering features like GPU acceleration and differentiability, both highly valuable for complex embodied systems. Although significant challenges persist, continued progress in these areas holds the promise of enabling researchers to finally settle these philosophical debates surrounding embodiment empirically.

## Verification

The history of technological advancement is often marked by the establishment of benchmarks. In artificial intelligence, the introduction of ImageNet ([Deng et al., 2009](#)) and its associated ILSVRC competition revolutionized the field by providing a standardized dataset and objective metrics for comparing models. While these benchmarks did not settle debates about the nature of "intelligence," they enabled measurable progress by focusing on a tangible proxy: classification accuracy. Similarly, the pursuit of emulating biological brains requires rigorous methods to evaluate success. The field risks stagnation without agreed-upon benchmarks, with progress hindered by subjective or incompatible criteria. Standardized evaluations – even imperfect ones – are crucial for aligning efforts, tracking advancements, and accelerating discovery, much as ImageNet catalyzed the rise of deep learning. The central challenge, therefore, is to design benchmarks that capture measurable aspects of neural or functional fidelity, driving progress until their limitations necessitate the next leap forward.

A foundational framework for evaluation was proposed in Sandberg and Bostrom's Whole Brain Emulation Roadmap ([Sandberg and Bostrom, 2008](#)). They distinguished between simulations, which replicate outputs, and emulations, which replicate internal dynamics, arguing that valid models must preserve the brain's causal structure – the step-by-step relationships between neural states. However, practical challenges abound. Even setting aside the brain's potential chaotic dynamics, minor discrepancies – stemming from noise in neural recordings ([Rupprecht, 2021](#)) or parameter inaccuracies – can compound over time, causing simulations to diverge from biological trajectories. To address this, Sandberg and Bostrom suggested tolerating deviations smaller than the brain's inherent "noise floor," the variability observed across repeated biological trials. This pragmatic approach underpins one of the field's most advanced benchmarks, ZAPBench ([Lueckmann et al., 2024](#)), which evaluates larval zebrafish emulations by predicting 30 seconds of neural activity from 10 seconds of observed calcium traces (sampled at 1 Hz). Models are scored on per-neuron mean absolute error (MAE) across over 70,000 neurons.

An alternative framework shifts the focus from neural activity to behavioral indistinguishability. Zador et al. propose an "embodied Turing test": if a virtual animal navigates novel environments, responds to threats, or learns tasks indistinguishably from its biological counterpart, the emulation succeeds – regardless of internal mismatches ([Zador et al., 2023](#)). This mirrors Turing's original vision, replacing conversation with sensorimotor behavior. For instance, an artificial beaver might be tested on dam-building, or a simulated fruit fly on evasive maneuvers during flight. However, this framework remains comparatively underdeveloped, existing more as a conceptual proposal than a fully fleshed-out benchmark.

The lack of consensus in brain emulation mirrors debates in AI. Just as no single metric – ImageNet accuracy, chess Elo ratings, or language modeling capability – fully captures "intelligence," no single benchmark will resolve what defines a successful brain emulation. A promising approach, then, is to develop benchmark suites that aggregate diverse tasks. Initiatives like BIG-Bench ([Srivastava et al., 2022](#)) and HELM ([Liang et al., 2022](#)) evaluate language models across hundreds of scenarios, recognizing that no individual task tells the whole story. A parallel and influential effort in computational neuroscience is Brain-Score.org ([Schrimpf et al., 2020](#)), which provides an integrative platform for evaluating how well computational models, particularly those of the visual system, align with a diverse array of neural and behavioral benchmarks. By scoring models on their ability to predict brain activity (e.g., in visual cortical areas) and match primate behavioral performance, Brain-Score aims to drive progress towards more neurally mechanistic models of specific intelligent functions. For whole brain emulation, a similar suite might combine neural activity prediction (as in ZAPBench), embodied behavior tests, causal perturbation experiments (e.g., optogenetic interventions), and new innovative metrics that the community will develop. Indeed, while benchmark suites and, in general, the portfolio approach to evaluation discussed here provide a more comprehensive evaluation framework, significant work remains to refine and expand the evaluation landscape. Nonetheless, by iteratively improving and expanding benchmarks, the field can advance objectively, grounded in empirical progress rather than abstract debates.

As part of this report, simulation attempts discussed for different organisms were rated on the following 0-3 point scale across 10 dimensions—the dimensions gesture at the breadth of sophisticated benchmarks that eventually need to be developed.

# Hardware requirements

Before running a brain emulation, models often need to be fitted or trained to match experimental data, which can be computationally intensive. The resources required for this fitting phase depend heavily on numerous factors, including the complexity of the model being fitted (e.g., number of parameters, compartments), the volume and type of data used as constraints, and the specific optimization algorithms employed. Due to this high degree of variability and a lack of standardized benchmarks or comprehensive references quantifying these costs across different scenarios, providing reliable general estimates is currently challenging. Therefore, although model fitting is a critical and potentially resource-intensive step, this report will focus on the computational demands of *running* brain emulations, rather than fitting them.

Turning to the demands of running brain emulations, hardware requirements remain radically uncertain, primarily because we don't yet know what level of biological detail is necessary for faithful emulation. The Whole Brain Emulation Roadmap (Sandberg and Bostrom, 2008) highlighted this uncertainty, with computational demands for a human brain ranging from $10^{15}$ to $10^{30}$ FLOPS: a 15-order-of-magnitude difference. Storage requirements similarly varied from $5 \times 10^{15}$ to $1.12 \times 10^{28}$ bits. Precise resource requirements will remain elusive until a consensus emerges on what constitutes a sufficient level of biophysical or functional fidelity.

What is certain is the dramatic advancement in computational power since the roadmap's publication in 2008. Total compute in supercomputers has been rising exponentially for decades, and supercomputers in the 2020s have been reaching exaflop territory, e.g., with El Capitan at more than 1.5 exaflops (Thomas, 2024), with typical supercomputer costs of $600M (Joseph, 2023). Further, it's worth noting that clusters used for machine learning applications typically do not appear in TOP500 and that, as a result, the computational capacity of state-of-the-art systems likely exceeds these values. For example, in 2023 Google demonstrated the ability to train language models on a cluster of 50,944 TPUv5e chips (for a peak performance of over 10 FP16 exaFLOPs) (Ananthamaran, 2023), and although no public benchmarks have been reported so far, the 100K H100 Colossus cluster recently assembled by xAI would at least theoretically be capable of over 100 FP16 exaFLOPs, with further plans to roughly double the size of the cluster in the coming months (Mantel, 2024).

What kind of emulations could such systems run? To provide concrete, though highly simplified, reference points, we analyze two scenarios: a "simplified lower-bound" scenario, characterized by leaky integrate-and-fire (LIF) neurons and alpha-function synapses, aiming to capture minimal spiking behavior, and an "illustrative moderately complex" scenario, characterized by five-compartment Hodgkin-Huxley (HH) neurons and Tsodyks-Markram synapses, likely capable of capturing dendritic computation ([Carlsmith, 2020](#)) and short-term synaptic plasticity, respectively. It is crucial to emphasize that our "moderately complex" scenario is still far from the level of detail that might ultimately be required for full biophysical realism; for instance, highly detailed models could necessitate hundreds of compartments per neuron, or potentially even more complex representations whose requirements are currently an open research question (Shuhersky, 2024, personal communication). Such per-neuron or per-synapse complexity increases would lead to proportionally larger storage and computational demands than estimated here. Furthermore, it's important to note that neither of our defined scenarios represents a true upper bound on computational requirements, as both omit other potentially crucial mechanisms like neuromodulation and glial interactions. They also do not involve long-term plasticity mechanisms like STDP and LTP, leading to emulations that are not capable of forming new memories. Nonetheless, these reference points bracket many modeling efforts today and can offer some insight into the computational requirements of emulations involving different organisms. We follow Skocik and Long's operation-counting approach ([Skocik and Long, 2014](#)), and extrapolate to whole brain compute requirements (using whole-brain estimates for mouse and human, though obtaining reliable whole-brain synapse counts can generally be challenging) – compute requirements based on known or estimated neuron and synapse counts (leaving larval zebrafish aside, for which we could not identify reliable synapse count estimates).

As part of these simplified estimates for time-based emulation compute requirements, assuming FP16 operations, we find that both *C. elegans* and the fly brain are within reach of modern GPUs, that the mouse brain would likely require a relatively small multi-GPU cluster, and that emulating the human brain would require a large-scale cluster – indeed, Lu et al. recently used 14,000 GPUs for their human-scale model ([Lu et al., 2024](#)) – though not beyond the capabilities of frontier systems like xAI's Colossus. It's important to emphasize, however, the limitations of these estimates. These FLOPS estimates represent the theoretical computational load of the core model equations. Actual performance on hardware will also depend on factors such as software implementation efficiency, memory bandwidth limitations, interconnect speeds, and hardware utilization, which can lead to practical simulation times differing from what raw FLOPS might suggest. Thus, these estimates neglect other potentially important mechanisms by focusing solely on neuron and synapse models. Further, even within this modeling paradigm, it's worth noting that the literature on computational requirements for synapse models is far less extensive than for neuron models, despite synapses often dominating computational costs. Finally, these estimates do not consider numerous other relevant factors in actual modeling efforts, including the specific simulation software used, the achieved hardware utilization rate, and more.

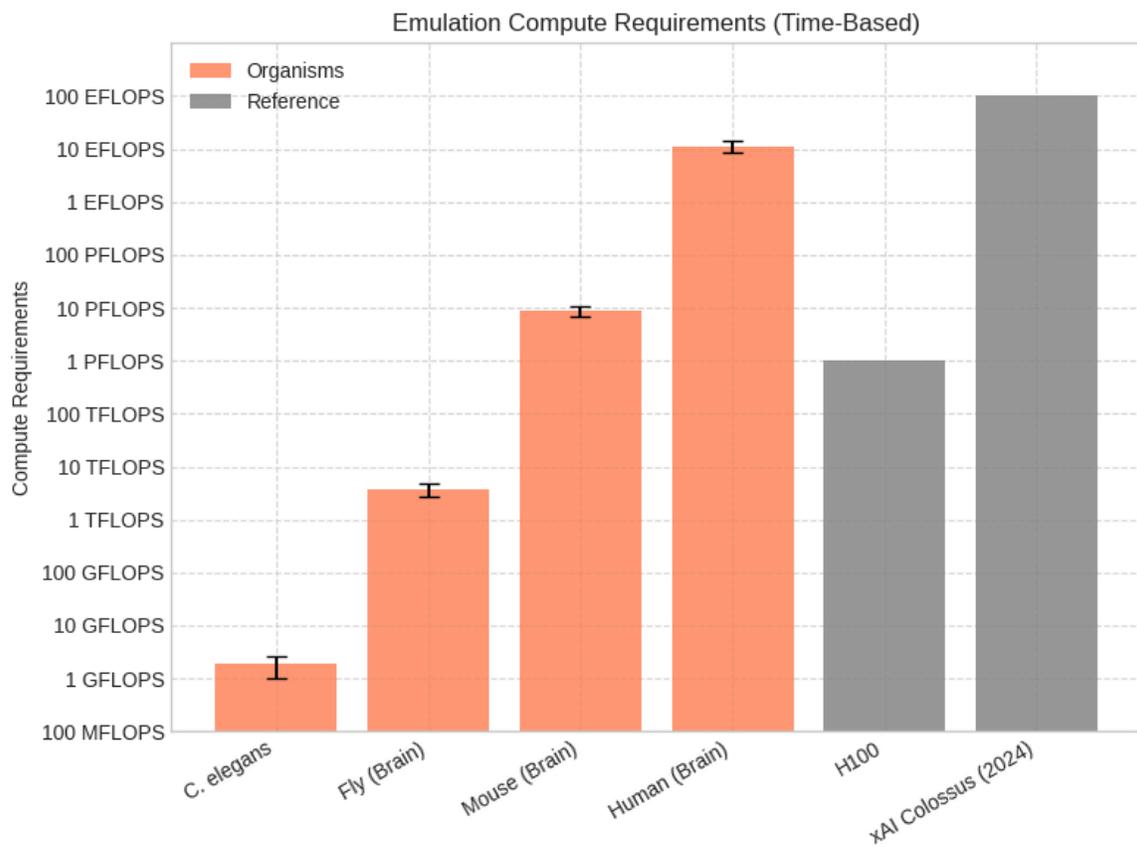

**Figure X: Estimated Compute Requirements for Time-Based Brain Emulation.** The computational power (FP16 FLOPS) required for *time-based* whole-brain emulations across four organisms (orange) compared with the peak performance of reference hardware (grey). The y-axis is on a logarithmic scale. Error bars reflect the range of requirements based on varying assumptions about the underlying biophysical model complexity (e.g., from simple point neurons to multi-compartment models). Calculations and underlying data for this figure are available in the [linked data repository](linked data repository).

Indeed, one reason why it can be hard to achieve high hardware utilization rates is storage requirements. The storage requirements estimated here refer to the memory needed to hold the dynamic state variables of the neurons and synapses during active simulation, typically residing in system RAM or the High Bandwidth Memory (HBM) of accelerators like GPUs. Storage requirements can exceed the high bandwidth memory of the system, and even when memory itself is sufficient, the interconnect between different chips can be a bottleneck. Over the past 30 years, the rate of improvement in processing power has far exceeded the rate of improvement in memory and interconnect development. Peak hardware FLOPS have improved by roughly 3x every 2 years, compared to only 1.6x and 1.4x for memory and interconnect bandwidth, respectively, leading to the so-called memory wall ([Gholami et al., 2024](); [An et al., 2024]()). Memory capacity has also undergone a similar trend: today's El Capitan boasts 1.74 exaFLOPS and over 5.4 petabytes of HBM3 memory ([Thomas, 2024]()); NEC's Earth Simulator boasted 41 teraFLOPS and 10 terabytes of DRAM memory in 2004 ([Sato, 2004]()); representing an approximately 42,400-fold increase in processing power and just a 540-fold increase in memory capacity. Like many other workloads, computational modeling of brain tissue has been affected by this trend: a 2014 study identified interconnect bandwidth as key bottleneck for spiking neural network simulations ([Kunkel et al., 2014]()) and a recent in-depth analysis found memory bandwidth and interconnect latency to represent key bottlenecks for all types of neuron simulations studied, from point neurons with current-based synapses to multicompartmental models with conductance-based synapses, particularly at high fan-ins and neuron counts ([Cremonesi et al., 2020]()).

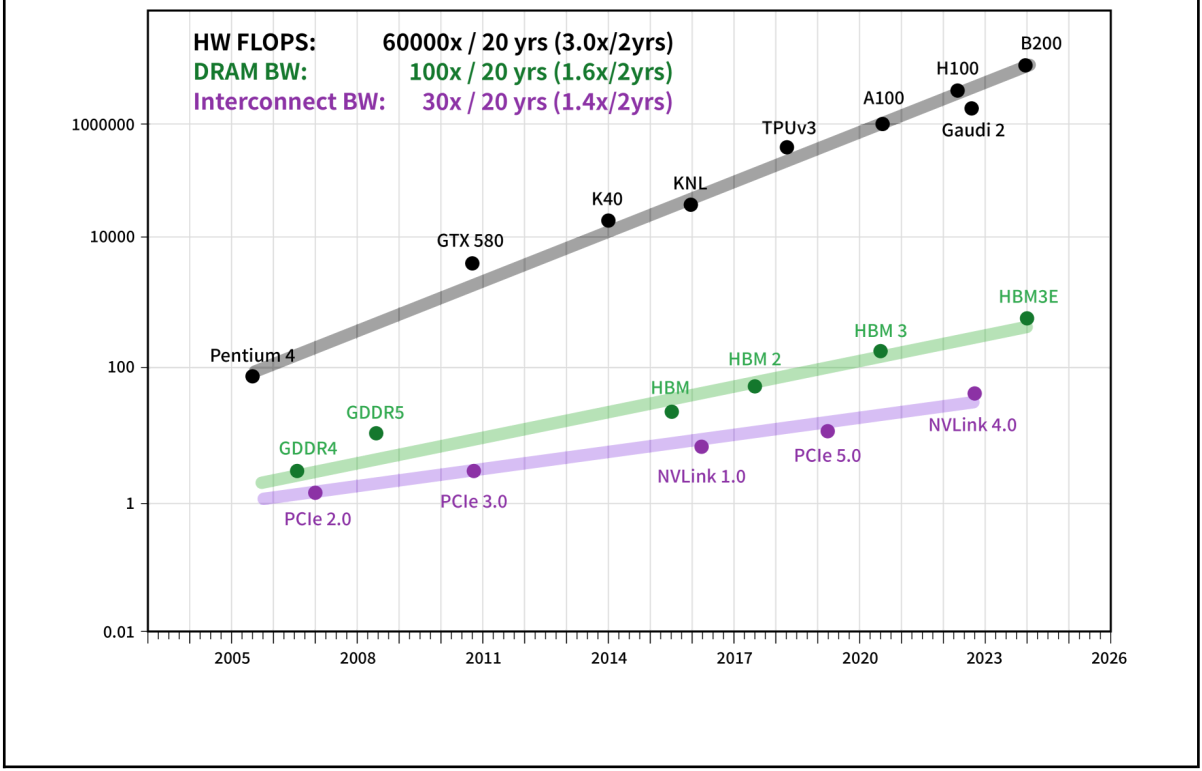

**Figure X: Divergent Growth Rates of Compute, Memory, and Interconnect Performance.** Exponential growth trends for hardware performance over the past two decades, plotting peak computational throughput (FLOPS, black), memory bandwidth (DRAM BW, green), and interconnect bandwidth (Interconnect BW, purple). The y-axis is logarithmic, with performance values normalized to the peak FLOPS of the [R10000](#) processor. The steeper slope for FLOPS (3.0x improvement every two years) compared to memory (1.6x) and interconnect (1.4x) bandwidth illustrates the growing gap between processing power and data access speeds, a phenomenon known as the "memory wall."

Similarly to the compute requirements discussed above, our simplified estimates suggest that, assuming 32-bit state variables, *C. elegans* and *Drosophila* brain emulations would fit comfortably within a modern GPU's available HBM. For an emulation of a mouse brain in our "illustrative moderately complex" scenario (5-compartment HH), it would likely require a somewhat small cluster, and an emulation of the human brain in the same scenario would require a large-scale cluster close in capability to ones such as xAI's Colossus. The precise storage per element for more complex models will vary based on the specific ion channels and state variables included. These estimates, however, make the important assumption that interconnects do not become the bottleneck, an assumption that may not hold in practice. Data movement and latency bottlenecks are now the primary constraint in training frontier AI systems, the main application of today's largest GPU/TPU clusters (Erdil and Schneider-Joseph, 2024). It's also worth noting the stark contrast in software maturity: considerable effort has gone into developing strategies for efficient distributed training of large-scale AI systems over the past decade, while computational neuroscience has only recently begun to leverage GPUs and TPUs at scale. Indeed, while a few pioneering efforts involved large-scale GPU clusters, most computational neuroscience research remains focused on single-GPU or small-cluster implementations where interconnect bottlenecks are less prominent.

**Figure X: Estimated Memory Requirements for Brain Emulation.** The runtime memory (Bytes) required to store the state of whole-brain emulations across four organisms (light blue) compared with the memory capacity of reference hardware (grey). The y-axis is on a logarithmic scale. Error bars reflect the range of requirements based on the same varying assumptions about biophysical model complexity used for compute estimates. Calculations and underlying data for this figure are available in the linked data repository.

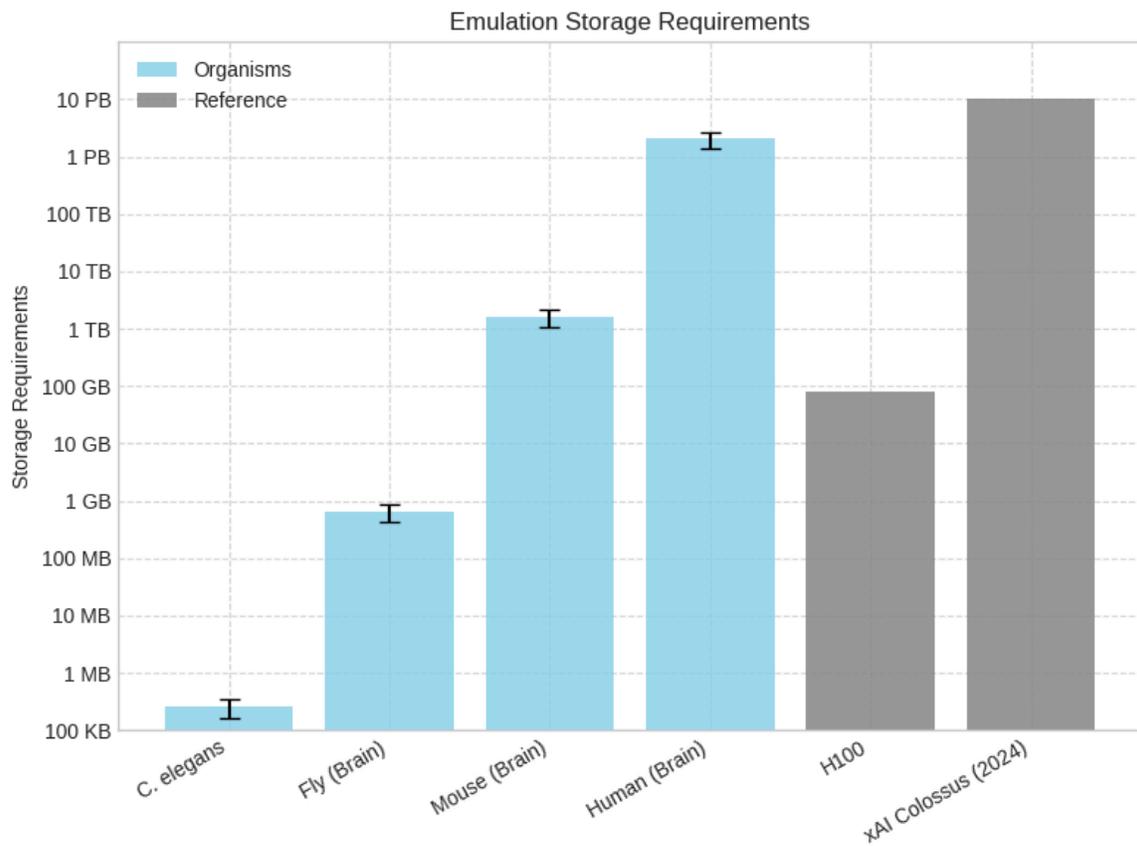

Interconnects can become the bottleneck partly because time-based simulations require constant updates for all neurons and synapses, typically every 0.1 milliseconds. An event-driven modeling paradigm can lower bandwidth demands by instead only requiring such computations when spikes and synaptic events happen, thus also decreasing compute requirements by a couple orders of magnitude. This is typical of neuromorphic systems like IBM's TrueNorth in 2014 (implementing 1 million neurons and 256 million synapses per chip) ([Akopyan et al., 2015](#)), Intel's Loihi platforms ([Davies et al., 2018](#), [Orchard et al., 2021](#)) and more recent developments like the Hala Point system at Sandia National Laboratories, with over 1.15 billion neurons and 128 billion synapses ([Intel, 2024](#)). Support for some event-driven operations is also present in some GPU/TPU-based frameworks like BrainPy ([Wang et al., 2023](#)), though this is still far from common. Should event-driven modeling be more widely adopted, not every model would necessarily see significant benefits. Indeed, some, like Hodgkin-Huxley neurons, can be difficult to adapt due to their continuous dynamics.

Furthermore, the efficiency gains of event-driven simulations can diminish in large, densely connected networks with high overall spike rates. In such regimes, the overhead of managing a vast number of concurrent events might negate the computational savings from only updating active elements. Nonetheless, as processing of synaptic events dominates computational requirements, the computational cost of emulating an entire brain can still decrease by a few orders of magnitude.

**Figure X: Estimated Compute Requirements for Event-Driven Brain Emulation.** The computational power (FP16 FLOPS) required for the same whole-brain emulations, but implemented using an *event-driven* paradigm (green bars), compared with reference hardware (grey). The y-axis is on a logarithmic scale. Error bars reflect the range of requirements based on the same varying assumptions about biophysical model complexity used for the other estimates. Calculations and underlying data for this figure are available in the [linked data repository](linked data repository).

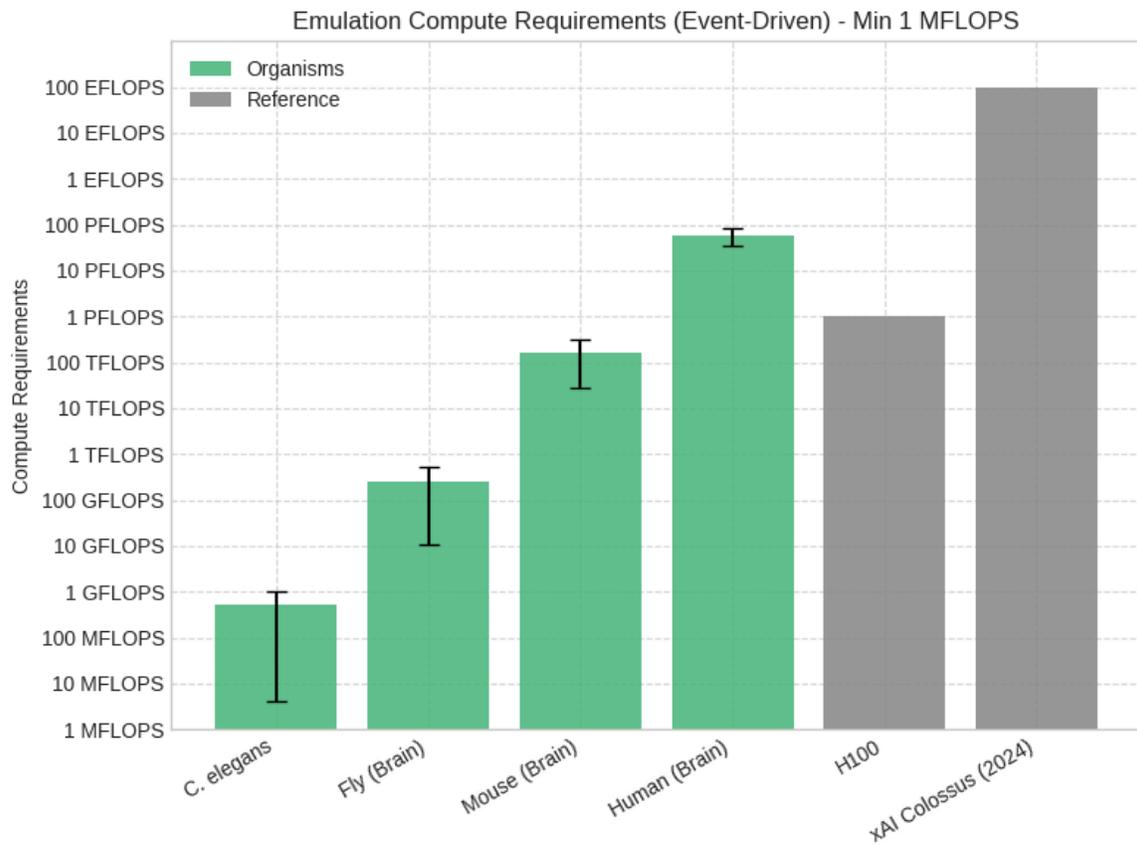

This can be seen in our estimates. Even in the more complex scenario, where the 5-compartment Hodgkin-Huxley neurons continue to be time-based, overall computational requirements still decrease by several orders of magnitude for all organisms considered, as a result of the more efficient event-driven modeling of synapses. It is essential, however, to contextualize these event-driven estimates within the current framework landscape. While GPUs and TPUs are now starting to be adopted by the computational neuroscience research community, enabling significant speedups over traditional CPU-based platforms like NEURON, and powering new frameworks such as ARBOR ([Akar et al., 2019](Akar et al., 2019)), BrainPy ([Wang et al., 2023](Wang et al., 2023)) and Jaxley ([Deistler et al., 2024](Deistler et al., 2024)), their adoption is a relatively recent phenomenon, and remains limited. Further, event-driven primitives continue to be mostly lacking. Nevertheless, these trends suggest that, with sufficient research and development, even real-time emulation of the human brain may be feasible in the not-too-distant future, although likely requiring, at least initially, large-scale clusters.

# Author Contributions

The initial draft of the report was based on research conducted by Maximilian Schons, Isaak Freeman and Niccolò Zanichelli. Further development through investigations by Maximilian Schons and Niccolò Zanichelli. Philip Shiu and Anton Arkhipov provided guidance throughout the process. All authors contributed research, writing, editing, or feedback across chapters. Maximilian Schons led the activities and coordinated the review and finalization of the report.

Niccolò Zanichelli and Maximilian Schons are shared First-Authors.

The authors acknowledge Fieldcrest Foundation and Foresight Institute for supporting the technical report through funding to Isaak Freeman, Maximilian Schons, and Niccolò Zanichelli.

## Competing Interests

Philip Shiu is an equity holder in Eon Systems PBC.

Data:  State of Brain Emulation Report 2025 Data Repository